\documentclass[a4paper,fleqn,usenatbib]{mnras}
\usepackage{times}
\usepackage[authoryear]{natbib}
\usepackage{url}
\usepackage{graphicx}
\usepackage[fleqn]{amsmath}
\usepackage{aas_macros}
\usepackage{afterpage}
\usepackage{xparse}
\usepackage{float}
\usepackage{subfig}
\usepackage{caption}

\usepackage{enumitem}

\NewDocumentCommand{\myproot}{O{} O{-2} O{2}  m}{\sqrt[\leftroot{#2}\uproot{#3}#1]{#4}}  

\def \simless {\mathbin{\lower 3pt\hbox{$\rlap{\raise 4pt
              \hbox{$\char'074$}}\mathchar"7218$}}}
\def \simgreat {\mathbin{\lower 3pt\hbox{$\rlap{\raise 4pt
              \hbox{$\char'076$}}\mathchar"7218$}}}
\def\ie{{\it i.e.,}}
\def\cf{{\it c.f.,}}
\def\eg{{\it e.g.,}}
\def\CSF{{\it Cool+SF}~}
    
\makeatletter

\newcommand{\Rmnum}[1]{\expandafter\@slowromancap\romannumeral #1@}
\makeatother
\usepackage{accents}
\usepackage{cite}

\title{The Growth and Enrichment of Intragroup Gas}
\author[L. Liang et al.]{Lichen Liang$^{~1}$, Fabrice Durier$^{~1}$, Arif Babul$^{~1}$, Romeel Dav$\acute{\mathrm{e}}$$^{~2,3,4}$, 
\newauthor  Benjamin D. Oppenheimer$^{~5}$, Neal Katz$^{~6}$, Mark Fardal$^{~6}$ \& Tom Quinn$^{~7}$\\
\\
$^{1}$Department of Physics \& Astronomy, University of Victoria, BC, V8X 4M6, Canada\\
$^{2}$Physics Department, University of Western Cape, Bellville, Cape Town 7535, South Africa\\
$^{3}$South African Astronomical Observatory, PO Box 9, Observatory, Cape Town 7935, South Africa\\
$^{4}$African Institute of Mathematical Sciences, Muizenberg, Cape Town 7945, South Africa\\
$^{5}$CASA, Department of Astrophysical and Planetary Sciences, University of Colorado, Boulder, CO 80309, USA\\
$^{6}$Astronomy Department, University of Massachusetts, Amherst, MA 01003, USA\\
$^{7}$Department of Astronomy, University of Washington, Box 351580, Seattle, WA 98195-1580, USA}
\begin{document}

\date{Accepted 2015 December 2. Received 2015 November 26; in original form 2015 July 07}

\pagerange{\pageref{firstpage}--\pageref{lastpage}} \pubyear{2015}

\maketitle

\label{firstpage}

\begin{abstract}
The observable properties of galaxy groups, and especially the thermal and chemical properties of the intragroup medium (IGrM), provide important constraints on the different feedback processes associated with massive galaxy formation and evolution. In this, the first in a series of studies aimed at identifying and exploring these constraints, we present a detailed analysis of the global properties of simulated galaxy groups with X-ray temperatures in the range $0.5-2$ keV over the redshift range $0\leq z \leq 3$. The groups are drawn from a cosmological simulation that includes a well-constrained prescription for galactic outflows powered by stars and supernovae, but no AGN feedback.  Our aims are (a) to establish a baseline against which we will compare future models; (b) to identify model successes that are genuinely due to stellar/supernovae-powered outflows; and (c) to pinpoint features that not only signal the need for AGN feedback but also constrain the nature of this feedback.

We find that even without AGN feedback, our simulation successfully reproduces the observed present-day group global IGrM properties such as the hot gas mass fraction, the various X-ray luminosity-temperature-entropy scaling relations, as well as the mass-weighted and emission-weighted IGrM iron and silicon abundance versus group X-ray temperature trends, for all but the most massive groups. We also show that these trends evolve self-similarly for $z < 1$, in agreement with the observations.  Contrary to expectations, we do not see any evidence of the IGrM undergoing catastrophic cooling.  And yet, the $z=0$ group stellar mass is a factor of $\sim2$ too high.  Probing further, we find that the latter is due to the build-up of cold gas in the massive galaxies {\it before} they are incorporated inside groups.   This, in turn, indicates that other feedback mechanisms must activate in real galaxies as soon as their stellar mass grows to $M_* \approx$ a few $\times10^{10}\;M_{\odot}$.  We show that these must be powerful enough to expel a significant fraction of the halo gas component from the galactic halos. Gentle ``maintenance-mode" AGN feedback, as has been suggested to occur in galaxy clusters, will not do; it cannot bring the stellar and the baryonic fractions into agreement with the observations at the same time.   Just as importantly, we find that stellar/supernovae-powered winds are vital for explaining the metal abundances in the IGrM, and these results ought to be relatively insensitive to the addition of AGN feedback.
\end{abstract}


\section{Introduction}

In current schema for the formation of observed cosmic structure, galaxies are often identified as the basic building blocks of cosmic structure. Early on, this phrase was meant to indicate that structure in the universe can be understood as gravitationally organized assemblages of galaxies in which individual galaxies are merely passive components, much like bricks in a wall.  However, accumulating multi-wavelength observations and increasingly detailed theoretical studies show that galaxies are anything but passive features of the cosmic landscape. The very processes underlying the formation and evolution of galaxies --- star formation, stellar nucleosynthesis, feedback and galactic outflows --- also impact the wider environment to such an extent that many of the observed properties of supra-galactic systems cannot be understood without reference to these processes. Understanding how these processes unfold and the extent of their impact on galactic and extragalactic scales is essential for constructing a self-consistent description of cosmic structure across the hierarchy as well as for accounting for their observed properties.

Over the years, numerous studies have advanced groups of galaxies as the best environments for studying the impact of galaxies on their surroundings \citep[][and references therein]{R97,F02,PS03,V06,D08,SN09,M10,M11,O14}. In the cosmic hierarchy, galaxy groups are the smallest aggregates of galaxies, with the least massive of these systems comprising only a few luminous galaxies.   What makes these systems especially interesting is that a significant fraction of the baryons attached to galaxy groups exists in the form of hot, diffuse gas that, at least in the case of the more massive groups in the nearby universe, is amenable to scrutiny via X-ray observations. Given the sizes and masses of groups, the expectation is that galactic processes will have affected much of this gas.

Of the various properties, the three features that have attracted the most attention are: 
\vspace*{-2.5mm}

\begin{description}[labelindent=\parindent,leftmargin=0.0cm, style=sameline]

\item[{\rm (i)}] The entropy of the hot diffuse gas within $R_{\rm{500}}$ as measured by the proxy variable $S=k_{B} T_{e}/n_{e}^{2/3}$ \citep[${\cf}$][]{BB99}: This quantity is much better than temperature or density when it comes to encapsulating the time-integrated history of heating and cooling {to which the gas has been subjected.}  \citet{PS03}, \citet{SN09} and \citet{P10} have found that within $R_{\rm{500}}$ the diffuse gas shows clear evidence of enhanced entropy and a growing body of work suggests that this {most likely is due to} non-gravitational heating induced by stellar-powered galactic outflows \citep{EV08,SD08,D08,HQM12,Z14} and/or 
active galactic nuclei (hereafter, AGNs) \citep{BB02,BMS04,M08,P08,SP08,M10,M11,T11,STY12,BMS14,PB14}.

\item[{\rm (ii)}] {The} hot gas fraction within the central regions of the groups: \citet{V06,G07,SN09} find that the hot gas fraction within $R_{\rm{2500}}$ is on the average much lower than that in the more massive clusters of galaxies. A lower hot gas fraction can arise as a result of a number of processes. The hot gas can be depleted by efficient cooling \citep[${\cf}$][]{L00,K05}. However,  this is not a viable explanation for the observations since efficient cooling would also result in stellar fractions that are much higher than observed \citep{DK02}.  A more likely explanation is that the gas, subjected to non-gravitational heating of the kind described in (i), exists in a more extended equilibrium configuration \citep[${\it e.g.}$][]{RC10,M10}.

\item[{\rm (iii)}] The metal content of the hot diffuse gas: The observed iron abundance of approximately $\sim 0.3$ solar, albeit with a large scatter \citep[${\it e.g.}$][]{ES91,TP03,DG04,DP07}, indicates that a significant fraction of the metals produced in galaxies escapes the interstellar medium in these systems. One way of affecting this transfer is via ram-pressure stripping \citep{VK06}. However,  \citet[][hereafter DOS08]{D08} show that this scheme, by itself,  is unable to {\it simultaneously} account for the observed iron abundance and the oxygen-to-iron ratio in the hot diffuse intragroup medium (hereafter, IGrM).  DOS08 conclude that the enrichment of the IGrM  is the outcome of metals being flushed out of the galaxies via powerful galaxy-wide outflows.   The outflows must necessarily be powerful because not only must the winds be carrying a significant fraction of  the metal-enriched gas but their velocities must be large enough to ensure that they ``slip the surly bonds'' of the galaxies' gravity.  Additionally, preliminary studies indicate that a fair fraction of the metals is ejected typically at epochs prior to the formation of the groups themselves \citep{O12,FD14}. 
\end{description}

\vspace*{-4mm}
Outflows are ubiquitous in both local as well as high-redshift galaxies \citep[see][and references therein]{MA05,MA06,SG11,OS12,B13,VM13,W14,T14,VM14,ST14}.   Observations suggest these outflows may be due to either stellar or AGN processes.   Winds powered by AGNs originate as high-velocity outflows on parsec scales \citep{P03b,P03a,T10a,T10b} and while a growing body of observational studies show that these outflows have a profound impact on the gas content in the central $\sim 1$ kpc of the host galaxies  \citep{SG11,VM13,VM14}, evidence suggesting that the AGNs can trigger galaxy-wide outflows capable of flushing the bulk of the metal-enriched, star-forming, interstellar medium (ISM) out of the galaxies remains elusive \citep{H14}.  Recent high-resolution simulation studies \citep[${\cf}$][]{Q12,GB14} {that track} the evolution of the high-velocity nuclear outflows  also find that the resultant winds  have very little impact on the extended galactic disk:  The nuclear outflow transitions into an expanding wind of shocked gas within the central $\sim$1 kpc  of the galaxy, which upon encountering the galactic disk,  follows the path of least resistance and escapes preferentially perpendicular to the disk.  Most of the metal-enriched ISM is generally left unaffected and in place.

In fact, when it comes to disrupting and  ejecting the metal-enriched ISM, stellar-powered outflows, driven by energy and momentum input from supernovae (SNe) as well as photo-heating and radiation pressure from massive stars (\citealp[hereafter MQT05]{NM05};  \citealp{MQT10,KT13}),  are much more promising.   For one,   the wind launch sites are themselves embedded in the ISM.  Additionally, recent high resolution simulations that explicitly account for the full set of stellar feedback processes \citep{HQM12,HK14} confirm that these are more than capable of launching powerful galaxy-wide winds.  Also,  a growing body of observational evidence is not only confirming that this is indeed happening \citep[see, for example,][]{B13,ST14,GH14}, but also find, in agreement with theoretical expectations, that such winds are metal-enriched, can reach velocities $>1000$ km/s, and imply a mass outflow rate that is comparable to the star formation rate.   These theoretical and observational results make a compelling case for stellar-powered outflows being the primary mechanism for the dispersal of metals beyond the galaxies and an integral feature of all realistic models for cosmic structure formation \citep{SD14}.

In a series of papers, Dav\'e and collaborators \citep*{OP06,DF06,FD08,D08} have carried out extensive numerical simulation studies to assess the impact of  the stellar-powered galactic outflows  \textendash{} based on the momentum-driven wind {scalings} of \citet{NM05} \textendash{}  over a wide range of cosmic environments and epochs.    
They find that this model is able to account for a host of observations of which three are especially noteworthy. {First, it} successfully reproduces the observed mass-metallicity relation in galaxies, along with its second-parameter dependence on star formation, both today and at higher redshifts \citep{FD08,RD11,HI13,SD14}.  {Second, it} also explains the observations \citep{O13} indicating the widespread enrichment of the intergalactic medium (IGM) as early as $z\sim 5$ \citep{OP06,O09}
and in fact is, as we will show in a follow-up paper (Durier et al., in preparation),  one of the more successful stellar feedback schemes at doing so.   {Third,} it also yields iron abundances and the oxygen-to-iron ratios ({\it i.e.}$\sim$[Fe/H] and [O/Fe]) in the hot intragroup medium of $z=0$ groups that broadly match the observations \citep{D08}.   An alternative to the momentum-driven model for the kinetic winds is the energy-driven model, the main difference between the two being that the mass outflow rate scales as 
$\dot{M}_{\rm wind} \propto \sigma_{\rm gal}^{-1}$ in the former case (see \S 2.1 for more details) and as $\dot{M}_{\rm wind} \propto \sigma_{\rm gal}^{-2}$ in the latter.
{Theoretical arguments \citep{MQT10, HQM12} favour the energy-driven wind scalings for the low to intermediate mass galaxies.  And recent simulation studies  \citep{FD14, CD15} suggest that the energy-driven wind model yield good agreement with the observed mass-metallicity relation, the stellar mass fraction versus galaxy mass trend, etc. for low to intermediate mass galaxies.   However, large-scale simulations that treat stellar-powered outflows from {\em all} galaxies as energy-driven winds, such as the Illustris simulations \citep{VGS14,VG14}, tend to produce mass-metallicity relations that are significantly steeper than observed \citep{SD14}.  Since the primary aim of the present work is to examine the emergence of galaxy groups, the massive stellar systems that populate such environments, and especially the growth and enrichment of the IGrM, we adopt the momentum-drive galactic outflows model\footnote{For completeness, we note that there are two distinct ways of implementing stellar feedback:  the ``kinetic'' approach, which we have adopted, and the ``thermal'' approach.  For further details about the latter, we refer the readers to \citet{DVS12,SBM13} (see also \citealt{S15,SM15} and references therein).}.}

In keeping with this goal, we identify  the formation times of the $z=0$ galaxy group population and compare these to the times when the hot diffuse IGrM in these systems is established as well as  when it is enriched with oxygen, silicon and iron.     We also explore whether the stellar-powered winds have any other direct or indirect impact on the properties of the IGrM or, for that matter,  any other group properties, beyond just  injecting the metals into the IGrM.     For instance, one can imagine that the metals, being very efficient coolants, can exacerbate the cooling of the IGrM and give rise to a greater build-up of cold gas and stars in the group central galaxies.   On the other hand, the outflows also heat the IGrM, and  if the resultant energy deposition significantly offsets the radiative losses, one {might expect a reduced IGrM cooling flow} onto the group central galaxy.   Since our present simulations do not include AGN feedback, we are especially interested in identifying robust trends that are not expected to change with the inclusion of AGN feedback.  One such result (see Section~\ref{Sec:5}) is our findings concerning the bulk metallicity of the IGrM and the extent of mass recycling between the IGrM and the galaxies,  which sets the stage for a more detailed analysis of how, where and when the iron and oxygen are introduced into the IGrM in a companion paper (Liang et al., in preparation).   The present analysis is useful in other ways also.  It provides detailed insights about where in the hierarchy of structures the present stellar-powered feedback model first starts to fail and {other feedback mechanisms, such as AGN feedback, might be needed, and provides clearer requirements for these additional feedback mechanisms.}  These results are informing our current efforts to develop new AGN feedback implementation that are more in keeping with both observations as well as theoretical expectations.

While on the subject of AGN feedback, we note that several collaborations who, like us, are working towards  self-consistent, holistic models for the formation and evolution of galaxies, groups and clusters, have started to incorporate AGN feedback into their simulations.  Of these, the two initiatives that are the furthest along are Illustris \citep{VG14} and Eagle \citep{S15,C15}.  Each collaboration has its own (very different) implementation of AGN feedback just as they each use very different approaches to modelling stellar feedback.  Both  fare reasonably well in terms of matching many of the observed trends and properties of galaxies.  However, each also has its own set of  challenges.  For example,   the AGN feedback scheme implemented in Illustris simulations tends to evacuate too much hot gas from the vicinity of massive galaxies \citep{G14} {while the Eagle simulations seem to suffer from insufficient expulsion of gas from the galaxies at early epochs, which results in the production of too many stars at early times and correspondingly, a reduced specific star formation rate at later times \citep{F15}.} Essentially, each approach has its strengths and weaknesses, leaving
considerable room for improvement and further development.

The present paper is organized as follows: In Section~\ref{Sec:2}, we provide a brief description of our simulation setup, discuss how we construct our catalog of simulated galaxy groups, and show some general properties of these groups at $z=0$.  In Section~\ref{Sec:3}, we discuss the global X-ray properties of our galaxy groups, focusing on three most commonly discussed group X-ray scaling relations: the (X-ray) luminosity$-$temperature, the luminosity-mass, the mass-temperature and the entropy-temperature relations.  We explore  the evolution of the model scaling relations over the redshift range $0 \leq z \leq 3$ and also compare the $z=0$ relations to the observational results. We discuss the baryonic content of the simulated groups in Section~\ref{Sec:4}, investigating the variation in the total baryon fraction as well as the stellar and the hot diffuse intragroup medium mass fractions with group mass over the redshift range $0 \leq z \leq 3$.   We compare these fractions with observational results from both low redshift groups as well as more recent results from groups at redshifts out to $z\sim 1$, and we provide a detailed analysis of the assembly of groups' total mass, IGrM, and stellar mass.  We then investigate the enrichment of the IGrM in Section~\ref{Sec:5}, looking at the sources of the iron, silicon and oxygen, the abundance ratios, etc.  Finally,  we summarize and discuss our findings in Section~\ref{Sec:6}.

\section{SIMULATING GALAXY GROUPS}
\label{Sec:2}

\subsection{Simulation Details}
\label{Sec:2.1}

We extracted galaxy groups from a cosmological hydrodynamic simulation of a representative comoving volume $(100 \;h^{-1}\; {\rm Mpc})^{3}$ of a $\Lambda$CDM universe with present-day parameters:  $\Omega_{\rm m,0}$ = 0.25, $\Omega_{\Lambda,0}$ = 0.75, $\Omega_{\rm b,0}$ = 0.044, $H_{\rm 0}$ = 70 km $\rmn{s^{-1} Mpc^{-1}}$, $\sigma_{8}$ = 0.83 and n = 0.95.   These are based on the WMAP-7 best-fit cosmological parameters  \citep{J11} and are in good agreement with the WMAP-9 results \citep{H13}.

{We initialized the simulation volume} with $576^3$ dark matter particles and $576^3$ gas particles, implying particle masses of $4.2 \times10^{8}$ $\rmn{M_{\odot}}$ and $9.0 \times10^{7}$ $\rm{M_{\odot}}$ for {the} dark matter and gas, respectively. {In the simulation we assumed a spline gravitational softening length of $5\; h^{-1}$ kpc comoving ($3.5\; h^{-1}$ equivalent Plummer softening).}

The initial conditions for the simulation volume were generated using an \citet{EH99} power spectrum in the linear regime, and the simulation was {evolved} from $z=129$ to $z=0$ using a modified version of GADGET-2 \citep{GA2}, a cosmological tree-particle mesh-smoothed particle hydrodynamics code that includes radiative cooling using primordial abundances as described in \citet*{NK96} and metal-line cooling as described in \citet*[][hereafter OD06]{OP06}.  Star formation is modelled using a multiphase prescription of \citet{SP03}.  In this prescription, only gas particles whose density exceeds a preset threshold of $n_{{\rmn H}}=0.13\;\rmn{cm^{-3}}$ are eligible to form stars. The star formation {rate} follows a Schmidt law (Schmidt 1959) with the star formation time-scale scaled to match the $z= 0$ Kennicutt law (Kennicutt 1998).  We assume that the mass function of forming stars is given by the Chabrier initial mass function (IMF) \citep{GC03}.  According to this IMF,  19.8$\%$ of the stellar mass goes into massive ($\it{i.e.}~\geq 10\;\rmn{M_{\odot}}$) stars that engender Type II supernovae \citep{OD08}.

Over the course of the simulation, we account for mass loss and metal enrichment from Type Ia and Type II SNe as well as the asymptotic giant branch (AGB) stars.  For a detailed discussion of how this was carried out, we refer readers to \citet*[][hereafter OD08]{OD08}. An overview is as follows:  In the case of Type II SNe, we use the instantaneous recycling approximation where the mass and the metals are returned immediately to the ISM. Type II SN metal enrichment uses the metallicity-dependent yields calculated from the \citet{LC05} supernova models. In the case of Type Ia SNe, we allow for both a prompt component as well as a delayed component using the model of \citet{SB05} in which, the former is tied to the star formation rate as in the case of Type II SNe, and the latter to the stellar mass. The mass loss and the enrichment by the prompt component is returned to the ISM instantaneously while that due to the delayed component commences after a delay of $0.7$ Gyrs.  As for the AGB stars, the corresponding mass and metal injection into the ISM commences after a delay of $30$ Myrs following a star formation event and extends over the subsequent $\sim 14$ Gyrs.  As discussed by OD08, the most significant impact of the AGB stars is to replenish the ISM. In general, a little more that 50$\%$ of the stellar mass in a Chabrier IMF is returned to the ISM over $\sim 10$ Gyrs and 30$\%$ of this comes from the AGB stars. The mass transfer from AGB stars also returns to the ISM the metals that were previously locked into these stars.Ê Primarily, the metallicity of this gas is the same as that of ISM from which they were spawned.Ê In detail, however, nucleosynthesis reactions during the AGB phase lead to a slight depletion of the oxygen abundance as well as an enhancement of the carbon abundance.  To conclude this brief overview, we note that we explicitly track the evolution in the abundances of four metal species --- carbon, oxygen, silicon, and iron. These four species are not only among the most abundant metals in the universe, they are also the species most often observed in quasar absorption line spectra probing the intergalactic medium, {in} the X-ray spectra of the hot intracluster and intragroup halo gas, as well as the spectra of the circumgalactic and the interstellar gas in galaxies.  In the present study, we will focus primarily on the oxygen, silicon and iron.

Our numerical implementation of the kinetic winds {powered by} stellar feedback is based on the scheme initially developed by \citet{SP03}, modified to conform to the momentum-driven wind model {scalings} of MQT05.   MQT05 describe a model where the energy and momentum injected into the ISM by the stars via stellar winds and supernova explosions, and by radiation pressure produced by the continuum absorption and scattering of photons on the dust grains that are collisionally coupled to the gas, propel galaxy-wide winds.   A detailed discussion of the model and the particular implementation that we are using has been discussed extensively in a series of papers starting with OD06, with updates described in OD08 and \citet{OD09}, and has recently been shown to be in general agreement with the results of recent high-resolution galaxy-scale simulations of \citet{HQM12,HK14,M15} that include explicit stellar feedback.  For completeness, we briefly summarize below the current implementation:

The mass outflow rate scales with the star formation rate as 
\begin{equation*}
\dot{M}_{\rm wind} = \eta \: \dot{M}_{\rm star}\ {\rm where}\ \eta = (150\; {\rm km/s})/{\sigma_{\rm gal}}
\end{equation*}
\noindent 
is the mass loading factor.
Here, $\sigma_{\rm gal}$ is the galaxy velocity dispersion.  It is used as a measure of the depth of the galaxy's potential and is computed using the total mass (baryons+dark matter) of the galaxies identified over the course of the simulation \citep[\cf][]{OD10}.  This means that when a gas particleÕs density exceeds the threshold for star formation, it is eligible to form stars with some probability $P_{\rm star}$ but at the same time, its probability for being incorporated into {a} wind is $P_{\rm wind}=\eta \: P_{\rm star}$.   If a particles is chosen to be in {an} outflow, it is given a velocity kick $v_{\rm wind}$, where 
\begin{equation}
 v_{\rm wind} = 4.3\sigma_{\rm gal} \sqrt{f_{L}-1}+ \beta \sigma_{\rm gal} ,
\end{equation} 
\noindent is orientated parallel to $\pm\mathbf{v}\times\mathbf{a}$, the cross product of the velocity and the acceleration vectors of the particle prior to entrainment.  

In the above equation, the first term on the right represents the wind launch velocity --- with $f_L$, the luminosity factor in units of the galactic Eddington luminosity (\ie~the critical luminosity for expelling gas from the galaxy via radiation pressure) given by equation (5) of OD08.  {As discussed in OD08, this wind launch velocity is capped by limiting the value of $\sigma_{\rm gal}$ to}  
\begin{equation}
 \sigma_{\rm gal,max} = 1400 \left({{50 {\rm Myrs}}\over \tau_{\rm SF}}\right)\; {\rm km/s},
\end{equation} 
{
where $\tau_{\rm SF}$ is the local star formation timescale,
to account for the fact that in the MTQ05 model, starburst luminosities need to reach the Eddington limit to expel the gas.   In practice, we find that this cap has no impact on winds from galaxies at $z > 1$ while at lower redshifts, the median  wind velocity from the group central galaxies is reduced by a constant factor that grows to $20\%$ by $z=0$.}

{The second term (with $\beta=2.9$) represents an additional velocity kick at launch, to simulate the continued dynamical pumping of the gas in the MQT05 momentum-driven wind model (see OD06 and OD08 for details).  Since our simulations neither resolve the detailed structure of the ISM nor the detailed hydrodynamics of the wind flowing through the ISM, this model assigns the ejected gas particles an appropriate initial velocity to ensure that the wind has roughly the expected velocity at large radii.
For the same reasons, 
we also decouple the wind particles hydrodynamically (but not dynamically) from their surroundings until the local gas density drops to 10\% of the star formation density threshold or for a time duration equal to $ 200\: (v_{\rm wind}/ 100\;{\rm [km/s]})^{-1}\; {\rm Myrs}$.  In real galaxies, this outflow would be expected to flow through the ISM and out of the disk along paths of least resistance --- \ie, along channels formed by overlapping supernova explosions.}

{We reiterate}  that our present simulation does not include the effects of the energy and momentum output of AGNs. As we will show in Section~\ref{Sec:4} of the present paper and in the follow-up paper (Paper-II),  this ÒdeficitÓ does not significantly alter our results about how the enrichment of the IGrM unfolds.   This is because (1)  metals are flushed out of the galaxies primarily by stellar-powered galactic outflows and (2) most of the metals in the IGM and in the hot IGrM are from lower mass galaxies whose evolution is expected to be only minimally impacted by AGN feedback, if at all.   The most massive galaxies, whose evolution is thought to be  strongly impacted by AGN feedback and which, in the absence of the latter, build up a much larger stellar mass than their observed counterparts, contribute approximately $25\%$ of the metals in the hot IGrM.   As we show in Section~\ref{Sec:4}, a reduced contribution from these galaxies due to quenching of star formation by AGN feedback should in fact improve the agreement between our simulation results and observations even further.

\begin{figure}
  \vspace*{0pt}
  \includegraphics[width=0.45\textwidth]{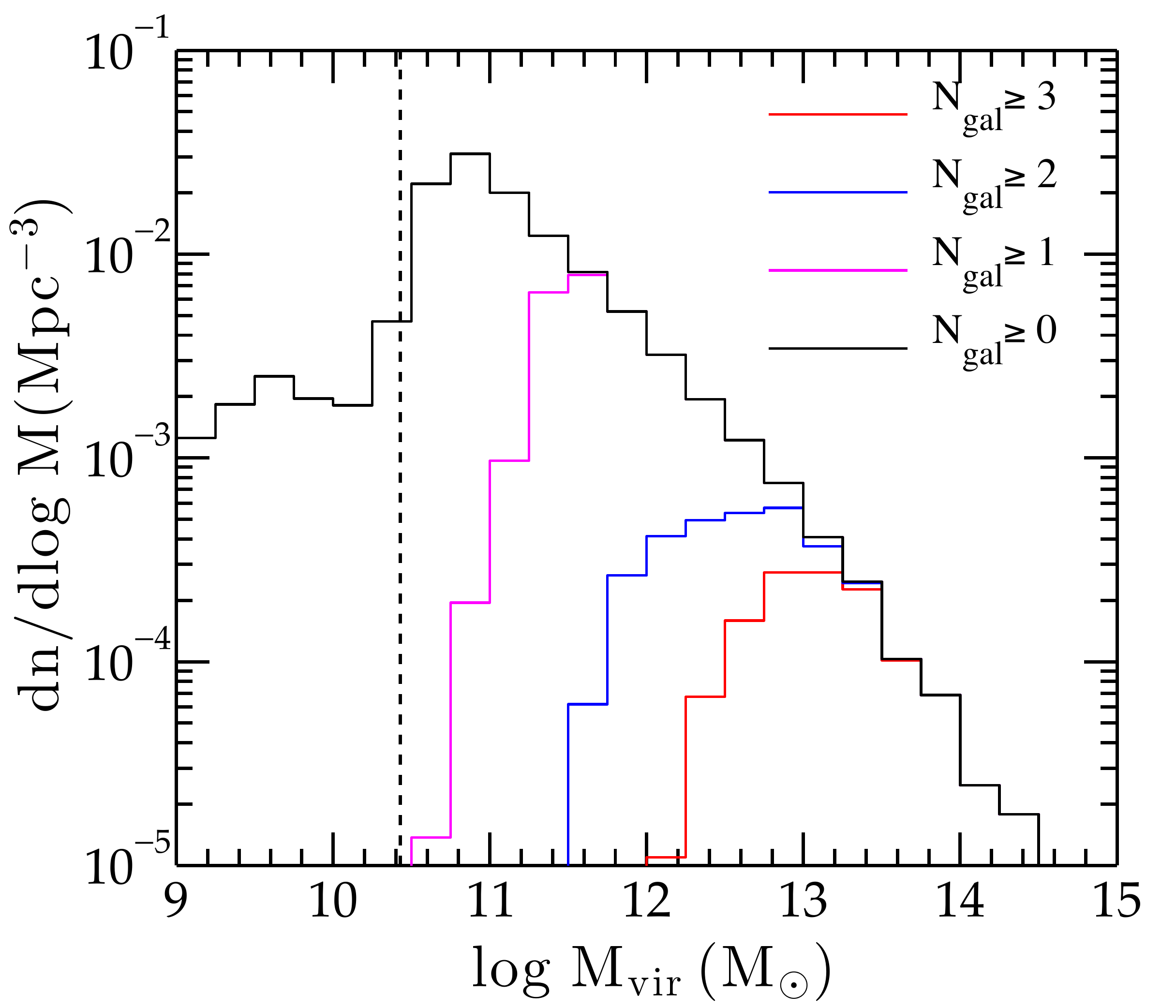}
  \caption{ {The} mass function of halos with at least three (red), two (blue), one (magenta) luminous galaxies, as well as of the complete halo population in the simulation volume (black) described in Section~\ref{Sec:2.2}. The dashed vertical line shows our halo mass resolution limit of $2.7\times10^{10} \rmn{M_{\odot}}$, corresponding to 64 dark matter particles.}
  \label{fig.1}
\end{figure}

\begin{figure}
  \vspace*{0pt}
  \includegraphics[width=0.48\textwidth]{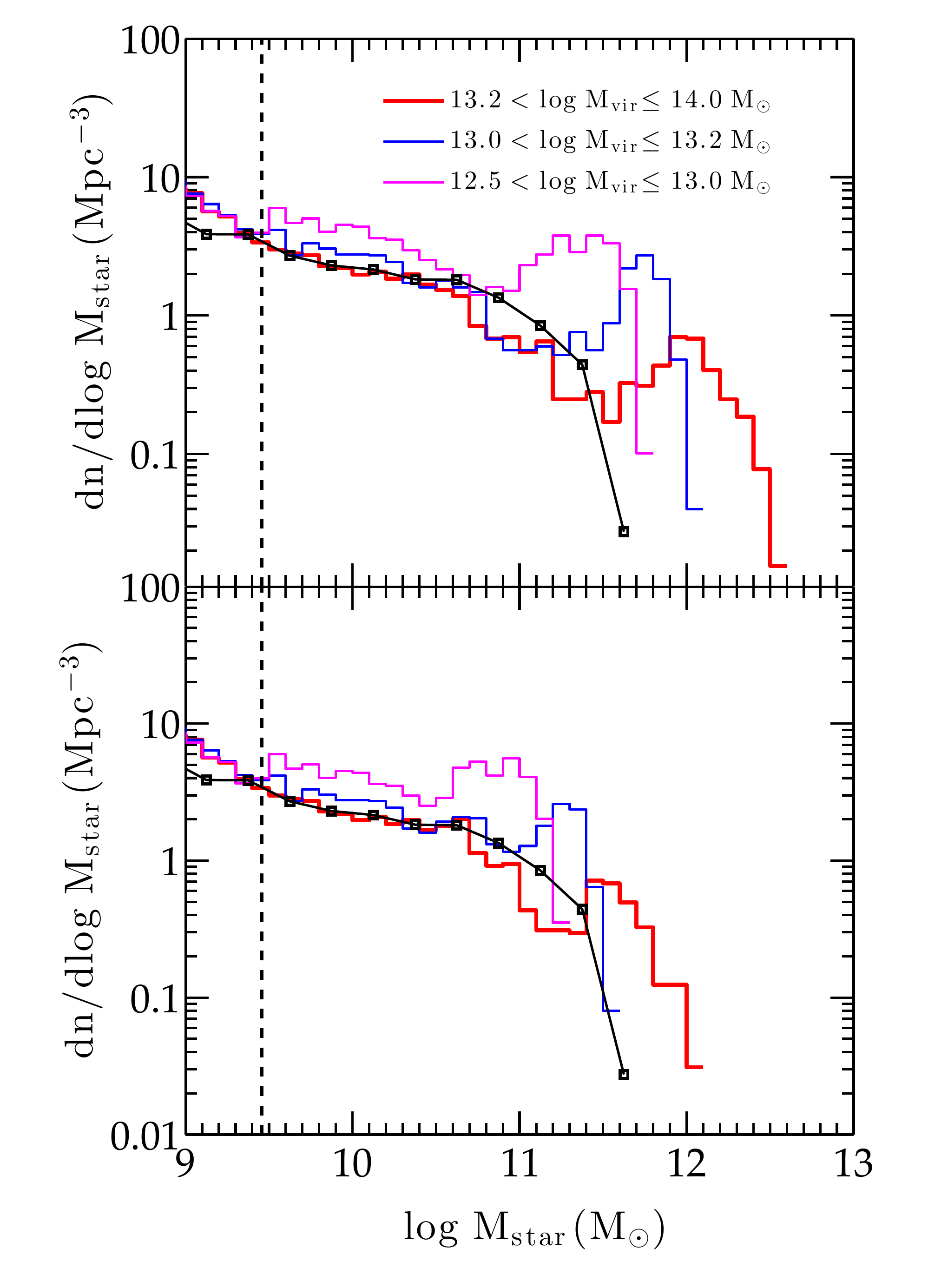}
  \caption{
  The top panel shows the $z=0$ galaxy stellar mass function (GSMF) of all luminous galaxies in the simulated groups, sorted into three mass bins:  $12.5 < {\rm log} \; M_{\rm vir} \leq 13.0 \;\rmn{M_{\odot}}$ (magenta), $13.0 < {\rm log} \;M_{\rm vir} \leq 13.2 \;\rmn{M_{\odot}}$ (blue), and $13.2 < {\rm log} \;M_{\rm vir} \leq14.0 \;\rmn{M_{\odot}}$ (red).  For comparative purposes, we also plot as connected black squares the GSMF for X-ray detected low mass groups spanning the mass range similar to that of our simulated groups \citep{GD12}.   The vertical dashed black line shows our luminous galaxy stellar mass resolution limit (see text).    In anticipation of the discussion in \S 4.1, the lower panel shows the same GSMFs as in the top except that the stellar mass of galaxies in the simulation with $M_\ast > 10^{11} \rmn{M_{\odot}}$ has been {\it artificially} reduced by a factor of 3.
}
  \label{fig.2}
\end{figure}

\subsection{Finding Galaxies and Galaxy Groups in the Simulation Volume}
\label{Sec:2.2}

Each simulation output is analyzed to identify both galaxies as well as bound halos. 
We identify bound halos using the spherical overdensity (SO) procedure described in \citet{DK05}:   First, associations of dark matter particles are found using a Friends-Of-Friends (FOF) percolation algorithm. Second, after finding the local potential minima of each association, a sphere is constructed around these particles and its radius is expanded until the mean enclosed total mass density equals the virial density for the assumed cosmology at the redshift under consideration: 

\begin{equation}
\overline{\rho_{\rm m, vir}}(z) = \Delta_{\rm vir} (z)\cdot E^{2}(z)  \rho_{\rm crit} (0),
\label{rhom}
 \end{equation}
\;

\noindent where $E(z)\equiv H(z)/H_0$ is the dimensionless Hubble parameter given by

\noindent \[ 
E^{2}(z)=1-\Omega_{\rm m,0}+(1+z)^{3}\Omega_{\rm m,0},
\]

\noindent  and the virial overdensity factor is well-described by the following fitting function \citep[$\cf$][]{BB02}:

\noindent  \[
\Delta_{\rm vir}(z)=49+96\Omega_{\rm m}(z)+\frac{200\Omega_{\rm m}(z)}{1+5\Omega_{\rm m}(z)},
\]

\noindent  \[
\Omega_{\rm m}(z)=\frac{\Omega_{\rm m,0}(1+z)^{3}}{1-\Omega_{\rm m,0}+\Omega_{\rm m,0}(1+z)^{3}}.
\]

\noindent We will refer to the radius of this sphere as the virial radius, $R_{\rm vir}$, and the mass enclosed as the virial mass of the halo, $M_{\rm vir}$.    Occasionally, we will reference $M_\Delta$ instead of $M_{\rm vir}$.   $M_\Delta$  is   the enclosed mass inside a sphere centered on the halo center within which the mean mass density is $\Delta\cdot\rho_{\rm crit}(z)=\Delta\cdot E^{2}(z)\rho_{\rm crit}(0)$.  Commonly used values of $\Delta$ are 200, 500 and 2500.    The mapping between these different quantities is redshift-dependent.  At $z=0$, 
\begin{align*}
&M_{\rm vir} \approx 1.2 M_{200}\; {\rm and}\; M_{500} \approx 0.7 M_{200},\\
&\,R_{\rm vir}\approx 1.36 R_{200}, \; R_{500}\approx 0.67 R_{200},\; {\rm and}\; R_{2500}\approx 0.27 R_{200}.
 \end{align*}

In Figure~\ref{fig.1}, we show the $z=0$ mass functions of all halos in the simulation volume (black curve), as well as halos with at least three (red), two (blue) and one (magenta) ``luminous'' galaxies (defined below). We identify groups (and clusters) of galaxies as halos containing $\geq 3$ ``luminous'' galaxies. On mass scales $\geq10^{13}\;\rmn{M_{\odot}}$, nearly all halos have at least 3 galaxies while in the mass range $10^{12}\;\rmn{M_{\odot}}  \simless M_{\rm vir} \simless10^{13}\;\rmn{M_{\odot}}$, only a fraction of the halos do.   
There are a total of 902 groups in our simulation volume at $z=0$.

Galaxies in the simulation volume are identified using the group-finding algorithm Spline Kernel Interpolative Denmax (SKID)\footnote{http://www-hpcc.astro.washington.edu/tools/skid.html} to locate gravitationally bound clumps of star particles and cold ($T< 3\times 10^{4}$ K) gas particles that are eligible to form stars \citep[$\cf$][]{OD10}. We identify a galaxy as ``resolved'' if the total mass in cold gas and stars is $\geq 2.9\times10^{9}$ $\rmn{M_{\odot}}$  and ``luminous'' if its stellar mass is $\geq 2.9\times10^{9}$ $\rmn{M_{\odot}}$, which is equivalent to $\geq 64$ star particles.

The top panel of Figure~\ref{fig.2} shows the average stellar mass function of  the luminous galaxies in the simulated groups at $z = 0$. We divide the groups into three mass bins: $12.5 < {\rm log} \; M_{\rm vir} \leq 13.0$ (magenta), $13.0 < {\rm log}\; M_{\rm vir} \leq 13.2$ (blue), and $13.2 < {\rm log}\; M_{\rm vir} \leq 14.0$ (red), 
with 393, 145 and 188 groups in each bin, respectively.  (The sum of these is short of the total number (902) quoted above because the balance of the groups have $  {\rm log} \; M_{\rm vir} \leq 12.5$ or $> 14.0 $.)  {The size of our mass bins were chosen to ensure that the intermediate and the most massive mass bins match the bins adopted by DOS08 to facilitate comparison with their results.}

For comparison, we also show, as connected black squares, the observed galaxy stellar mass function (GSMF) for the X-ray detected low mass groups 
in the COSMOS survey \citep{GD12} spanning the similar mass range ($10^{13}\;\rmn{M_{\odot}}< M_{200} < 2\times 10^{14}\;\rmn{M_{\odot}}$)  to that of our simulated groups \citep{GD12}.   It is readily apparent that the simulation shows a clear excess of galaxies with large stellar masses.   In the lower two mass bins, there is typically only one `super-sized' galaxy per group, the central galaxy, whose stellar mass exceeds $10^{11}\; \rmn{M_{\odot}}$ and it is these galaxies that are responsible for the excess.    In addition to a `super-sized' central galaxy, the most massive groups also host one (and sometimes, two)  `super-sized' satellite galaxies within $R_{\rm vir}$.   On closer inspection, we have confirmed that (1) these  massive satellite galaxies were incorporated into the present-day groups via mergers within the past 6 Gyrs, (2) they were all central galaxies prior to the merger, and (3) these massive satellites are steadily sinking to the center and will eventually merge with the existing dominant central galaxy in the group.   It is  expected that AGN feedback will quench the growth of these `super-sized' galaxies.   From the resolution limit to approximately $M_{*} \approx 10^{11} \rmn{M_{\odot}}$,  the GSMF of the group galaxies in our simulation volume agrees very well with the observations.   AGN feedback is believed to play a minimal role in these galaxies; their evolution is principally governed by stellar feedback.


\subsection{Computing Group Properties}
\label{Sec:2.3}

To compare the characteristics of our groups to the observations, we compute various X-ray properties of our group halos.   Unless explicitly noted, we focus exclusively on the properties of the hot ($T>5\times 10^5\; {\rm K}$), diffuse  IGrM particles.

The first of these properties is the rest-frame 0.5-2.0 keV X-ray luminosity within $R_{500}$: $L_{X, 0.5-2.0\;{\rm keV}}$.   This is computed by summing over the luminosity of the individual IGrM gas particle within a distance $r \leq R_{500}$ of the halo center.   The emission characteristics of gas particles is computed using the Astrophysical Plasma Emission Code\footnote{http://cxc.harvard.edu/atomdb/sources apec.html}  (APEC) from \citet{SM01} assuming that the gas is optically thin and in collisional ionization equilibrium.   APEC uses the particleÕs mass, SPH-weighted density, temperature and the metallicity as input, and  outputs X-ray spectra, from which the luminosity is computed by summing {the} intensities over the required range of photon energies (\eg~0.5-2.0 keV).  Contributions to line and continuum emission associated with each of the individually tracked elements (iron, oxygen, silicon and carbon, along with hydrogen and helium) are computed separately and summed.   The abundances of all the other elements are specified assuming that their abundance ratios relative to iron is solar \citep*{AG89}.   We note, however, that our results are not sensitive to the latter because in the IGrM temperature and density regime, the line emission is dominated by iron, oxygen and silicon.

To study the metal content of the IGrM, we consider two complementary measures: the mass-weighted and emission-weighted abundances, defined respectively as

\begin{equation}
\label{MetW}
Z^{\rm mw}_{\rm q} = \frac{\Sigma_{\rm i} Z_{\rm q, i}m_{\rm i}}{\Sigma^{N}_{\rm i}m_{\rm i}} \ \ {\rm and}\ \ Z^{\rm ew}_{\rm q} = \frac{\Sigma_{\rm i} Z_{\rm q, i}L_{\rm i}}{\Sigma^{N}_{\rm i}L_{\rm i}},
\end{equation}
  
\noindent In these equations, ``q" is the metal species under consideration, $Z_{\rm q, i}$ is the SPH kernel-weighted abundance of the i$^{\rm th}$ particle, $m_{\rm i}$ is its mass, and $L_{\rm i}$ is the X-ray luminosity of that particle.  The sums run over all IGrM particles within the volume under consideration.    Additionally, to facilitate comparison between numerical and observational results, we scale and report all metal abundance estimates in terms of the solar ``photosphere abundance" values from \citet{AG89}: that is, 
  $Z_{\rm O,\odot} = 0.009618$, $Z_{\rm Si,\odot} = 0.0007109$ and $Z_{\rm Fe,\odot} = 0.001875$.

We also compute two different temperature measures of the hot gas in our groups.  Observationally, the temperature of the hot gas is determined by fitting its observed X-ray spectrum.  Generally, the spectrum within a beam is a composite of the continuum and the line radiation from a range of gas phases with varying temperatures and metallicity. {Since the vast majority of the observed galaxy group spectra have been measured using CCDs that do not allow the emission from individual components to be spectrally distinguished}, and the statistical quality of the observations is insufficient to detect all the features in the spectrum, and since the spectrum itself is only available within a limited range of frequencies, the current convention is to compare the composite spectrum with single-temperature, single-metallicity thermal plasma models, and assign the temperature of the best-fit model to the observations.   The question for theorists is, and has been for years now: what measure best corresponds to this temperature?  

\begin{figure}
  \vspace*{0pt}
  \includegraphics[width=0.45\textwidth]{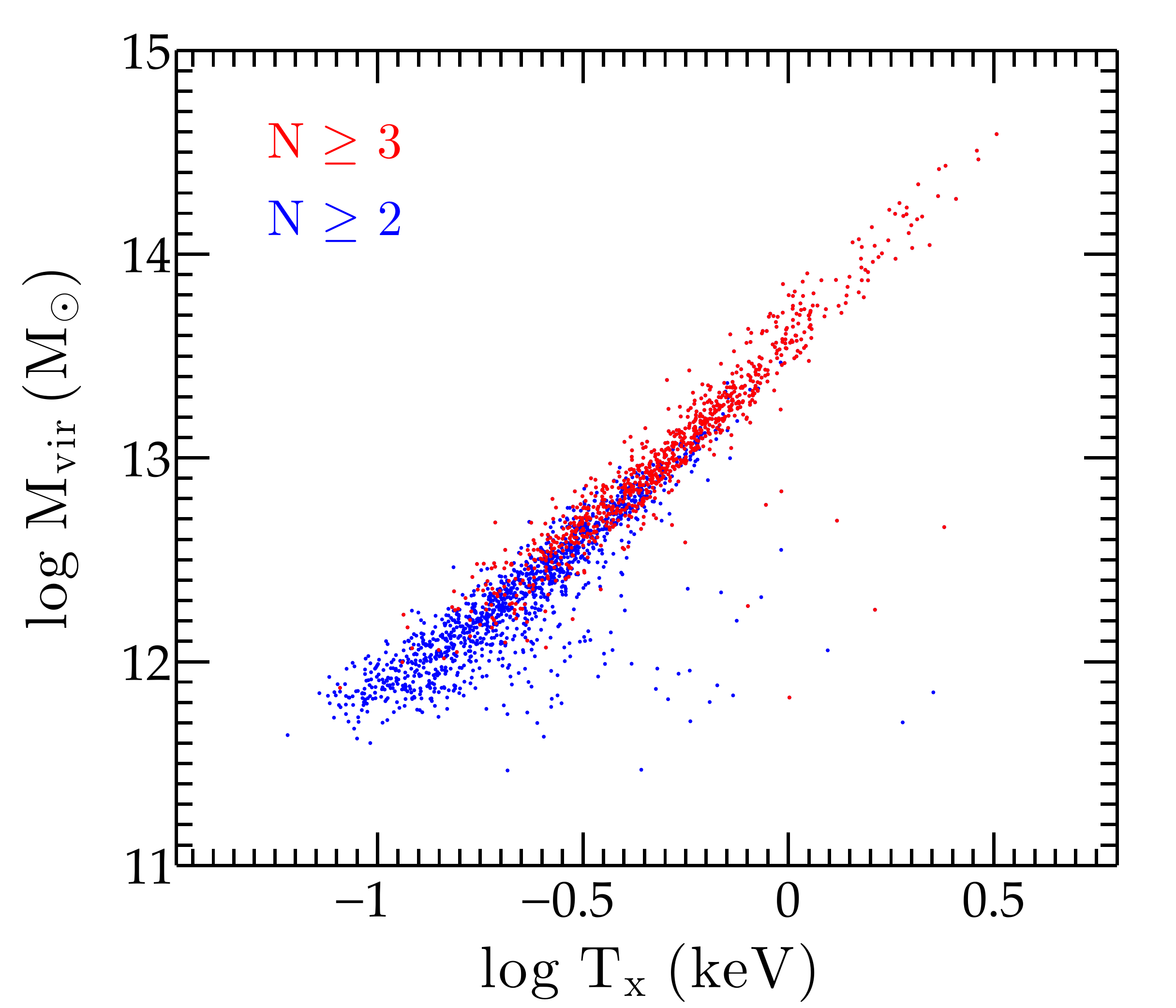}
  \caption{ $M_{\rmn{vir}} - T_X$ relation of galaxy groups with at least three (red) and two (blue) luminous galaxies. $T_X$ is tightly correlated with $M_{\rm vir}$ and follows the scaling relation: $M_{\rmn{vir}}\propto T_{X}^{1.7}$.  Groups that lie significantly off this relationship are located near larger systems and are ``contaminated" by the latters' hot diffuse gas. Excluding groups with fewer than three luminous galaxies, eliminates most of these ``contaminated" halos.}
  \label{fig.3}
\end{figure}

The mean emission-weighted temperature ($T_X$), which is an average of the temperatures of the individual components weighted by their radiative emission contributions is one possibility and in fact, is the most commonly used measure in theoretical work.    We too compute this temperature but we restrict the weighing to X-ray emission within a relatively narrow energy range.    In other words,  our emission-weighted temperature is defined as the weighted average temperature of the gas particles, where we use the particles' rest-frame 0.5-2.0 keV X-ray luminosity as the weighting factor.  As shown in Figure~\ref{fig.3}, $T_X$ is tightly correlated with group mass and we will use the mass ($M_{\rm vir}$, $M_{\rm 200}$ or $M_{\rm 500}$) or the mean emission-weighted temperature ($T_X$) interchangeably when referring to or categorizing the group halos. 

Observationally, group and cluster hot gas temperatures are determined by identifying a single-temperature thermal model whose spectrum best matches the observed spectrum.  Reproducing this procedure is sufficiently involved, especially if the observed spectrum is an integrated output of gas that spans a range of temperatures, that most theoretical studies have, until relatively recently, tended to rely on $T_X$ as a stand-in.   A decade ago,  \citet{MZ04} showed that for clusters of galaxies, $T_X$ is generally biased high by as much as $\sim$25\%  compared to  the observationally determined integrated X-ray temperatures and introduced an alternate weighting scheme leading to a new temperature measure, which we will refer to as the  ``spectroscopic" temperature ($T_{\rmn{spec}}$), {which} is intended to be directly comparable to the observationally determined temperature, at least in the case of the hot ($T>2$ keV) intracluster gas.  \citet{AV06} has since extended the approach to cooler groups and here we will use their algorithm.   As demonstrated by  \citet{AV06}, $T_{\rmn{spec}}$ is an accurate  estimate (to within a few percent) of the fitted temperature of a multiphase plasma with components whose temperatures and metallicities span the range expected in group and cluster environments. 

{We compute the  two temperatures mentioned above ($T_X$ and $T_{\rmn{spec}}$)  using either all {the} hot diffuse particles within $R_{500}$ or only those within the radial range $0.15R_{500} \leq r \leq R_{500}$.  The latter measures will be referred to as ``core-corrected" and identified with subscript ``corr" (short for `corrected').   Observational studies of groups typically quote ``core-corrected" temperatures.}

Comparing $T_{\rmn{spec}}$ to $T_X$, we find that in the small  ($\leq 1$ keV) groups, the two agree with each other to within $\sim$3\%, in agreement with results shown in \citet{AV06}. {However,} in hotter systems the divergence is significant enough to be a cause for concern.  We discuss this further in the next section.  We also note that neither $T_{\rmn{spec}}$ nor $T_X$ is an unbiased measure of the actual mean temperature of the gas, which is much better approximated via a mass-weighted average.

{Finally, we point out that to compare the X-ray scaling relations for our simulated groups across different redshifts, we adopt the common convention in the literature and plot quantities motivated by the self-similar model of group and cluster halos \citep{K86}. In this model, the scaling relations are preserved when using the 
quantities $L_X (z)\, E(z)^{-1}$, $M_\Delta (z) \,E(z)$ and $S_\Delta (z) \, E(z)^{4/3}$, where  $E(z)\equiv H(z)/H_0$ is the dimensionless Hubble parameter (\cf ~equation \ref{rhom}),  instead of $L_X (z)$, $M_\Delta (z) $ and $S_\Delta (z) $.  The self-similar model assumes that the profiles describing the internal structure of all groups and clusters have the same  functional form, with the gas properties created through gravitational collapse alone, so that these properties scale only with the system mass and the critical density at the time of observation.  Strictly speaking, the self-similar model is an anachronism in that it does not account for processes like radiative cooling or heating of the gas by stellar (or AGN) winds and jets, or for a variation in the IGrM fraction with halo mass.   As a result, the observed scaling with mass deviates from the predictions of this model.   Nonetheless, the observed scaling with redshift (for moderate redshift values) agrees surprisingly well with the self-similar predictions, suggesting that over limited periods of time the systems may well be evolving in a self-similar fashion.  For further details, we refer the readers to  \S 3.9 of \citet{KB12} and to \citet{E15}.}

\section{GLOBAL X-RAY PROPERTIES OF GALAXY GROUPS}
\label{Sec:3}

In this section, we discuss some of the global properties of galaxy groups, focusing on the observed X-ray scaling relations, of the simulated $z=0$ groups and compare these to those of simulated groups at earlier epochs ($z=0.5$ to $3.0$), and to available observations.   As we shall show, even in the absence of AGN feedback, the present model does remarkably well in accounting for the observations.

\subsection{The Mass-Luminosity-Temperature Scaling Relations}

\begin{figure}
  \vspace*{0pt}
  \includegraphics[width=0.48\textwidth]{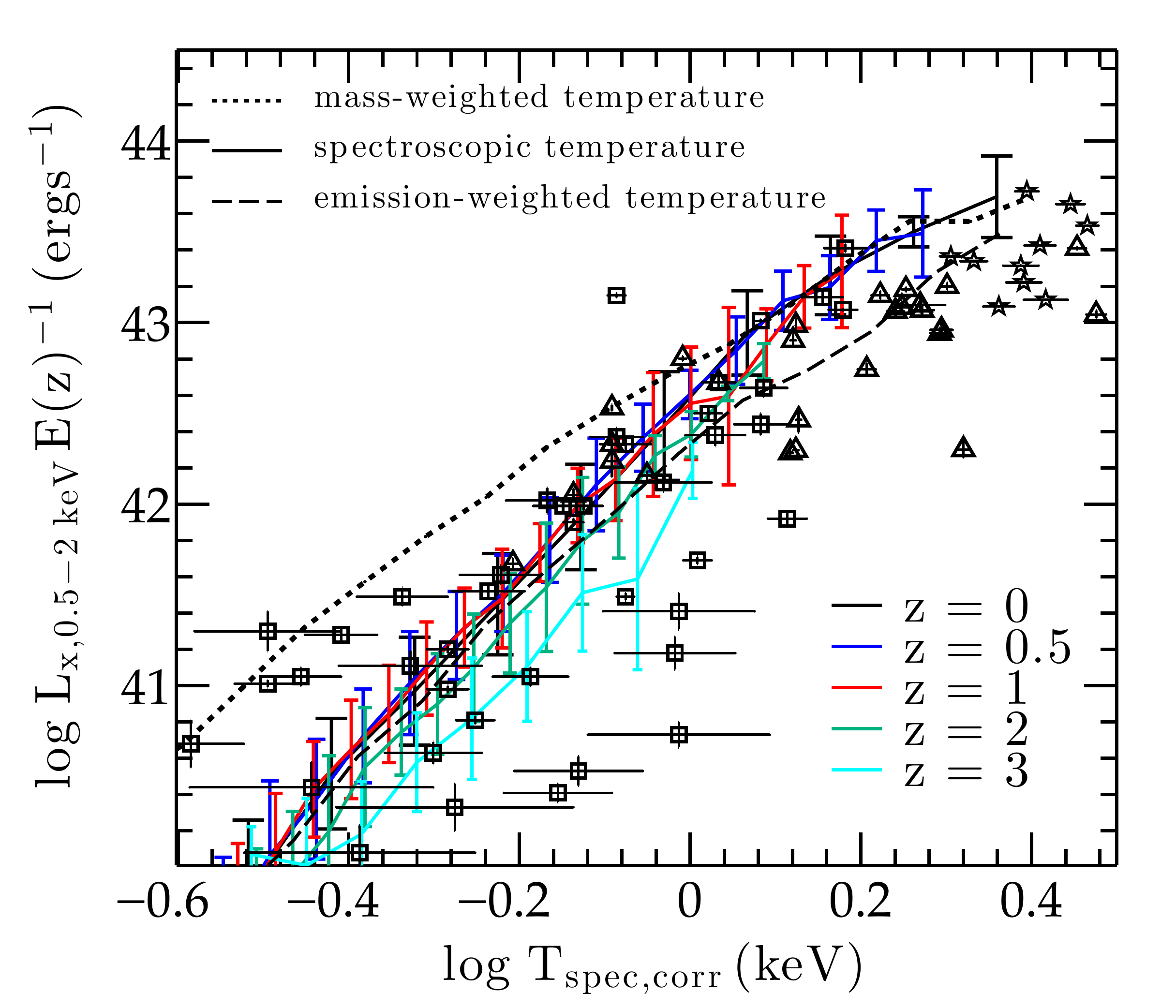}
  \caption{ X-ray luminosity$-T$ relation for simulated groups at $z = 0$ (black), $z = 0.5$ (blue), and $z = 1$ (red),  $z = 2$ (green), and $z=3$ (cyan).  The solid lines show the scaling relationship between the X-ray luminosity that is emitted by gas within $R_{500}$ and the core-corrected spectroscopic temperature. The error bars indicate 1-$\sigma$ scatter. The dotted and the dashed curves show the mean $L_X-T$ for the $z = 0$ simulated groups, where $T$ is the mass-weighted and {emission}-weighted temperature (both core-corrected), respectively.  Squares, stars and triangles show observed low redshift group data from \citet{OP04}, \citet{P09} and \citet{EK11}, respectively. We plot all the groups in \citet{OP04} including those with a small radial extent in observable X-rays (i.e. their ÕH sampleÕ). Luminosity in the \citet{P09} and \citet{EK11} data is corrected to the $0.5-2$ keV band.}
  \label{fig.4}
\end{figure}

\begin{figure}
  \vspace*{0pt}
  \includegraphics[width=0.47\textwidth]{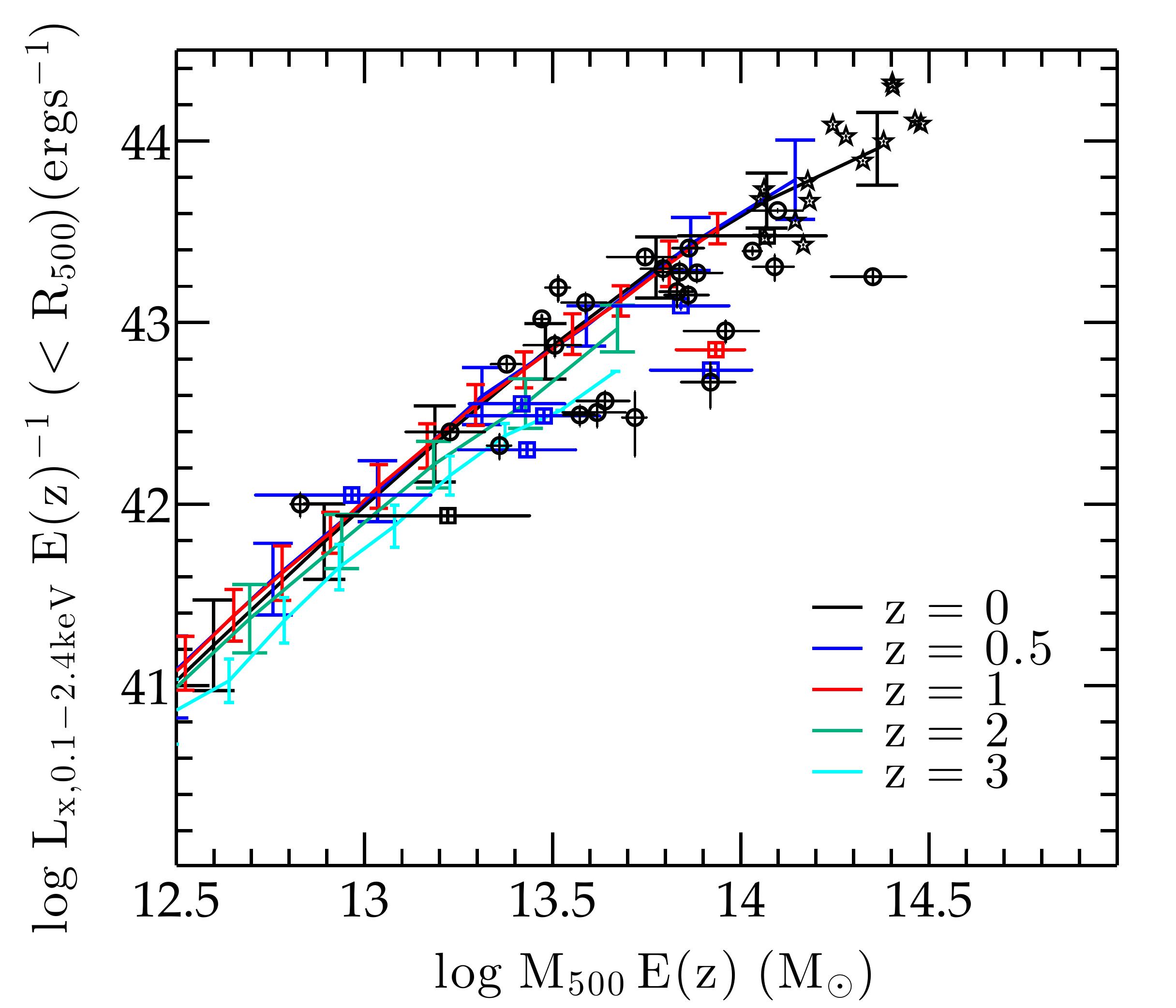}
  \caption{ $L_X-M$ relation for simulated groups at $z = 0$ (black), $z = 0.5$ (blue), $z=1$ (red),  $z=2$ (green), and $z=3$ (cyan). The error bars show 1-$\sigma$ scatter.  The circles, stars and squares show data from  \citet{EK11}, \citet{P09}, and \citet{L10}, respectively.  The hydrostatic mass estimates from the first two studies have been corrected for the hydrostatic bias \citep{H15} and $L_{X,{\rm bol}}$ from \citet{P09} have been converted to $L_{X,0.1-2.4\,{\rm keV}}$.  We also convert the weak-lensing $M_{200}$ values from \citet{L10} to $M_{500}$ using an NFW profile,$^2$
  and we scale their luminosities using the median value of  $L_{X,0.1-2.4\, {\rm keV}}(<R_{200}) / L_{X,0.1-2.4\, {\rm keV}}(<R_{500})$ for our simulated groups.  The observed groups at $z\leq 0.25$, $0.25<z\leq 0.75$, and $z > 0.75 $ are plotted as black, blue and red symbols, respectively.
 }
  \label{fig.5}
\end{figure}

In the absence of feedback and cooling flows, the X-ray luminosity on the group scale ought to scale with the mean gas temperature of the IGrM as $L_X\propto T$  \citep{BB99,BB02}.
This relationship is {\it flatter} than the more familiar  $L_X\propto T^2$ scaling in situations where bremsstrahlung dominates the X-ray emission because at temperatures less than 1 keV recombination radiation is as important, if not more, than bremsstrahlung.   The observed scaling relationship for groups, $L_X\propto T^{~4}$, however, is much steeper \citep{HP00}, indicating that either heating  \citep{BB99,BB02} and/or cooling \citep{VB01} has significantly altered the hot X-ray gas distribution.    Both processes eliminate the denser, lower entropy, X-ray bright, gas.

{Figure~\ref{fig.4} shows the rest-frame $0.5-2.0$ keV X-ray luminosity emitted within the central $R_{500}$ versus the mean core-corrected spectroscopic temperature (solid lines) for the simulated groups at redshifts $z=0$ (black), $0.5$ (blue), $1$ (red), $2$ (green) and $3$ (cyan).   Once cosmic evolution is taken into account,  the group $L_X-T_{\rm spec,corr}$ curves at $z\leq 1$  essentially lie on top of each other.   At higher redshifts, however,  the groups \textendash{} at a given temperature \textendash{} are less luminous than predicted by the self-similar evolution model.  These trends can be understood in terms of the  behaviour of the hot gas fraction in the groups  that we will discuss in the next subsection.}

Focusing{\let\thefootnote\relax\footnote{$^2\;$~For consistency, we use the same mass-concentration relationship adopted by \citet{L10}.}}
 on the mean $L_X-T_{\rm spec,corr}$ relationship for our simulated $z=0$ groups, we note that this is in good agreement with the observations for $ T_{\rm spec,corr} \simless 1$ keV.   In this low temperature regime, the luminosity scales as $L_X\propto T_{\rm spec,corr}^{5.5}$.  The steep nature of this relationship is partly due to the use of spectroscopic temperature.  The spectroscopic temperature can differ considerably from the true temperature.   To illustrate this, we plot in Figure \ref{fig.4} the $L_X-T$ relationship for our $z = 0$ simulated groups using the core-corrected mass-weighted temperature ($T_{\rm mw}$ \textendash{} black dotted curve).  For  $ T \simless 1$ keV, $L_X \propto T_{\rm mw}^{3.8}$.  The difference between this relationship and the self-similar expectation ($L_X \propto T$) suggests that the groups have been subjected to some process that has a greater impact on the IGrM in groups with the shallowest potential wells (low temperatures) and less so on the gas in groups with the deepest potential wells (high temperatures).    Both heating (or preheating) of the IGrM by the galactic outflows, which will cause the gas to expand out of (or resist falling into) the shallowest potential wells, as well as the removal of the IGrM by radiative cooling, are plausible mechanisms.   At this point, we cannot distinguish between these two.

For $T > 1.2$ keV, the $L_X-T$ based on the spectroscopic and the mass-weighted temperatures converge, implying that in massive groups, the former is a good measure of the latter.  Both scaling relations also start to flatten as bremsstrahlung grows in importance.  However, we note that compared to the observations, the  high mass simulated groups are more luminous, and/or   a bit cooler.  This suggests that the hot gas in these systems is denser than in real systems.  In the scenario where the galactic outflows really do impact the gas density in shallow wells, the emergence of overluminous groups as their halo mass approaches (and exceeds) $M\approx 10^{14}\; \rmn{M_{\odot}}$ suggests that the outflows have become ineffectual and another mechanism is necessary.

For illustrative purposes, we also plot $L_X-T$ relationship for the $z = 0$ groups using the mean emission-weighted temperature ($T_X$ \textendash{} black dashed curve).  Until relatively recently, this was the relationship used to compare model $L_X-T$ to observations. For the lowest temperature groups, the emission-weighted and the spectroscopic temperatures are nearly equal, and the two $L_X-T$ curves track each other. For $T > 0.7$ keV, the emission-weighted $L_X-T_X$ deviates from that based on the spectroscopic temperature and remains in good agreement with the observations. This agreement, however,  is spurious since the observations are based on the spectroscopic temperature, and indicates a need for caution: The comparison between the $T_X-$based relation for simulated groups and the $T_{\rm spec}-$based observations masks the need for an additional heating or redistribution mechanism in the more massive groups.

It is instructive to compare our results to those of the ``stellar feedback only'' run from the OWLS collaboration  (referred to as the ``Reference Model'' in \citealt{M10} and
as the ``ZCool+SF+SN model'' in \citealt{M11}).  This model (hereafter referred to as OWLS-stars) primarily accounts for only SNe feedback, which is implemented via the kinetic  wind model of \citet{DVS08}, where the mass loading factor is fixed to constant  ($\eta=2$)  instead of varying inversely with galaxy velocity dispersion as in our model and the wind velocity {is also set to a constant,} $600$ km/s.   Comparing the resultant $L_X-T_X$ for the simulated groups
\citep[\cf~right panel of Figure 6 in][]{M10}, we find that the groups in the OWLS-stars run are about a factor of $\sim 10$ more luminous than our groups at $T\approx 0.5$ keV, and about a factor of $\sim 2$ more luminous than our groups at $T\approx 1$ keV.    While our results match the observed $L_X-T_X$ relationship, theirs lie systematically above the observations and define a shallower trend.   A comparison with the results shown in \citet{M11} indicates that this difference is due to a larger IGrM component within $R_{500}$ in the OWLS-stars groups and possibly that the IGrM is more centrally concentrated and hence, denser.  We will comment on this further in the next section where we discuss the IGrM and the baryon fractions in our groups.

Finally, we note that the error bars on the simulation results  indicate 1-$\sigma$ scatter above and below the mean.   The scatter in the observational data is significantly larger.   There are a number of possible reasons for this {difference}.  First, our simulations are missing AGN feedback and one can imagine that the variations introduced by yet another heating/redistribution mechanism could increase the dispersion in the X-ray luminosity of the simulated groups at a fixed temperature \citep[see, for example,][]{M10}.   The observations, however, are also not homogeneous.  For instance, we plot all groups in \citet{OP04}, including those in which the X-ray emission is only detected to a relatively small radial extent.   This can artificially enhance the scatter.   In fact, most group samples, including the ones plotted, are not statistically representative \citep{O14} and the inhomogeneities and biases will also be reflected in the scatter.

Figure~\ref{fig.5} shows the $L_X-M$ trends for the simulated groups at redshifts $z=0$ (black), $z=0.5$ (blue), $z=1$ (red), $z=2$ (green) and $z=3$ (cyan). This time we plot the rest-frame 0.1-2.4 keV X-ray luminosity to facilitate comparison with available observations.  Like the $L_X-T_{\rm spec,corr}$ curves, once cosmic evolution is taken into account through the dimensionless Hubble parameter $E(z)$, the $L_X-M$ curves for the $z\leq 1$ group populations scale as $L_x\propto M_{500}^{1.7}$ and essentially lie on top of each other, implying a self-similar evolution over this redshift range.   Like the $L_X-T_{\rm spec,corr}$ relations, the higher redshift curves lie off {those at} $z\leq 1$ and {the} deviation goes in the same direction.   The groups  at a given value of  $M\:\! E(z)$  \textendash{} note that temperature and $M \;\! E(z)$ are equivalent measures of the depth of the gravitational potential well \textendash{}  are less luminous than predicted by the self-similar evolution model and the underlying reasons, {on} which we will elaborate when we discuss the behaviour of the hot gas fraction in the groups, are also the same.

There are two different types of observational data shown in the plot:  One set  (black circles and stars), where the group masses are in fact hydrostatic mass estimates derived from the X-ray measurements  by \citet{EK11} and  \citet{P09}, respectively, and another  (black, blue and red squares), where the masses are derived using weak lensing methods \citep{L10}.    It is well known that the hydrostatic masses typically underestimate the true mass and to facilitate a fair comparison with our {\it actual} mass determinations, we have corrected the \citet{P09} and \citet{EK11} masses  using the  bias factor determined by \citet{H15} \citep[see also][]{MH13}.   Our numerical results are in good agreement with the observational results.    Moreover,  {the fact} that the $L_X-M$ trends delineated by the black, blue and red points show no significant {offsets} from each other  suggests that the evolution of the observed group $L_X-M$ relation is consistent with that predicted by the self-similar model over the redshift range $0 \leq z < 1$, which is also in agreement with our numerical results.
 
 \begin{figure}
  \vspace*{0pt}
  \includegraphics[width=0.48\textwidth]{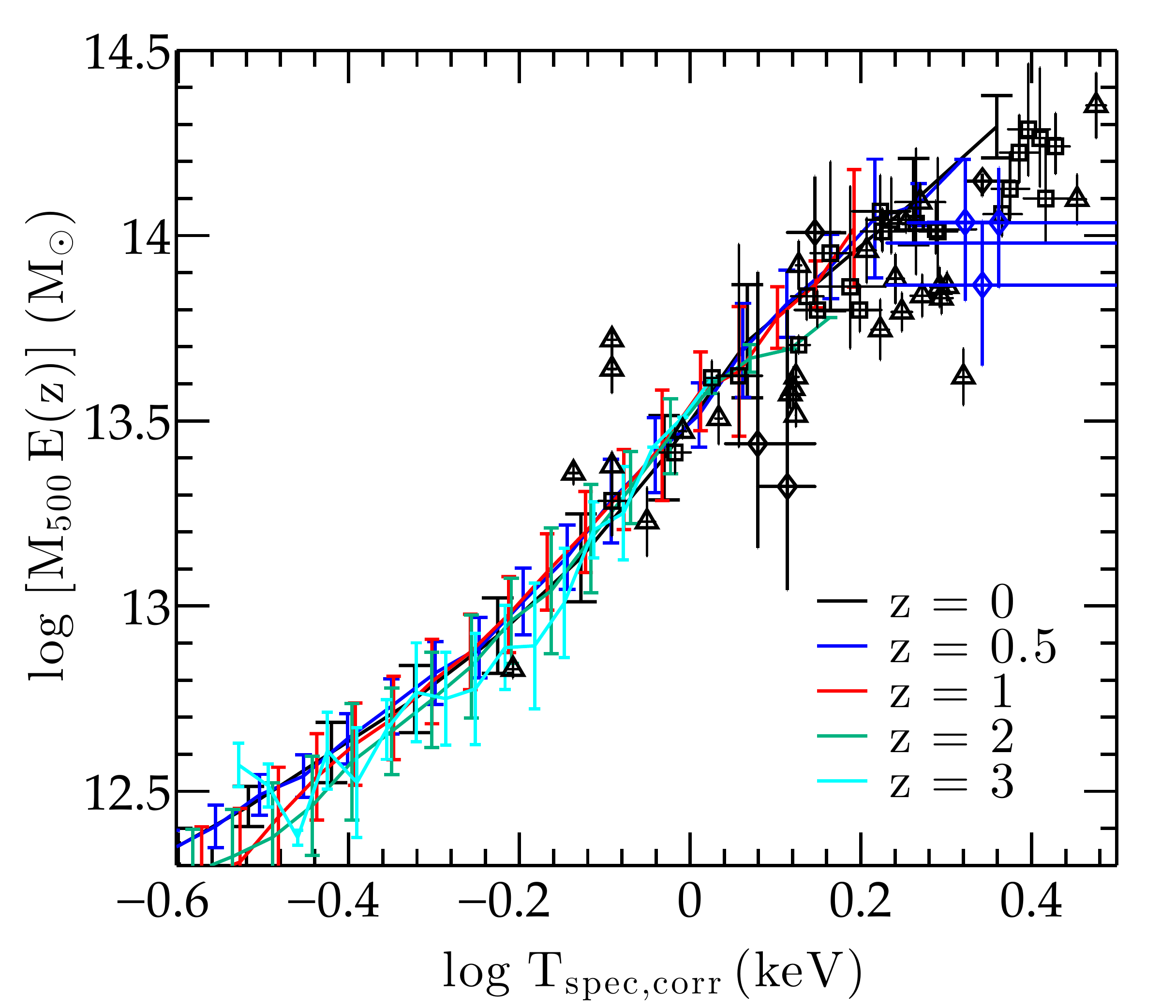}
  \caption{ $M-T_{\rm spec,corr}$ relation for simulated groups at $z = 0$ (black), $z = 0.5$ (blue), $z=1$ (red),  $z=2$ (green), and $z=3$ (cyan). The error bars show 1-$\sigma$ scatter.  
The black squares and triangles show the results from \citet{SN09} and \citet{EK11}.  The hydrostatic mass estimates given in these two studies have been corrected for the hydrostatic bias  \citep{H15}.  We also note that the temperatures in the latter study are not always extracted in a consistent, systematic fashion.  The  diamonds show results from \citet{K13};  their masses are weak-lensing estimates. The observed groups at $z\leq 0.25$ and $0.25<z\leq 0.75$ are plotted as black and blue symbols, respectively.}
  \label{fig.6}
\end{figure}

Figure~\ref{fig.6} shows the $M-T_{\rm spec,corr}$ trends for the simulated groups at redshifts $z=0$ (black), $z=0.5$ (blue), $z=1$ (red), $z=2$ (green) and $z=3$ (cyan).   Once cosmic evolution is taken into account, all the curves, even those for $z  > 1$ groups, lie on top of each other.   The shape of these curves differs from that shown in Figure~\ref{fig.3} because there we had plotted mass versus emission-weighted temperature whereas here we are using the core-corrected {\it spectroscopic} temperature.   Comparing {the} $M-T$ relation for our simulated $z=0$ groups to that of OWLS-stars groups \citep[\cf~Figure 5 of][\textendash{} this figure shows mass versus emission-weighted temperature whereas we plot mass versus spectroscopic temperature]{M10}, we find that the two are in excellent agreement with each other once the differences between the two temperatures at $T_{\rm spec,corr} \simgreat 0.8$ keV are accounted for.

As in Figure~\ref{fig.5}, we plot corrected  hydrostatic masses (squares and triangles) and masses derived from weak lensing analyses (black and blue diamonds).   Within the scatter,  all the data (weak lensing and X-ray based, present-day and at moderate redshift) are consistent with each other.   Focusing on measurements with $T_{\rm spec,corr} > 0.8$ keV, the scaling of mass with temperature can be well described by $M_{500}\propto T_{\rm spec,corr}^{1.6}$, which is consistent with the self-similar scaling relationship.  The numerical trends, on the other hand, are  steeper in the neighbourhood of $~1$ keV but flattens at both higher and lower temperatures.    In spite of this, the numerical results are consistent with the observations except perhaps {for} the most massive groups with $M_{500} \simgreat 10^{14}\; \rmn{M_{\odot}}$.  In these massive systems, the temperature in the simulated systems seems to be a bit cooler than that in the observed systems.    We note that this divergence at the high mass end is also present in the $L_X-T_{\rm spec,corr}$ plot.

At various points in the preceding discussion, we have noted that the trends exhibited by the simulated groups in the $L_X-T_{\rm spec,corr}-M$ space are the consequences of two different types of processes: (1) the structure of the intergalactic medium that collapses to form the IGrM and, therefore, the properties of the IGrM itself,  {are} altered by the galactic outflows from {an} earlier ($z>2$) generation of galaxies; and (2) heating and/or cooling processes occurring once the groups form can convert the cooler, denser, X-ray luminous gas into either  hotter, more diffuse (and therefore, less luminous) gas, or into cold, dense gas that is effectively dark in the X-rays.   {To gain insights into the relative importance of these effects, we examine the entropy of the hot X-ray emitting gas.}

\subsection{The Entropy-Temperature Scaling}

Entropy is a very useful physical quantity to consider when the IGrM is subject to cooling and heating processes because the former typically lowers the entropy while the latter always raises it.   This is not the case with either density or temperature because heating can cause the gas to expand, potentially lowering both quantities \citep[see][for a more detailed discussion]{M08}.  Additionally, the gas distribution will tend to organize itself so that the lowest entropy gas is at the group centre and the highest is at the group periphery.  On the other hand, in the present situation, there is an additional complication to keep in mind: If cooling is able to cause the low entropy IGrM in the center to drop out and settle in the central galaxy, higher entropy gas from further out will flow in to take its place  and the entropy in the centre will appear ``enhanced''  unless cooling is able to erode the entropy of this inflowing gas as quickly as it flows in.   This effect was first seen in the numerical simulation by \citet{L00} and discussed more fully by \citet{VB01}. 

\begin{figure}
  \vspace*{0pt}
  \includegraphics[width=0.48\textwidth]{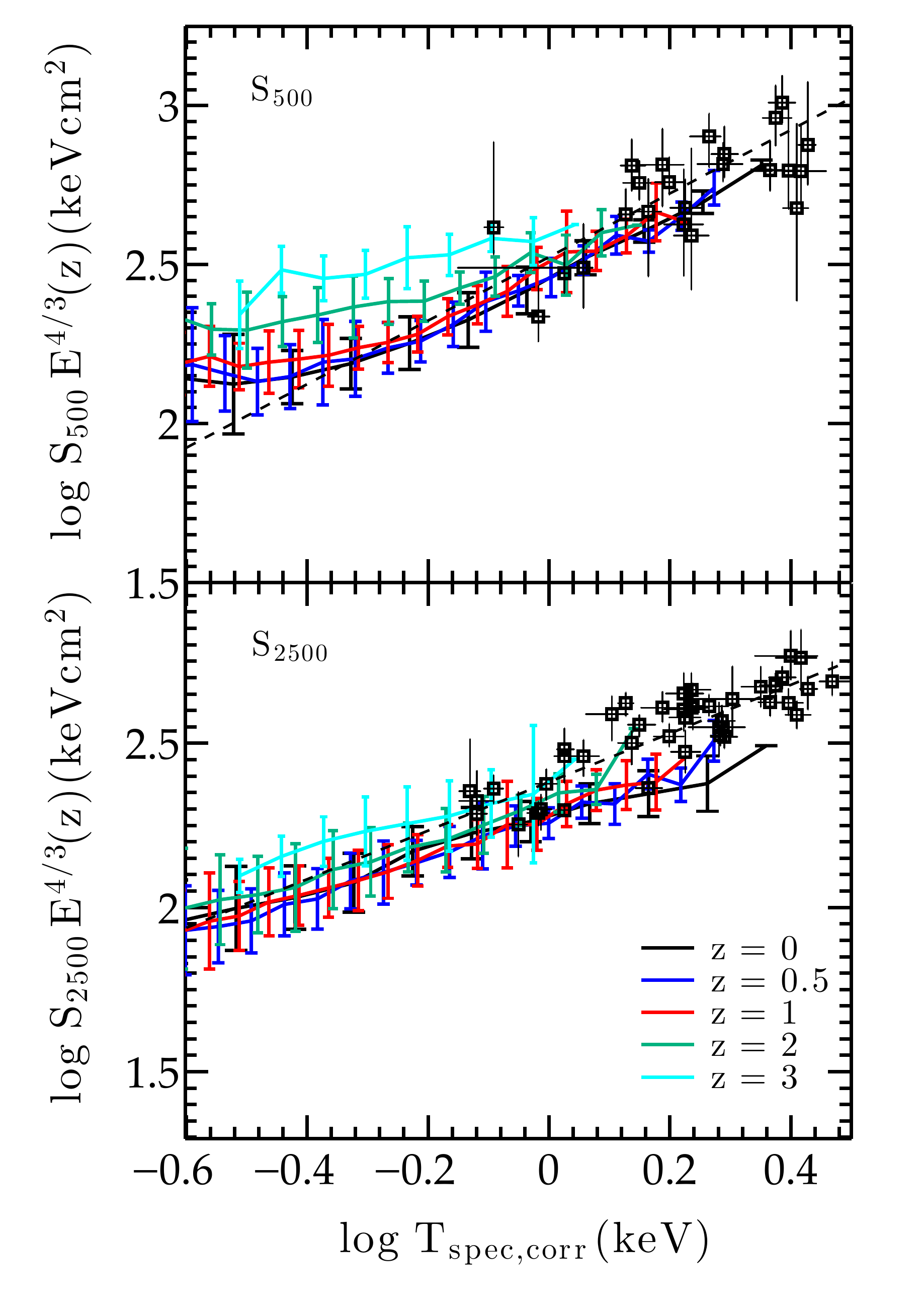}
  \caption{ Gas entropy at $R_{500}$ (top panel) and $R_{2500}$ (bottom panel) of the simulated groups at $z=0$ (black), $z=0.5$ (blue), $z=1$ (red), $z=2$ (green) and $z=3$ (cyan), as a function of core-corrected spectroscopic temperature.  The error bars show 1-$\sigma$ scatter. The observational data of the low redshift sample from \citet[][hereafter S09]{SN09} is shown by black squares. The dashed lines in the top and bottom panels represent the power-law fits to the $S-T$ relation at the two different radii for the full group+cluster sample from S09, with a power law index of 1 and 0.74, respectively.}
  \label{fig.7}
\end{figure}

In Figure~\ref{fig.7}, we show the cosmic expansion corrected gas entropy at $R_{500}$ (top panel) and $R_{2500}$ (bottom panel), in present-day groups (black solid curve) as well as for groups at $z=0.5$ (blue), $z=1$ (red), $z=2$ (green), and $z=3$ (cyan), as a function of the integrated core-corrected spectroscopic temperature of the IGrM within $R_{500}$ .  We also plot entropy estimates from {the} X-ray observations of \citet{SN09}, computed using the deprojected temperature and electron density profiles.  To be precise, we are {\em not} plotting the actual thermodynamic specific entropy, but rather its widely accepted proxy\footnote{The `entropy proxy' (S) and the thermodynamic specific entropy (s) are related to each other as $ds\propto d\ln{S}$ \citep[see][]{BB99}.} given by

\begin{equation}
S(r) = \frac{k_B T_{\rmn{spec}}(r)}{n_e(r)^{2/3}}\:,
\end{equation}  

\noindent where $k_B$ is the Boltzmann constant, and $T_{\rmn{spec}}(r)$ and $n_{e}(r)$ are, respectively, the spectroscopic temperature and the electron number density within a thin spherical shell at radius $r$.  We shall, hereafter, refer to $S(r)$ as `entropy'.   

Focusing first on the $R_{500}$, we see that scaled entropy systematically declines with time.  The change is much greater at high redshifts ($z > 1$) than at low redshifts ($z < 1$) and also, much more stronger in groups with shallow potential wells than in those with deep potential wells.    Both results are primarily driven by the evolution in the IGrM density at $R_{500}$.   For group halos of a given temperature, if we take the present-day value of IGrM density at $R_{500}$ as a reference, then we are led to conclude that the IGrM density at $R_{500}$ at an earlier epoch is not  $E(z)^2$ times the present-day value, as would be expected if the IGrM density were evolving self-similarly.   Rather, it is lower than the expected value and hence, the cosmic expansion corrected entropy is higher.   Put another way, the halos at $z=3$ are IGrM poor (relative to the total matter).   As the halos grow, they encompass additional dark and baryonic matter but the ratio of baryons-to-dark matter is larger than the universal value.   In effect, the groups are recapturing some of the baryons that were ejected from their member galaxies at earlier times, in addition to the usual complement associated with the accreting dark matter (reminiscent of the ``outside-in'' IGM enrichment scenario of \citealt{O12}).   The  higher temperature groups are able to recapture a greater fraction of the previously expelled baryons at an earlier time ($z > 2$) than the lower temperature groups.   The outcome of this differential accretion of baryons and dark matter is that the density of hot gas in the groups evolves differently from self-similar expectations  \textendash{} \ie~until $z\approx 1$, after which the evolution of the entropy and the density for all the group halos is consistent with self-similar evolution.  We will examine the density and entropy profiles in detail in a follow-up paper.   However, this general behaviour also explains the self-similar evolution of the $L_X-T_{\rm spec,corr}$ and $L_X-M$ from $z=0$ to $z=1$ and then, the decrease in the amplitude of the (cosmic expansion corrected) scaling relation from $z=1$ to $z=3$ (\cf~Figures~\ref{fig.4} and~\ref{fig.5}).  At higher redshifts, the groups of a given temperature or ${\rm M}\:\! E(z)$ (both are equivalent measures)  are not as luminous as they ought to be if they were evolving self-similarly because the IGrM is not as dense as self-similarity would predict.

In contrast with $S_{500}$,  $S_{2500}$ evolves somewhat differently.  The scaled entropy evolves self-similarly from $z\approx 2$ to the present and even between $z=3$ and $z=2$, the change is relatively mild.   The core regions tend to form earlier and we expect that they will settle down into a steady-state configuration at an earlier epoch.  DOS08 compared the core entropies of groups in a simulation with and without outflows and found that the gas in the former {case} had higher entropy.   This is a non-trivial result.    As we mentioned above, it is not unexpected to see elevated entropies at $R_{2500}$ in simulations with cooling, star formation and inefficient stellar feedback \citep[hereafter \CSF simulations \textendash{} for more details about such simulations, see ][and DOS08]{L00, K05}, relative to a non-radiative {simulation}  (\cf~Figure 10 of \citealt{L00}).  The inclusion of outflows, however, could just as easily have led to still higher core entropies, or lower core entropies.  As DOS08 explain, the latter can occur if heating from the outflows just balances cooling, and the low-entropy gas remains in place {unchanged, \ie~}it is neither removed via cooling nor raised to a higher adiabat.   The outflows in our simulations, on average, heat the IGrM at least close to the centre of the groups.

Comparing $S_{500}$ and $S_{2500}$ entropy results for our simulated groups to the observations, we find that the IGrM entropy at $R_{500}$ in our massive groups is consistent with the observations. \citet{SN09} find that the observed $S_{500}$ values scale with the integrated core-corrected spectroscopic temperature of the gas within $R_{500}$ as $S_{500} \propto  T_{\rm spec,corr}$ (dashed line) across their full sample of groups and clusters. Our simulation results also follow the same scaling for $T_{\rm spec,corr} > 0.3$ keV. Below this threshold, $S_{500}$ in the simulated groups flattens. This flattening {is because of a decrease in the IGrM density} with decreasing group mass. We will discuss this further in the next subsection.   At $R_{2500}$, \citet{SN09} find that the observed entropy scales as $S_{2500}\propto T_{\rm spec,corr}^{0.74}$ (dashed line).   The $S_{2500}$ of the groups in the simulated sample is less steep, scaling with temperature as $S_{2500}\propto T_{\rm spec,corr}^{0.50}$, and the simulated groups at the high mass end have approximately $40\%$ lower core entropies compared to the observations.  One explanation is that the density of IGrM in the cores of the massive simulated groups is slightly higher than in real groups and this, in turn, would explain why massive simulated groups seem to be somewhat more X-ray luminous than their real counterparts.

Comparing our entropy results  to those of groups in the OWLS-stars run \citep[see Figure 2 in][]{M10}, we find that entropy at $R_{500}$ in the two simulations is very similar.  At $R_{2500}$, however,  the entropy in the OWLS groups is lower than in our groups by approximately $25-30\%$.   This supports our previous conjecture that the IGrM in OWLS groups is more centrally concentrated than in our groups, which in turn would explain the differences in the X-ray luminosities of the two group populations.


Overall our stellar-powered wind model fares remarkably well when it comes to matching the observed group $L_X-T_{\rm spec,corr}-M$ {scaling relations} except perhaps in the most massive groups.  The behaviour of the IGrM entropy at $R_{2500}$ and $R_{500}$ suggests that the principal variable governing the behaviour of these relationships in our simulated groups is the IGrM density.

\section{THE BARYON CONTENT OF GALAXY GROUPS}
\label{Sec:4}

\begin{figure*}
  \vspace*{0pt}
  \begin{center}
  \includegraphics[width=0.95\textwidth]{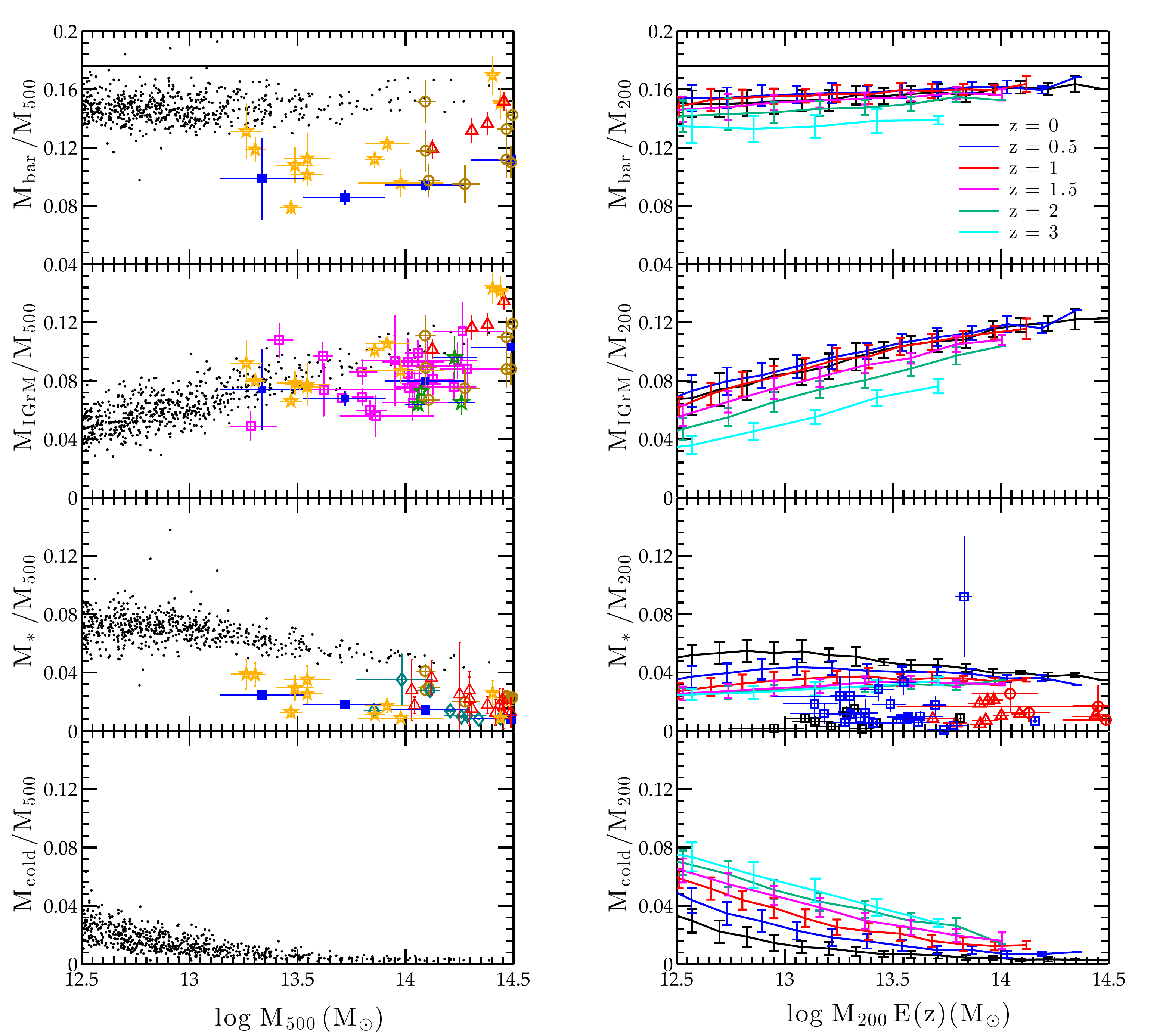}
  \end{center}
 \caption{Left column:  Stellar and gas mass fractions within $R_{500}$ in simulated $z=0$ groups. {\it Top panel}: Total baryonic fraction. The black line indicates the cosmological value, $\Omega_{b}/\Omega_{m}$ = 0.176.  The symbols (see text for details) show observational estimates for hot gas + stars. Error bars depict 1-$\sigma$ scatter. {\it Second panel}: Hot gas fraction. {\it Third panel}: Stellar mass fraction.   The simulation results include stars in the galaxies as well as those comprising the diffuse intragroup stars [IGS]) component.  Of the observational estimates, only the golden circles \citep{AG13} account for the IGS.   {\it Bottom panel}:  Cold gas fraction ({\it i.e.}~diffuse gas with $T < 5\times10^{5}$ K and the galactic ISM).   Right column:  The same mass fractions for simulated groups at $z = 0$ (black), $z = 0.5$ (blue), $z = 1$ (red), $z=1.5$ (magenta), $z = 2$ (green) and $z=3$ (cyan) computed within $R_{200}$ to facilitate comparison with observations. Triangles, circles and squares are observational results from \citet{M13}, \citet{V14} and \citet{C12}, respectively. Data for $z \simless 0.25$ groups are in black, $0.25 < z \simless 0.75$ in red, and $0.75 < z \simless 1.25$ groups in blue.  These do not account for the IGS.}
  \label{fig.8}
\end{figure*}

We now turn to direct determinations of the total baryon fraction as well as the stellar and the IGrM gas mass fractions in the simulated groups to confirm whether the IGrM density behaves as we have argued above, and try to understand {the reasons behind its behaviour}.   The partitioning of the baryons between {stars and hot gas} is interesting in and of itself.   We know that our model is not able to suppress the overproduction of stars in the largest galaxies in the groups (\cf~ Figure~\ref{fig.2}) but by looking into the issue more carefully, we hope to understand when and where the model starts to fail.   Additionally, we also quantify the assembly of the groups using several different measures.

\subsection{Stellar, Gas and Total Baryon Fractions}

The sequence of plots in the left panel of Figure~\ref{fig.8} shows the total baryon,  $T>5\times 10^5\; {\rm K}$, hot, diffuse gas (IGrM), stellar, and cold gas fractions in simulated groups at $z=0$, as function of the total group mass.   All {the} quantities are computed with $R_{500}$ to facilitate comparisons with the observations.    The plots in the right column show the same mass fractions within $R_{200}$ to illustrate their behaviour across group populations at different redshifts as well as beyond just the inner {regions} of the halos.  The coloured lines show the results for groups at $z=0$ (black), $0.5$ (blue), $1.0$ (red), $1.5$ (magenta), $2.0$ (green) and $3.0$ (cyan).  On the $x$-axis, we plot $M_{200}\;\! E(z)$ so that we can compare populations of halos with potentials of similar depths across the {different} epochs.  Baryon properties, as well as the impact of feedback and radiative cooling, both tend to be strongly correlated with the depth of the halos' gravitational potential well.    For completeness, we note that we have tracked the evolution of individual groups in the right column of Figure~\ref{fig.8} and find that once formed, they {\em do not} simply evolve vertically in these plots.  Rather, they move both vertically up or down (depending on the quantity under consideration) with decreasing redshift, transitioning from one coloured line to the next,  while also generally sliding to the right along the $x$-axis.  This is because the masses of individual groups generally grow faster than expected under the self-similar growth model; {\ie}~they increase faster than $M_{200}(z_{\rm group}) \left[E(z_{\rm group})/E(z) \right]$, where $z_{\rm group}$ is the redshift at which the most massive progenitor halos of the present-day groups first acquire three luminous galaxies and as per our definition, become `groups'.  Consequently, the potential well of individual groups tend to deepen towards the present.   

In the first panel of both columns, we investigate the total baryon fraction.   The solid line indicates the cosmological baryon fraction of $\Omega_b/\Omega_m=0.176$ for the simulation.  The red triangles, blue squares, golden circles and orange stars in the left panel show observational estimates of the fraction of mass in hot gas and stars, which for massive groups is essentially the same as the total baryon fraction, from \citet{LY03}, \citet[revised\footnote{The results for \citet{GD09} shown in Figure~\ref{fig.8} differ from those in their paper because they have been revised as suggested by \citet{AL12}.  See also discussion in \citet{GD12}.   The stellar masses results presented are derived using {the} Chabrier IMF.}]{GD09},  \citet{AG13}, and \citet{LM13}, respectively.   Of these, only \citet{AG13} (golden circles) explicitly account for the extended, diffuse intragroup stellar component (hereafter referred to as the intragroup stars or IGS).   The simulation results include all stars, those in the galaxies as well as those that belong to the extended population.   In spite of the large scatter in the observed values, the total baryon fraction within $R_{500}$ in our simulated present-day groups  is generally greater than that in real groups by approximately $35-40\%$.  

\begin{figure*}
  \vspace*{0pt}
  \begin{center}
  \includegraphics[width=0.95\textwidth]{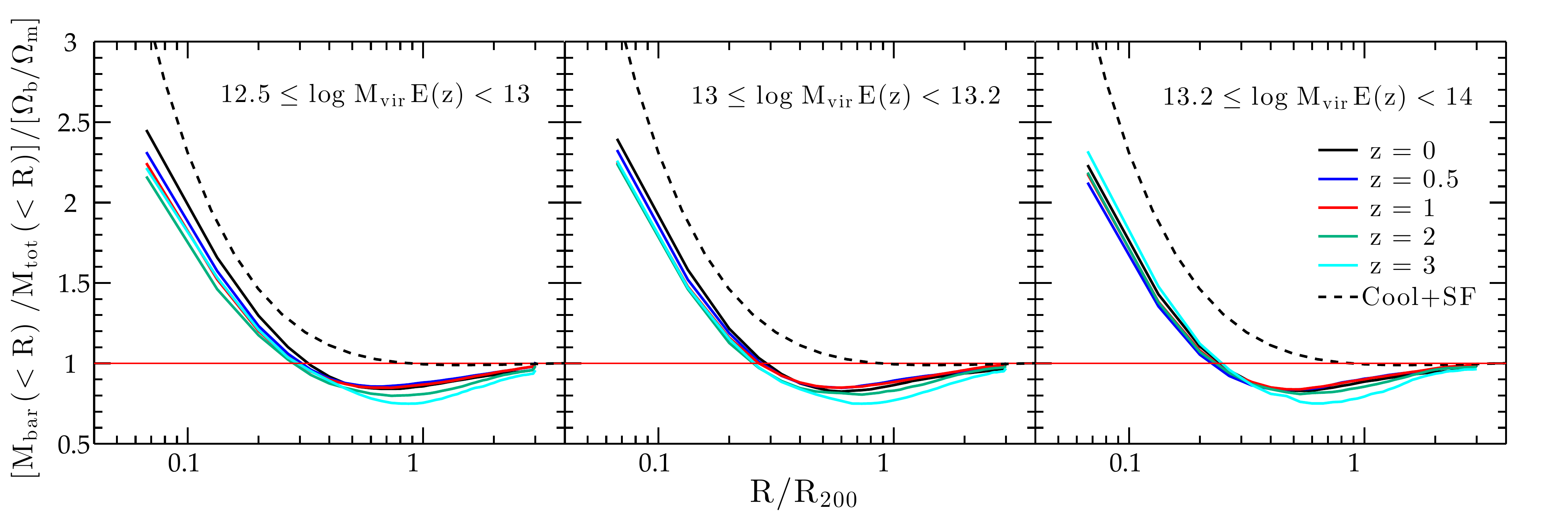}
  \end{center}
 \caption{The mean baryon fraction within radius $R/R_{200}$ in simulated groups at $z=0$ groups (black curve),  $z = 0.5$ (blue), $z = 1$ (red), $z=1.5$ (magenta), $z = 2$ (green) and $z=3$ (cyan), normalized to the cosmic baryon fraction $\Omega_b/\Omega_m = 0.176$ for the simulation.  We have sorted the  groups into three bins according to the depth of their potential wells:  
 In each panel, the dashed black curve shows the $z=0$ mean baryon fraction profile for the \CSF simulation from \citet{L00}, {which had no galactic winds.}
 }
\label{fig.9}
\end{figure*}

This is the first indication that powerful stellar-powered galactic outflows, in and of themselves, are not capable of preventing the over-concentration of baryons within $R_{500}$ in the simulated groups, in comparison to the observations.   This is not to say that the outflows have no impact on the galaxy groups.    The baryon fraction of our simulated groups is comparable to that seen in non-radiative simulations \citep[see for instance][]{C07} \textendash{} \ie simulations that do not allow for cooling.  It is most definitely lower than the values seen in the \CSF simulations \citep[][DOS08]{L00, K05},  where the baryon fraction within $R_{500}$ is typically equal to or even slightly larger than the universal value.   Below we try to understand precisely how the galactic outflows affect the baryon fraction in our simulated groups.

{
Examining the simulation results more closely, we note that the median value of the baryon fractions within $R_{500}$ in the $z=0$ groups  exhibits a gradual rise with increasing halo mass, going from $\sim 82\%$ of the universal cosmological value in the least massive groups to $\sim 90\%$ of the universal value in the most massive groups.  Comparing these results against those for the OWLS-stars groups (in this case ``ZCOOL+SF+SN" {simulation}; \citealt{M11}), we find {good agreement.  The median baryon fractions in the latter groups range from $88\%$ of the cosmological value in the lower mass groups to $92\%$ in the high mass systems.}   The median baryon fractions within $R_{200}$ of the $z=0$ groups  also increases gradually with halo mass,  from $\sim 86\%$ of the universal value in our lowest mass groups to $\sim 93\%$ in the high mass groups.   The baryon fractions within $R_{500}$ are slightly but systematically lower than within $R_{200}$ because, in our simulated groups, the principal source of outflows is the dominant central galaxy and as noted while discussing the IGrM entropy, these preferentially heat the gas in the central regions.}


The trend highlighted above, of lower mass systems being more baryon depleted, is a common feature of the strong stellar feedback models.  
In the case of {the momentum-driven wind model}, \citet{D09} has shown that the baryon fraction in present-day Milky Way-sized halos ($\sim 10^{12}\;\rmn{M_{\odot}}$) drops to $60\%$ of the cosmic value  and decreases further to  below $50\%$ of the cosmological value in
lower mass galactic halos ($\sim 10^{10}\;\rmn{M_{\odot}}$).   On these galactic scales, the reason is fairly clear.   The winds physically carry away a large fraction of the baryons from such systems and in fact, can drive down the baryon fraction over regions that extend well beyond the virial radius of the galactic halos.   Similar results are also seen in other ``no AGN'' simulations where stellar feedback is implemented via a thermal prescription \citep[see ][and references therein]{SM15}.




As one {moves} up in halo mass, the deepening gravitational potential wells engender a transition from a state where the bulk of the baryonic matter within a halo is in the form of stars and cold gas localized in the galaxy ({or} galaxies), to one where the hot diffuse gas component that suffuses the entire halo eventually dominates the baryon budget.    This transition has been discussed in detail by \citet{K09} and \citet{GD15}, and references therein.  Following \citet{GD15}, we define the transition point between these two states as one where the hot gas mass exceeds 50\% of the total {\it gas} mass.   In our simulations, this changeover group mass occurs at  $M_{200}\approx 1.8\times 10^{12}\;\rmn{M_{\odot}}$ at $z=0$, $M_{200}\approx 3\times 10^{12}\;\rmn{M_{\odot}}$ at $z=1$ and $M_{200}\approx 6.3\times 10^{12}\;\rmn{M_{\odot}}$ at $z=2$.  These masses are comparable to the transition mass cited in \citet{GD15} although in detail our transition masses are slightly larger because we impose an additional constraint that the halos must host at least three luminous galaxies.   The presence of a pervasive IGrM alters the wind dynamics:  {Galactic} winds flowing through such a medium {are} subject to hydrodynamic drag, the relative importance of which grows as the density of the medium increases.   In the case of our group halos, the combination of the deeper gravitational potential wells and the higher likelihood of strong hydrodynamic interactions results in the winds being almost completely confined within the halos.   So the reduced baryon fraction, relative to the cosmic mean, is {\it not} due to the outflow of the baryons.

Instead, the reduced baryon fraction in the group halos {is the result of} the following two effects.  First, groups typically form in regions that have been rendered somewhat baryon deficient by winds from their member galaxies at earlier times.   Consequently, when the groups first form, their baryon fraction can be as low as $\sim 75\%$ of  the cosmic mean, as illustrated by the baryon fraction curve for $z=3$ groups in the top right panel of Figure~\ref{fig.8}.   This ``baryon depletion'' is illustrated much more clearly in Figure~\ref{fig.9}.     At $z\approx 3$, for example,  the baryon fraction within $R_{200}$ is $75\%-77\%$ of  the mean value for the simulation, and one would have to go out to $\sim 3 R_{200}$ before the fraction returns to the cosmic mean.  (For comparison, we also show the baryon fraction profile for a \CSF {simulation} from \citet{L00}; the baryon fraction never really drops below the mean value and converges to the mean value by $R_{200}$.)  This depletion is slightly less pronounced in the more massive groups and diminishes towards the present.  

Returning to the top right panel of Figure~\ref{fig.8},  we have noted previously that individual groups, once formed, tend to grow in mass.   And as the halos grow, their physical reach extends further out.  As a result, in addition to the inflow of the usual baryonic complement of the accreting dark matter,  they are also able to recapture an increasing fraction of the expelled gas previously associated with the group galaxies \citep[\cf][for a discussion of {a} similar phenomenon in galactic halos.]{FD14}, and the overall baryon fraction increases with halo mass.

This, however, is not all.   There is a second effect at play, otherwise we would expect the baryon fraction to continue rising with decreasing redshift and approach the mean cosmological value.   Instead, we observe a sharp increase in the halo baryon fraction between $z= 3$ and $z= 2$, a much more tempered rise from $z= 2$ to $z= 1$, and very little change, if any, thereafter.     This second effect is the result of the winds ejected from the group galaxies interacting with and heating the hot halo gas,  which not only reduces the rate at which the halo gas cools and accumulates in the group central galaxies but also causes its distribution to {\it remain} more extended.  

In the preceding discussion, we use the word ``extended'' deliberately.   Hot gaseous halos generally extend beyond the virial radius \citep[\cf][]{BM13,GD15} but in simulations without winds, radiative cooling inside the halos leads to the loss of pressure support, which then results in a denser, more compact baryon distribution.   In the case of our wind model, the hot halo gas density starts out lower than usual and it is easier for heating by the winds to compensate for a significant fraction of the radiative cooling losses and drastically slow down its collapse.   As indicated in the top right panel,  halos with gravitational potential wells of a given depth (\ie at a fixed $M_{200} E(z)$) establish an equilibrium distribution by $z=1$, and the baryon fraction remains essentially constant from $z=1$ to the present.

However, the efficacy of the galactic winds to maintain an extended hot gas distribution via heating drops with deepening potential wells because the characteristic temperature  ($T_{\rm wind}$) corresponding to the complete thermalization of the kinetic energy in the outflows from any one galaxy in a group halo, even the dominant galaxy, does not grow as quickly as the group halo's virial temperature ($T_{\rm vir}$).   The reason for this is that $T_{\rm wind}$ scales as $M_{\rm gal}^{2/3}$, where $M_{\rm gal}$ is the mass of individual galaxies, whereas $T_{\rm vir}$ grows as $M_{\rm 200}^{2/3}$ and as we explain below,  the fraction of the baryons condensing into the cold gas$+$stars phase decreases with increasing group halo mass.  Moreover, by virtue of being multi-galaxy systems, even the baryons that have condensed out are distributed over 3 or more ``luminous'' galaxies.  In the end,  the IGrM is not able to withstand gravitational compression despite being heated.

The second set of panels in Figure~\ref{fig.8} show the diffuse hot ($T>5\times 10^5\; {\rm K}$) IGrM gas fraction in the simulated groups.  The IGrM fraction within $R_{500}$ (left column) nearly doubles, from $\sim 0.05$ to $0.1$, in going from $M_{500}\approx 3\times 10^{12}\;\rmn{M_{\odot}}$ to $M_{500}\approx 10^{14}\;\rmn{M_{\odot}}$.   {This increase with group mass is the result of the larger mass systems having deeper potential wells and higher virial temperatures.  As a result, more of the diffuse gas is shock-heated} to {constitute the IGrM.  Additionally, the deeper potential wells are also better able to compress and confine this gas.}

Comparing our results against Figure 4 of \citet{M10}, we find that the IGrM gas fraction within $R_{500}$ of comparable groups at the low mass end of the group distribution in our simulation is about $30\%$ lower (\ie~$0.053$ for our groups versus $0.07$ for the OWLS-stars groups) and about $10\%$ lower for the groups at the high mass end.   The lower IGrM fraction is a key reason why our simulated groups are less X-ray luminous than the OWLS-stars groups.  We do note that \citet{M10} define the IGrM using the temperature cut of $T>1\times 10^5\; {\rm K}$ instead of $T>5\times 10^5\; {\rm K}$, as we have.  However, we have recomputed the IGrM fraction for our groups using this lower threshold and find {negligible changes to our} results.

Comparing the simulation results for the $z=0$ IGrM fraction within $R_{500}$ (left panel)  against observations (red triangles, open magenta squares, green stars , filled blue squares, golden circles, and orange stars are data from \citet{LY03}, \citet{SN09}, \citet{SA09}, \citet{GD09}, \citet{AG13}, and \citet{LM13}, respectively), we find that the two are in reasonable agreement.  In detail, there is a hint that the IGrM fraction in the simulated groups is rising slightly faster with increasing group mass but it is difficult to be more definite given the large scatter in the observations.   Such a trend would, however, be consistent with our previous finding that the most massive simulated groups are slightly more X-ray luminous than real systems and that the stellar-powered winds are unable to keep the baryon fraction from creeping upwards.   Together, all of these suggest the need for another gas heating/redistribution mechanism.

The right panel in the second row of Figure~\ref{fig.8} shows the IGrM fraction in the groups at different redshift.   Like the results for the baryon fraction  (top right panel),  the IGrM fraction in halos with comparable potential wells increases between $z\approx 3$ and $z\approx 1$ and then, stabilizes.    Since most of the freshly accreted baryons directly contribute to the IGrM over the epochs and in the halo mass regime being considered, that the total baryon and the IGrM fractions behave similarly is not surprising.

{To} investigate the make-up of the present-day IGrM within $R_{200}$ in detail, we have tracked all the IGrM gas particles back in time and tagged all those that, at any point in the past, were bound to a galaxy {\it and} enriched while bound.  We refer to this component of the present-day IGrM as ``processed'' and the rest of the gas as ``unprocessed.''\footnote{``Unprocessed'' material is essentially gas that has never passed through a galaxy and has entered the group halos via diffuse accretion directly from the intergalactic medium.  We emphasize that ``unprocessed'' should {\it not} be interpreted as un-enriched.   A significant fraction of the present-day ``unprocessed'' IGrM has non-primordial metallicity {owing} to enrichment by the diffuse IGS component.}  {We find  that the fraction of the present-day IGrM that  is ``processed''  gas ranges from $\sim15\%$ in the highest mass groups to $3-4\%$ in the lowest mass groups.   (We relaxed the ``enriched while bound'' condition and repeated the analysis, and got identical results.)  This increase in the fraction of ``processed'' IGrM or equivalently, the decrease in the fraction of ``unprocessed'' IGrM, with increasing halo mass is the continuation of the trend observed by \citet{FD14}.   On galactic scales ($M_{200}\sim 10^{11}\;\rmn{M_{\odot}}$), \citet{FD14} find that the ``unprocessed'' component, which they call ``ambient gas'', makes up nearly all of the hot gas.   That the fraction of processed gas in the IGrM is  relatively  small even in the most massive groups may seem surprising but only a small fraction of the metal-rich wind material ejected from central galaxies, for example, thermalizes at $T > 5\times 10^5\; {\rm K}$ and remains in the IGrM.}  Acting more like a galactic fountain, the most of the wind lifts off from the galaxy, transfers its kinetic energy to the ambient gas, and falls back into the galaxy.   {This behaviour has been discussed in detail in \citet{OD10} and \citet{FD14}} 
 
The fraction of the gas that is heated to  $T>5\times 10^5\; {\rm K}$ and is  lost to cooling is relatively small.  Over a Hubble time, it ranges from $\sim$10\% at $M_{200}= 3\times 10^{12}\,\rmn{M_{\odot}}$, to $\sim$2.5\% at $M_{200}= 3\times 10^{13}\;\rmn{M_{\odot}}$, to
$\sim$0.2\% at $M_{200}= 3\times 10^{14}\;\rmn{M_{\odot}}$.  Most of the gas that drops out of the IGrM is initially heated only to $5\times 10^5\;{\rm K} < T < 3\times 10^6\;{\rm K}$.   Metal-enriched IGrM in this temperature range sits on the broad peak of the cooling curve and is subject to efficient cooling, which even heating by stellar-powered galactic winds/fountains cannot fully offset.

The third set of panels in Figure~\ref{fig.8} show the stellar mass fraction in our simulated groups.   We include both stars in the galaxies as well as stars belonging to the diffuse intragroup component (IGS) {when we compute} the stellar mass. Comparing the stellar fraction within $R_{500}$ in our $z=0$  simulated groups to the same from the OWLS-stars simulation \citep{M11}, we find that the two are similar, ranging from $0.07$ in the low mass groups to $0.05$ in the high mass groups.  Additionally, this stellar fraction, like the baryon fraction, is a significant improvement over those seen in simulations of \citet{L00} and \citet{NKV07}, {which} do not include this type of stellar feedback, confirming that galaxy-wide outflows indeed do suppress excessive star formation.   This improvement, however, is not sufficient to bring the simulation results into agreement with the observations.   The red triangles, blue squares, green diamonds, orange stars, and golden circles in the left panel show results from \citet{LY03}, \citet{GD09}\footnote{The results for \citet{GD09} shown in Figure~\ref{fig.8} differ from those in their paper because they have been revised as suggested by \citet{AL12} \textendash{} see also discussion in \citet{GD12}.   We show the corrected results based on the Chabrier IMF.}, \citet{MB11}, \citet{LM13}, and \citet{AG13}, respectively.  Of these, only the latter  account for the IGS component.   Compared to the observations, the simulated groups have, on the whole, a factor of  $\sim$2 more mass in stars within $R_{500}$. 

In the right panel, we show how the group stellar fraction within $R_{200}$ changes with redshift.   We also plot the observational estimates of the stellar fraction in groups at $z\sim 0.5$ and $z\sim 1$ as blue and red symbols, respectively.  These data points should be compared to curves of the same colour.  These observational results are among the first estimates of the stellar fraction in groups at higher redshifts and are subject to considerable uncertainty \citep[{\cf} discussion in][for example]{AL12, AG13}.  This renders a detailed comparison between {the simulations and the observations} difficult.  Nonetheless, the general trend seen in the left panel ({\it i.e.} at $z = 0$) \textemdash{} that the stellar fraction in the simulated groups is generally higher than in the observed groups \textendash{} seems to hold out to $z = 1$.  

Examining the $z=0$ simulated groups in a bit more detail, we note that 
the `super-sized' galaxies (\ie~the galaxies with $M_* > 10^{11}\;\rmn{M_{\odot}}$ that we mentioned when discussing Figure \ref{fig.2}) contain $\sim 85\%$ of the  stellar mass within $R_{500}$ in the lowest mass groups, and the fraction drops with group halo mass to $\sim 65\%$ in the most massive groups.   The average stellar mass of  these galaxies ranges from $4\times 10^{11}\;\rmn{M_{\odot}}$ in the lowest mass groups to $2\times 10^{12}\;\rmn{M_{\odot}}$ in the most massive groups.    Artificially reducing the stellar mass of just these super-sized systems by a factor of 3 resolves the discrepancy between the observed and model stellar mass fractions across the entire mass range over which this fraction has been observationally determined.  {It also goes a long way towards improving the agreement with observed galaxy stellar mass function (\cf~bottom panel of Figure~\ref{fig.2}).}


In all our groups, the group central galaxy is always a ``super-sized'' galaxy.   In the lowest mass groups,  $\sim 80\%$ of the stars within $R_{500}$ reside in the central galaxy.   Examining the stellar build-up in these central galaxies, we find that about $25\%$ of the stars formed elsewhere and were subsequently incorporated into the central galaxies through galaxy-galaxy mergers;  $15\%$ of the stars formed {\it in-situ} from cooled IGrM;\footnote{{We remind the reader that ``cooled IGrM'' refers to gas in the MMP that is heated to $T > 5\times 10^5\; {\rm K}$ at some point in the past and cools directly onto the central galaxy.}}  and the balance (approximately $60\%$ of the total stellar mass at $z=0$) formed {\it in-situ} either from $T<5\times 10^5\; {\rm K}$ {gas that was either originally funnelled onto the central galaxies via cold mode accretion \citep{K09},  or from cold gas that was deposited in the central galaxies by mergers.} The bulk of the mergers affecting the central galaxies in low mass groups occur either before or just after the groups  \textendash{} that is, systems with at least three ``luminous'' galaxies \textendash{} formed.

In {the} most massive groups, the contribution of the central galaxy to the total stellar mass {within} $R_{500}$ drops to about $40\%$ and 
as for the stars that comprise these central galaxies,  about $58\%$ were brought in by mergers, $40\%$ formed {\it in-situ} from cold gas, and only about $2\%$ formed out of cooled down IGrM.   These percentages are important in two respects.  First, the fraction of the stellar mass in the central galaxies that is deposited by mergers increases with overall halo mass.  This trend has been noted previously by  \citet{HI13} in their study of galaxy-scale halos.   Our results show that the trend continues on the group-scale.   Second, and perhaps much more importantly, these results show that the overabundance of stars in our simulated groups is {\it not} primarily due to the cooling of the hot diffuse IGrM despite the fact that we do not have AGNs in our simulation.   Unchecked cooling of the IGrM contributes only {a} small fraction of the excess.

The {plots} showing the stellar fraction in groups at different epochs offer some idea of what is going on.   We have previously noted that once individual groups {form}, they slide to the right along the $x$-axis in this plot because the stellar mass in fact grows faster than the actual mass of the group halos in all except the most massive groups.  This star formation is fuelled by an excess of cold gas that has accumulated in the group galaxies while these galaxies are at the centres of their own halos either {\it before} the groups form, in the case of the central galaxies, or {\it before} they are incorporated into the groups, in the case of the satellite galaxies. 

We can see evidence for the presence of significant cold gas in the group galaxies in the bottom two panels of Figure~\ref{fig.8}.  The panels show the total mass fraction of ``cold" gas in the groups, where ``cold gas'' includes both the diffuse gas with $T<5\times 10^5\; {\rm K}$ as well as the dense gas that comprises the ISM in group galaxies.  In practice, the former is negligible because diffuse gas with temperatures $T<5\times 10^5\; {\rm K}$  lies on the broad peak of the cooling curve, experiences very efficient cooling and ends up flowing into the central galaxy.   The gas in groups is typically either hot and diffuse or cold and dense.   At any redshift, the lowest mass groups, which are also typically the youngest within the population, have the most amount of cold gas.   In the same vein, the earliest groups have the highest cold gas fraction.    Taken jointly, these results show that the galaxies, especially the more massive galaxies, that first come together to form the groups contain a significant cold gas reservoir.    As the groups grow and age, the cold gas reservoir is not replenished as rapidly as it is consumed by star formation, and the cold gas fraction drops.   We note that there is also considerable merger activity, especially early in the history of the groups,  during which some of the massive satellites sink to the centre and are cannibalized by the group central galaxies, and while this impacts the distribution of the stars and the gas within the groups, it does not affect the curves in Figure~\ref{fig.8} because we are considering the total stellar and cold gas fractions within $R_{200}$.

Demonstrating that the group galaxies host a significant fraction of cold gas is not the same as asserting that the group galaxies have an excess of cold gas.   We therefore turn to two recent studies  to compare the cold gas content of the most massive of our $z=0$ group galaxies  to real galaxies:  (1) The \citet{S11} sample that comprises 350 nearby massive ($M_* > 10^{10}\, \rmn{M_{\odot}}$), of which 222 have both CO and HI measurements.   We focus on the latter subset and compute the ``cold gas'' mass of each galaxy as $M_{\rm cold gas} = (M_{\rm HI} + M_{{\rm H}_2})/X$, where the division by $X=0.75$, the hydrogen mass fraction, corrects for the helium mass.   (2)
The \citet{CS13} study that lists HI measurements for 800 galaxies with stellar masses $10^{10}$ $\simless$ $M_*$ $\simless$ $10^{11.5}\, \rmn{M_{\odot}}$ and redshifts $0.025 \leq z \leq 0.05$.   We convert the HI masses to total cold gas mass using a constant ratio of $M_{{\rm H}_2}/M_{\rm HI} = 0.295$, based on the results of \citet{S11}, and $X=0.75$.    

\begin{figure*}%
         \centering
\captionsetup[subfigure]{labelformat=empty}
         \subfloat[][]{%
           \label{fig.10-a}%
\includegraphics[width=0.49\textwidth]{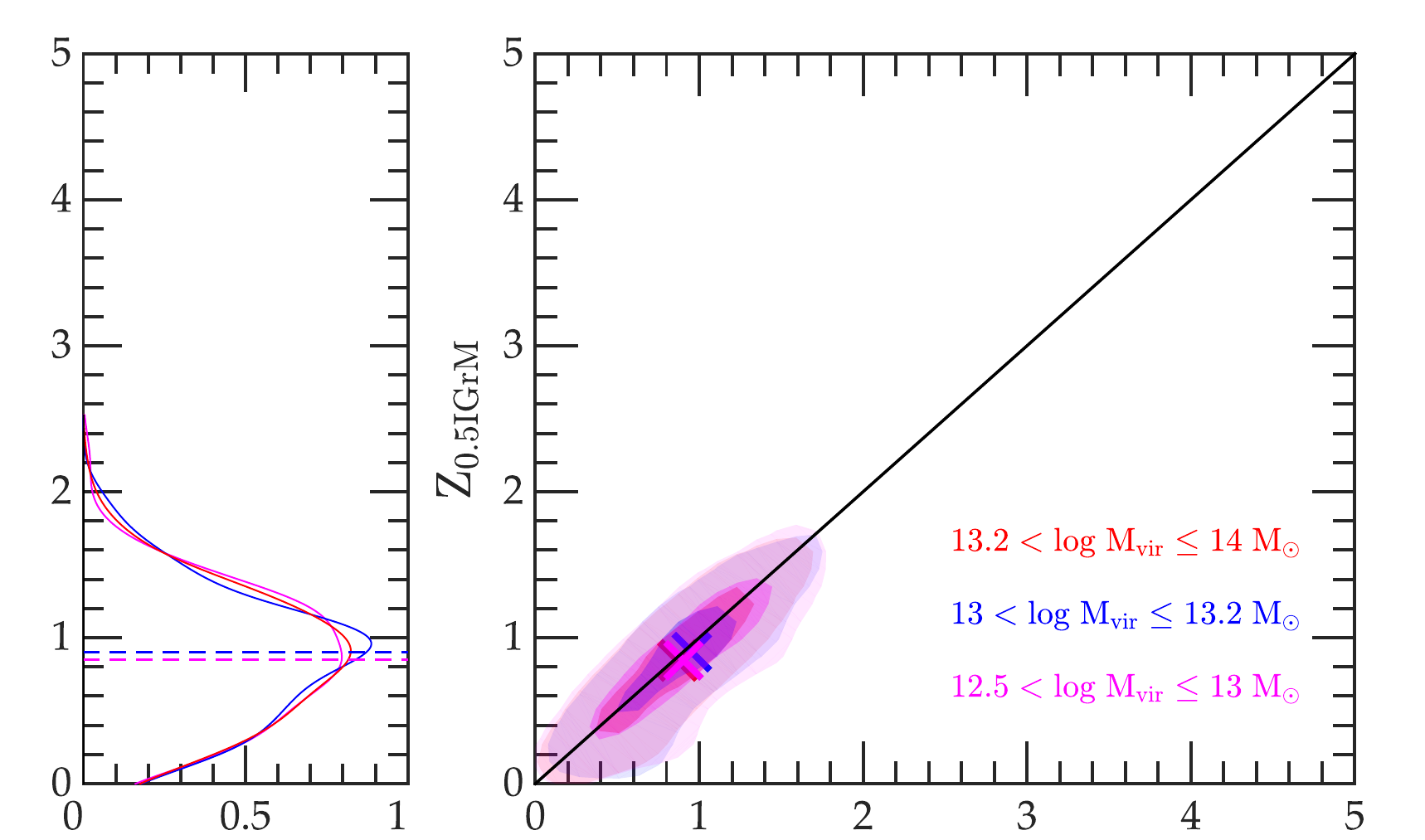}}%
         \hspace{5pt}%
         \subfloat[][]{%
           \label{fig.10-b}%
\includegraphics[width=0.49\textwidth]{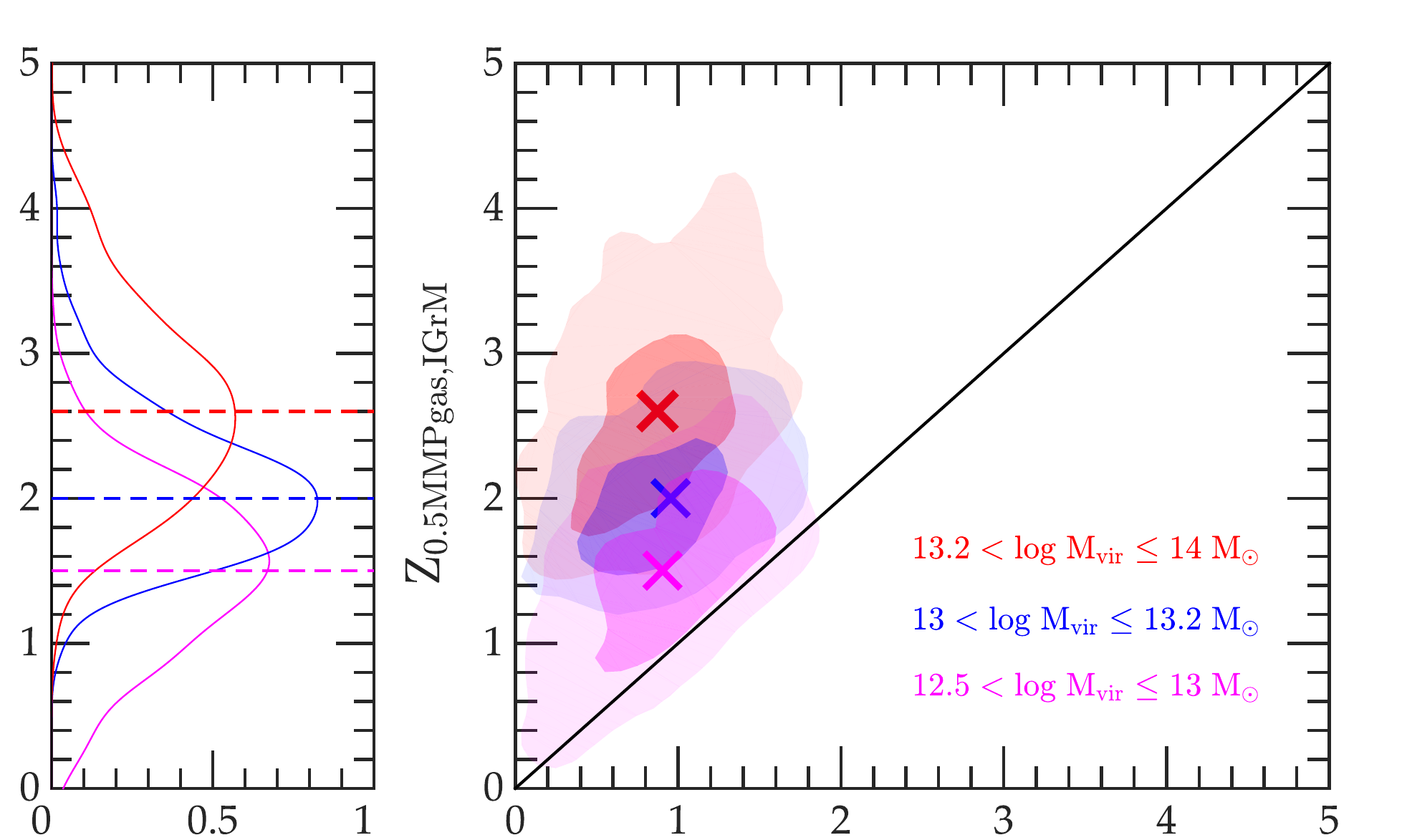}}\\
\vspace{-25pt}%
\captionsetup[subfigure]{labelformat=empty}
         \subfloat[][]{%
           \label{fig.10-c}%
\includegraphics[width=0.48\textwidth]{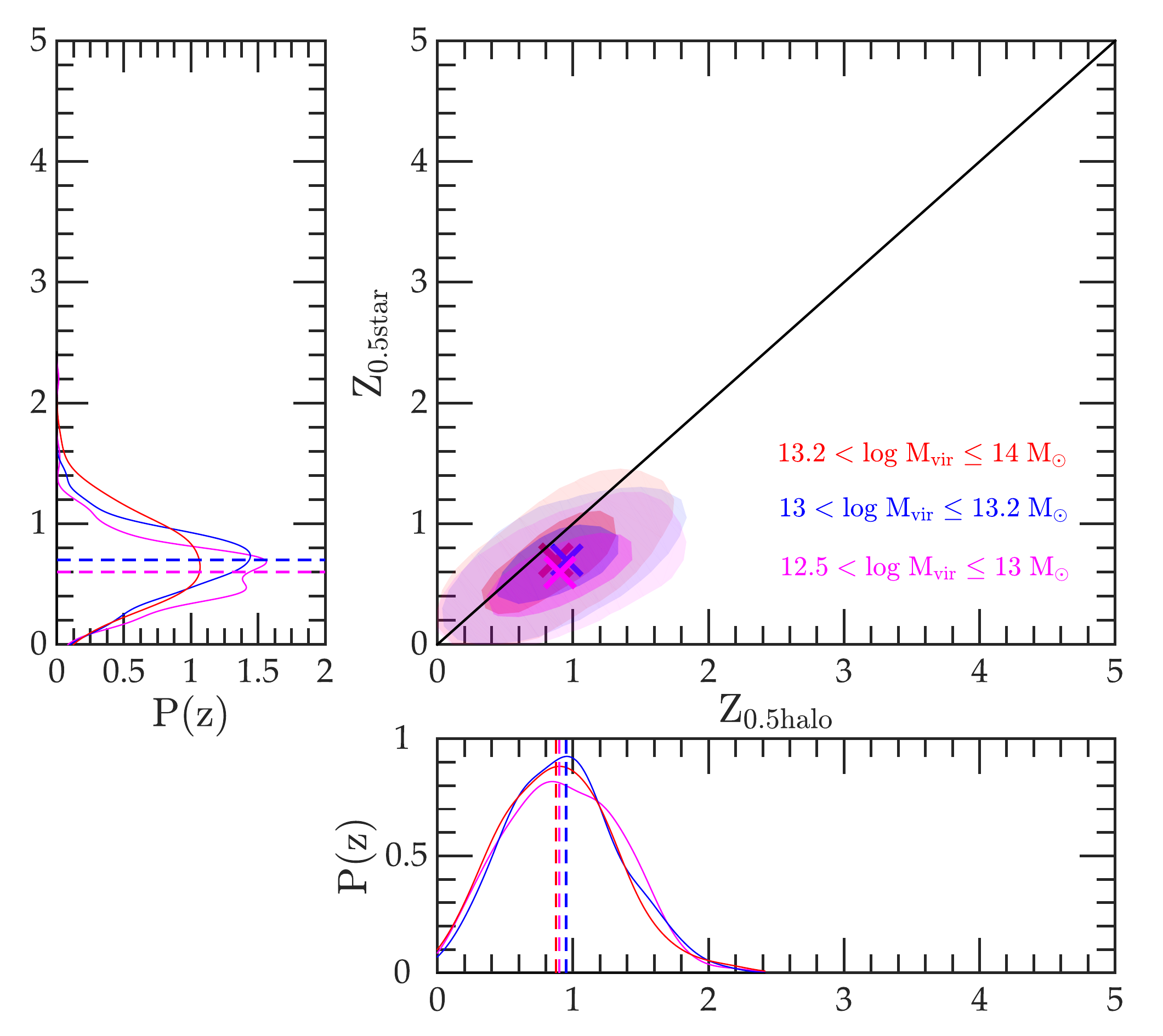}}%
         \hspace{5pt}%
         \subfloat[][]{%
           \label{fig.10-d}%
\includegraphics[width=0.48\textwidth]{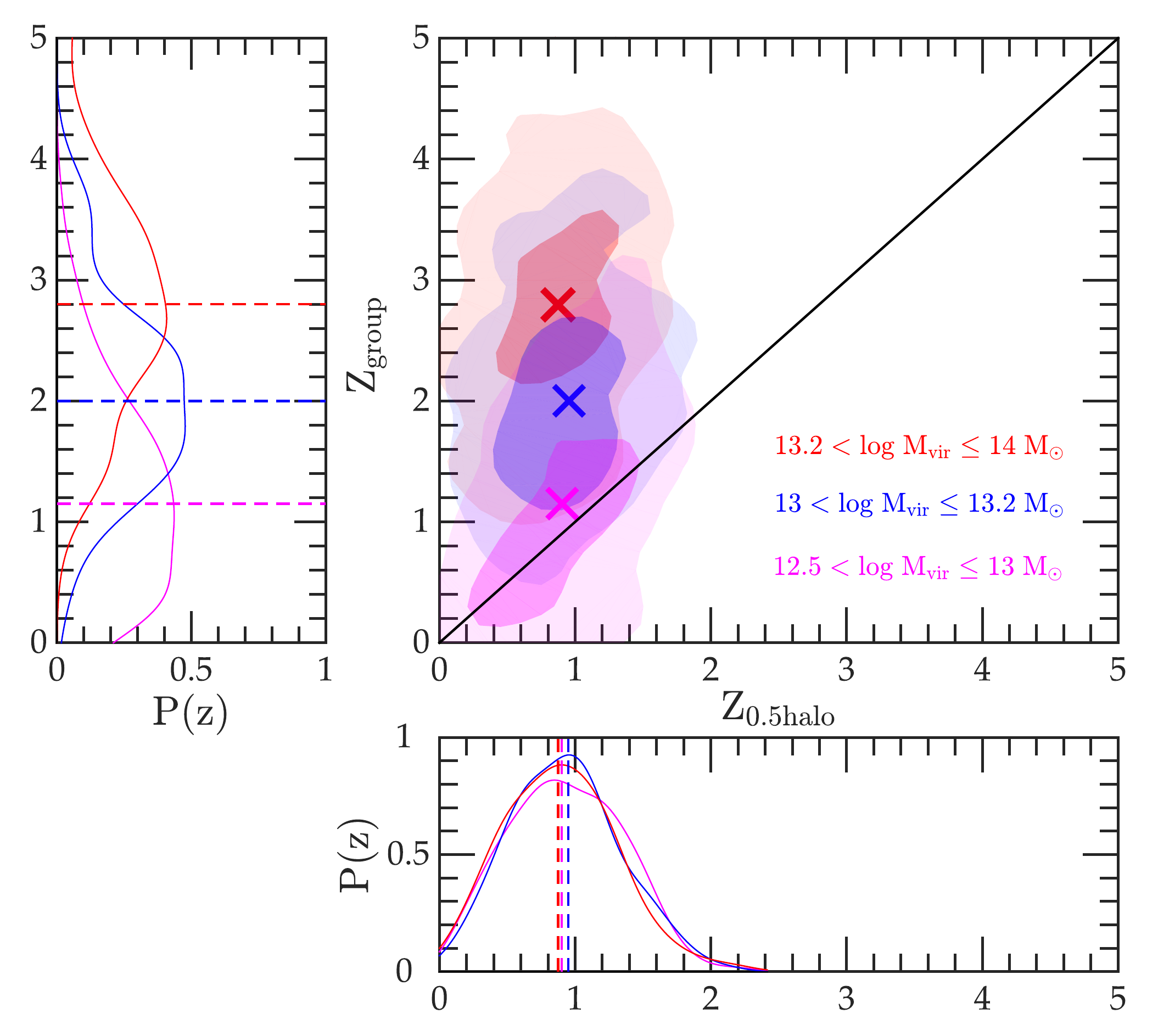}}%
\vspace{-10pt}%
         \caption{A set of four plots showing the distribution of the five key redshifts that summarize the groups' formation histories, {defined in Section~\ref{Sec:4.2}}, and the relationships between them:
     $Z_{0.5\;  {\rm IGrM}}$ vs. $Z_{0.5\;  {\rm halo}}$ (top left);     $Z_{\rm 0.5\;  MMP_{gas}, IGrM}$ vs. $Z_{0.5\;  {\rm halo}}$ (top right); $Z_{0.5\;  {\rm star}}$ vs. $Z_{0.5\;  {\rm halo}}$ (bottom left); and $Z_{\rm group}$ vs. $Z_{0.5\;  {\rm halo}}$ (bottom right).
In the main plot of each set, the different colored regions show the 2D distribution of the redshifts for the low, intermediate and high mass groups
\textendash{} \ie~$12.5 < {\rm log} \; M_{\rm vir} \leq 13.0 \;\rmn{M_{\odot}}$ (magenta), $13.0 < {\rm log} \;M_{\rm vir} \leq 13.2 \;\rmn{M_{\odot}}$ (blue), and $13.2 < {\rm log} \;M_{\rm vir} \leq14.0 \;\rmn{M_{\odot}}$ (red)  \textendash{} separately.   The inner and the outer contours of the shaded regions of each colour correspond to 1-$\sigma$ and 2-$\sigma$, while the $\times$ marks the median for all the galaxies within each mass bin.   The panels to the left and below the main plots show the normalized marginalized distributions of  $y$-axis redshift (left) and  $x$-axis redshift (below): $P(z) = (dN/dz)/N_{\rm tot}$.    The different coloured curves show the redshift distributions for the  three mass bins
and the dashed lines indicate their median:           
}
         \label{fig.10}%
       \end{figure*}


Both studies yield a mean cold gas mass of approximately $5\times 10^{9}\, \rmn{M_{\odot}}$ for galaxies with $M_* > 10^{11}\, \rmn{M_{\odot}}$.   We specifically restrict ourselves to such massive galaxies because the vast majority of the  galaxies that populate the two most massive bins in $M_*$ in the Catinella and Saintonge samples are, in fact, group galaxies.  For the simulated galaxies, we follow \citet{RD11} and define ``cold star-forming gas'' as gas within the galaxies whose density exceeds the star formation threshold of $n_{\rm H} > 0.13$ cm$^{-3}$ (\cf~Section~\ref{Sec:2.1}).   The resulting cold gas mass in our simulated $M_* > 10^{11}\, \rmn{M_{\odot}}$ group galaxies is 5-6 times larger.

Further investigation indicates that our hierarchical structure formation model, in which feedback is entirely due to stellar-powered galactic outflows, first breaks down not on the group scale but rather in the giant galaxies (with stellar mass $\simgreat 4\times 10^{10}\;\rmn{M_{\odot}}$) that precede the formation of the groups.    {These are the first systems in the hierarchy where the deepening gravitational potential wells and a higher cross-section for hydrodynamic interactions between the galactic outflows and the shock-heated halo gas component, once the latter starts to form, are able to confine the outflowing gas within the circumgalactic region around the galaxies.  Most of this material, being extremely metal-rich, cools down and falls back into the galaxy.   This is discussed at length in \citet{OD10} and their Figure 2 shows that in galaxies with stellar mass $ \simgreat 4\times 10^{10}\;\rmn{M_{\odot}}$, the median time between the launching of a wind particle, and it falling back into the galaxy and is either converted into a star or launched for a second time is  $\simless 1$ Gyr.   In effect, the winds power galactic fountain flows rather than galactic outflows and consequently,  the $\geq L_\ast$ galaxies can no longer moderate their star formation rates by depleting their cold gas mass via expulsion.   The high cold gas mass in our group central galaxies  is a consequence of this, and the overproduction of stars is a byproduct.    This is nicely illustrated in Figure 4 of \citet{OD10}, which shows that $\simgreat 70\%$ of the stars in massive galaxies at $z=0$ have formed out of re-accreted wind material.}

{There are two potential ways of resolving the above problem: (1) Increase the wind launch velocities so that even in the giant galaxies, the winds are not confined  within the circumgalactic regions and when they thermalize, the ejected material heats up to IGrM temperatures.
This may help reduce both the cold gas and the stellar masses of the massive galaxies in our simulated groups but at  a cost of making the IGrM mass fractions in these groups, and their corresponding X-ray properties,  potentially discrepant with the observations, and it will not improve our baryon fraction results.  Or, (2) $M_\ast \approx 4\times 10^{10}\;\rmn{M_{\odot}}$ is the transition mass scale where an alternate feedback mechanism, like AGN feedback, must come into play.  The main requirement of this alternate feedback mechanism is that it must be sufficiently potent that it, either by itself or in combination with the galaxy-wide stellar powered outflows, can drive down the total baryon fractions and the cold gas mass fractions in giant galaxy or group halos with  $M_{500} < 2\times 10^{14}\;\rmn{M_\odot}$.}



\subsection{Assembly of the Present-day Groups}
\label{Sec:4.2}

Having discussed the evolution of various baryonic components comprising the groups in some detail above, we conclude our discussion of the group baryonic properties by  considering the five redshifts that encapsulate the key features of the groups' formation histories.   To determine these, we reconstructed each present-day group's merger history by stepping back in time from the present and identifying, at each epoch, all the individual halos that are the present-day group's ancestors.  We label the largest of these the most massive progenitor (MMP).    Our five redshifts are based on the properties of the MMP.   These redshifts are:  
 \vspace{-6pt} 
\begin{description}[labelindent=0.3cm,leftmargin=1.4cm, style=sameline]
\item[$\boldsymbol{ Z_{0.5\;  {\rm halo}}}$:]{ The redshift at which the total mass of a present-day group's MMP is half of the group's final mass; \hfil\break\ie~$M_{{\rm MMP},200}(z) = \frac{1}{2} M_{200}(z)|_{z=0}$.}
\item [$\boldsymbol{Z_{0.5\;  {\rm IGrM}}}$:]{ The redshift at which the hot ($T>5\times 10^5\;$K) gas mass in the MMP is half of the group's final IGrM mass;\hfil\break \ie~$M_{{\rm MMP, IGrM},200}(z) = \frac{1}{2} M_{{\rm IGrM},200}(z)|_{z=0}$.}
\item [$\boldsymbol{Z_{0.5\;  {\rm star}}}$:]{ The redshift at which the total stellar mass in the MMP is half of the group's $z=0$ total stellar mass; \ie~$M_{{\rm MMP, star},200}(z) = \frac{1}{2} M_{{\rm star},200}(z)|_{z=0}$.}
\item [$\boldsymbol{Z_{\rm group}}$:]{{The highest redshift at which the MMP hosts at least three luminous galaxies and can be considered a group.}\hfill\break
\ie~$N_{\rm MMP, gal}(z) \geq 3$}
\item [$\boldsymbol{Z_{\rm 0.5\;  MMP_{gas}, IGrM}}$:]{ {the highest redshift at which the hot diffuse IGrM mass in the MMP exceeds 50\% of the total gas mass.} \hfill\break
\ie~$M_{{\rm MMP, IGrM},200}(z) / M_{{\rm MMP, all gas},200}(z)  > 0.5$.}
\end{description}

In Figure~\ref{fig.10}, we show the individual distribution of these redshifts as well as the relationship between them.     We have chosen $Z_{0.5\;  {\rm halo}}$, the redshift commonly referred to as the ``formation redshift'' of the present-day groups, as the common reference for four cross plots.   To start with, we consider this redshift by itself first.  The bottom panel in each of column of plots {shows} the normalized distribution of {the} formation epoch.

The distribution of formation times for groups in all three mass bins {are} similar and the median formation epoch is $z\approx 0.9$.   If the halos populating each of the mass bins were a representative (\ie~unbiased) subset of all the dark matter halos in the simulation volume with masses  $12.5 < {\rm log} \; M_{\rm vir} \leq 13.0 \;\rmn{M_{\odot}}$ (low), $13.0 < {\rm log} \;M_{\rm vir} \leq 13.2 \;\rmn{M_{\odot}}$ (intermediate), and $13.2 < {\rm log} \;M_{\rm vir} \leq14.0 \;\rmn{M_{\odot}}$ (high), we would expect the ``low mass halos'' to be slightly older than the ``intermediate mass halos'' and the ``high mass halos'' to be younger.    The groups in the intermediate  (median formation redshift is indicated by the blue dashed line) and the high mass bins (median formation redshift is indicated by the red dashed line) conform to these expectations.   This is perhaps not surprising.   As shown in Figure~\ref{fig.1}, nearly all dark matter halos with masses $M_{\rm vir} > 10^{13}\;\rmn{M_{\odot}}$ are groups and, therefore the group halos in the intermediate and the high mass bins form a representative sample.   The median formation redshift of the groups in the lowest mass bin (median formation redshift is indicated by the magenta dashed line), however, {breaks the expected} trend: Their median formation redshift is lower than that of the intermediate mass halos.   This is because the group halos that populate the low mass bins are {\it not} an unbiased sample of all dark matter halos with masses in the range $12.5 < {\rm log} \; M_{\rm vir} \leq 13.0 \;\rmn{M_{\odot}}$.   Rather, these groups form a very special subset with at least three luminous galaxies and  as discussed by \citet{Z06}, this type of constraint on the galaxy occupation number results in the selection of a relatively younger subset of halos, which is indeed what we {find}.   

The $y$-axis of the top left panel of Figure~\ref{fig.10} shows the distribution of $Z_{0.5\;  {\rm IGrM}}$, the epoch when the hot diffuse IGrM mass in the MMP exceeds 50\% of the IGrM mass in the final group halo.    The redshift distributions for the groups in the three mass bins are very similar.     Much more interestingly, the halo formation time and $Z_{0.5\;  {\rm IGrM}}$ are very tightly correlated.   This suggests that the MMPs of the present-day groups have already built up a substantial reservoir of hot diffuse gas by the time the group halos form at $Z_{0.5\;  {\rm halo}}$.  These results are consistent with the trends seen in Figure~\ref{fig.8}, which show that post-formation, the growth of dark matter mass and IGrM mass proceeds in lock-step.

In the top right panel of Figure~\ref{fig.10}, we show the joint and the marginal distributions of $Z_{\rm 0.5\;  MMP_{gas}, IGrM}$, the redshift when the hot diffuse IGrM begins to dominate the total gas mass in the MMP, and $Z_{0.5\;  {\rm halo}}$.   The plot suggests little or no correlation between these two redshifts.  However, the normalized distribution of $Z_{\rm 0.5\;  MMP_{gas}, IGrM}$ confirms our earlier assertion that the progenitors of the most massive $z=0$ groups (red curve) build up a substantial reservoir of hot diffuse X-ray emitting gas fairly early on; the median value of  $Z_{\rm 0.5\;  MMP_{gas}, IGrM}$ for these systems is $z=2.6$.  The distribution for the intermediate mass halos is shifted to lower redshifts, with a median of $z=2$, and the distribution of the lowest mass groups is shifted to lower redshifts, with a median of $z=1.5$.  For the majority of the groups, the gas content of the MMPs is dominated by hot gas well before the MMP mass reaches 50\% of the corresponding present-day group's final mass.

In the bottom left panel, we show the joint and the marginal distributions of $Z_{\rm 0.5\; star}$, the redshift at which half of the total $z=0$ stellar mass within $R_{200}$ is in place within the MMP, and $Z_{0.5\;  {\rm halo}}$.   The main plot shows that these two redshifts are strongly correlated.   During the early phases of group formation, the MMP grows principally via mergers, which add to both the dark matter mass as well as the stellar mass of the system.   However, if the two grow in perfect lock-step, we would expect their joint distribution to define a narrow ellipse whose major axis lies along the $1:1$ line, but they don't and even the median $Z_{\rm 0.5\; star}$ is slightly lower than $Z_{\rm 0.5\; halo}$, with $\Delta z \approx 0.2\textendash{}0.3$.   As discussed previously, the mergers not only contribute stars and dark matter, they also bring in cold gas.   In Figure~\ref{fig.8}, we discussed the conversion of this cold gas into stars, especially at late ($z < 1$) times.   This {\it in-situ} star formation breaks the 1-to-1 mapping between halo assembly and the establishment of the stellar mass.    As we have noted previously, any additional feedback mechanism that is added to these simulations must be able to prevent the build-up of cold gas in the smaller systems because once this cold gas reservoir is established, it is unlikely that any mechanism acting solely on group or cluster scales can prevent this gas from being delivered to the central galaxies.   An additional point of interest is that the median value of $Z_{\rm 0.5\; star}$ indicates that half of the stellar component of today's galaxy groups was already in place about 6 billion years ago.
 
{Finally, the bottom right panel shows the joint and the marginal distributions of ${Z_{\rm 0.5halo}}$ and ${Z_{\rm group}}$, the highest redshift at which the MMP first incorporates three or more luminous galaxies and meets our definition of a group.}   There are two features worth noting. {First, the distribution of redshifts at which the MMPs of the present-day groups first qualify as ÒgroupsÓ is fairly broad, much broader than the distribution of group halo formation times. Second, the MMPs of the present-day high and intermediate mass groups generally acquire a third luminous galaxy well before half of the groups' final mass is assembled and hence, there isn't a clear relationship between $Z_{\rm group}$ and $Z_{\rm 0.5halo}$ of these systems.   Only in the case of the lowest mass groups do a significant fraction of the halos form first and then become groups, and the two epochs, $Z_{\rm group}$ and $Z_{\rm 0.5halo}$, appear linked.  The median $Z_{\rm group}$ for the high, intermediate and low mass systems are $z=2.7$, 2.0 and 1.2, respectively, as compared to the median formation epoch of $z \approx0.9$ for the same three categories of groups.}

\begin{figure}%
         \centering
\includegraphics[width=0.49\textwidth]{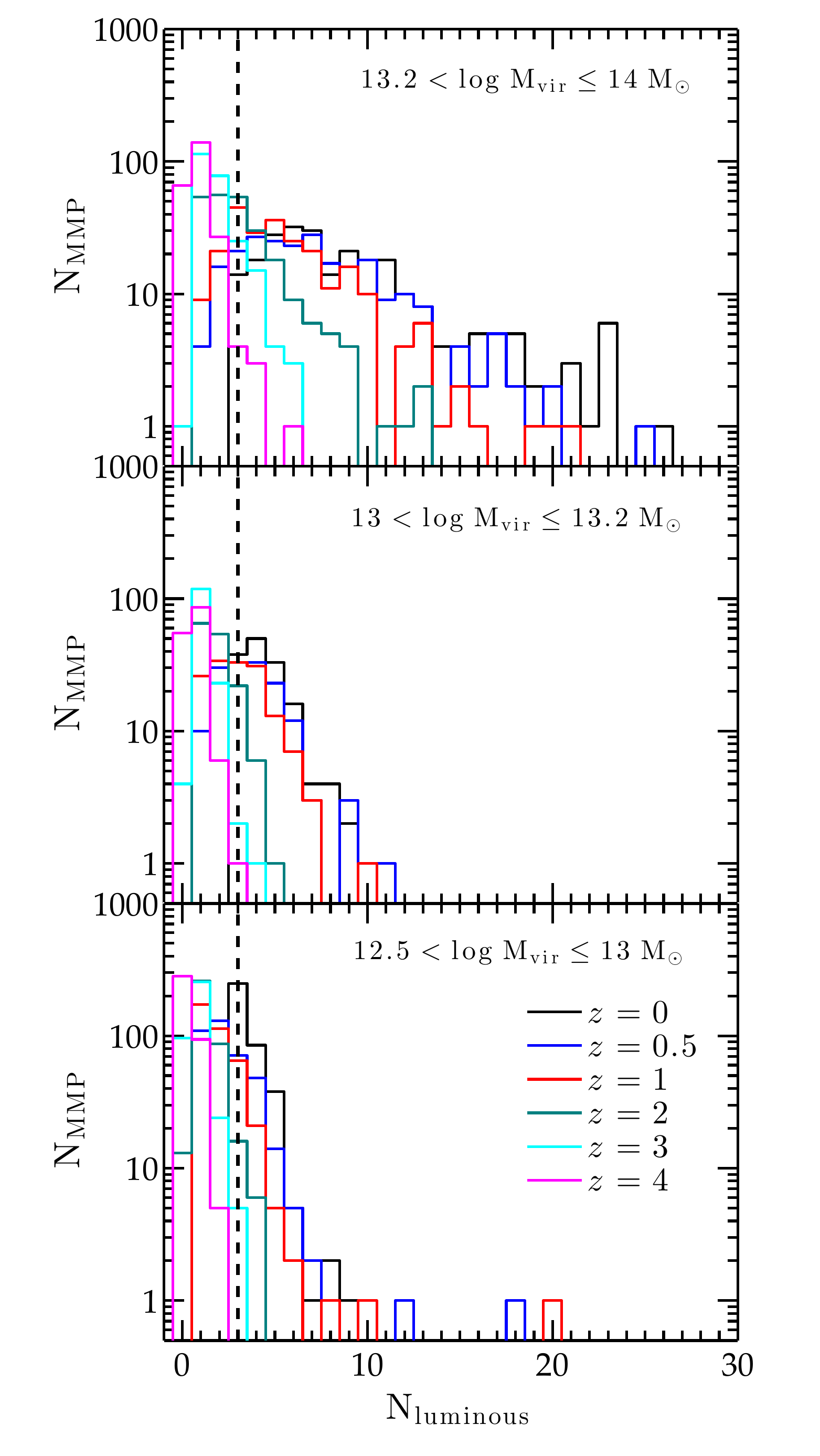}
\vspace{-15pt}%
         \caption{{Histograms showing the number of ``luminous" galaxies (\ie~$M_*\geq 2.9\times10^{9}\;M_{\odot}$) in the $z=0$ groups (black), as well as in their MMPs at $z=0.5$ (blue), $z=1$ (red), $z=2$ (green), $z=3$ (cyan) and $z=4$ (magenta). The top, middle and bottom panels show the results for the present-day low, intermediate and high mass groups, respectively. The vertical dashed line corresponds to $N_{\rm luminous}=3$, the threshold above which a halo is defined as a `group'. }}%
         \label{fig.11}%
	\end{figure}

\begin{figure}%
         \centering
\vspace{-4pt}%
\includegraphics[width=0.48\textwidth]{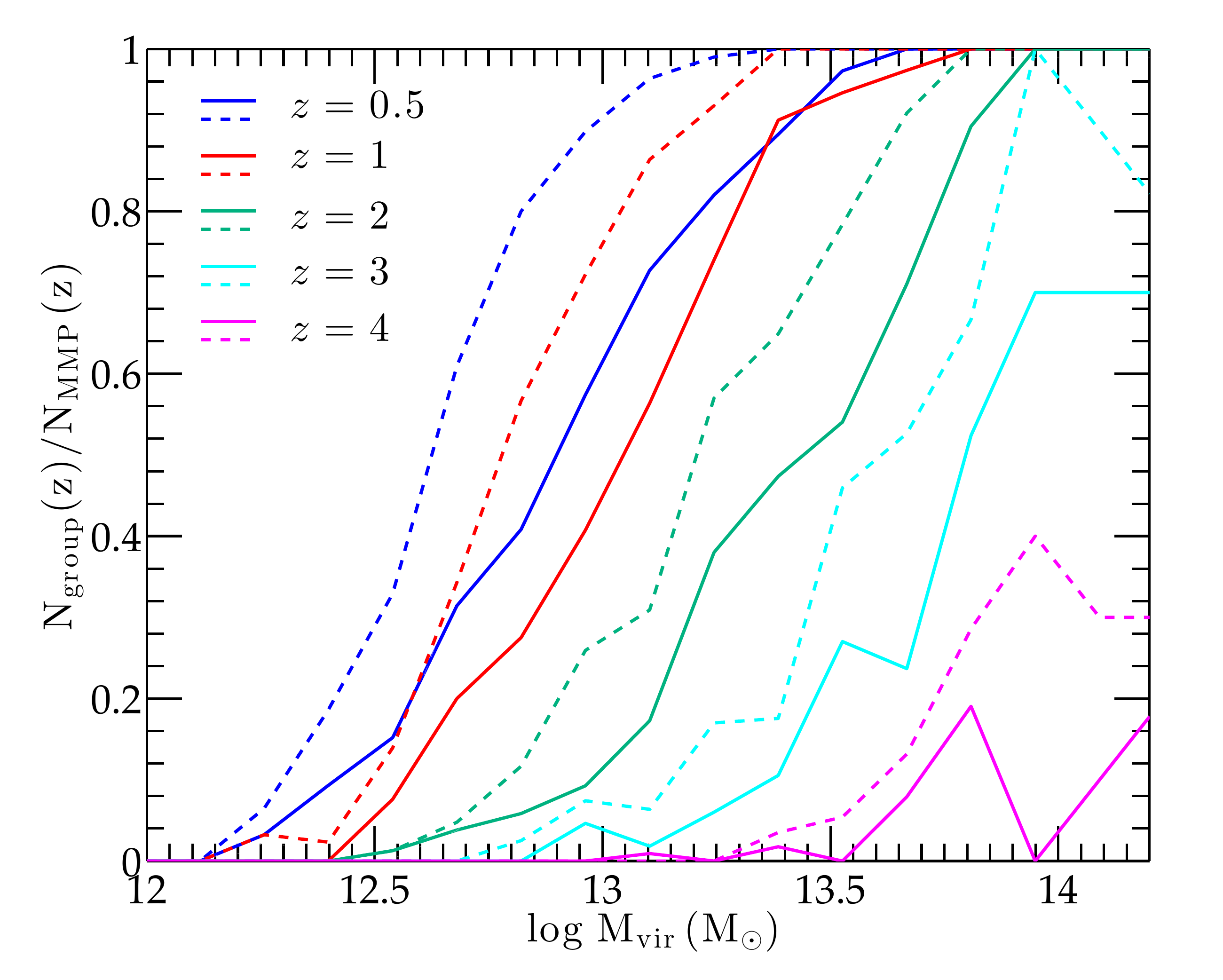}
\vspace{-5pt}%
         \caption{{The solid curves show the fraction of the MMPs of the selected $z=0$ groups that qualify as groups at various redshifts as labelled, against their present-day virial mass. The corresponding dashed curves show the fraction of their MMPs that had once qualified as group before a given redshift, regardless of the number of ``luminous" groups in the MMP at that redshift. }}%
         \label{fig.12}%
      \end{figure}
      
{To better understand the behaviour of $Z_{\rm group}$, we plot in Figure~\ref{fig.11} the distribution of the number of luminous galaxies in the MMPs of the present-day groups at several different redshifts. The results show that the MMPs of the most massive systems today qualify as groups in their own right (by acquiring three luminous galaxies) much earlier than the MMPs of the lowest mass groups.  Moreover, in keeping with the ``downsizing" picture usually discussed in the context of galaxy formation \citep{N06, F09}, the most massive groups also grow the fastest, with the number of luminous galaxies in these systems increasing five-fold from $z=4$ to $z=0$. In contrast, nearly two-thirds of the MMPs of the present-day groups in our lowest mass bin host only one or two luminous galaxies even at redshifts as low as $z=0.5$.}

{We note that the number of luminous galaxies in the groups does not grow monotonically with time. The satellite galaxies can sink down to the group centre due to dynamical friction and merge with the central galaxies, resulting in a decline in the number of group galaxies.  This is illustrated in Figure~\ref{fig.12}, where we plot the fraction of MMPs of the $z=0$ groups that host at least 3 luminous galaxies at the redshifts under consideration (solid curves), and the fraction of MMPs that qualified as groups at some earlier redshift (dashed curves).  In terms of $Z_{\rm group}$, the latter corresponds to the fraction present-day groups with $Z_{\rm group}> z$. }

{Focusing first on the dashed curves in Figure~\ref{fig.12}, we see that $50\%$ of the MMPs of the present-day low (\ie~$12.5 < {\rm log} \; M_{\rm vir} \leq 13.0 \;\rmn{M_{\odot}}$), intermediate (\ie~$13 < {\rm log} \; M_{\rm vir} \leq 13.2 \;\rmn{M_{\odot}}$) and high (\ie~$13.2 < {\rm log} \; M_{\rm vir} \leq 14.0 \;\rmn{M_{\odot}}$) mass groups first crossed the ``group threshold"  by $z\approx1$ (red), $z\approx2$ (green) and $z\approx3$ (cyan), respectively. These values are consistent with the results for $Z_{\rm group}$ shown in Figure~\ref{fig.10}.   However, achieving group status and maintaining that status at subsequent times are two different concerns and this is illustrated by the differences between the dashed and the solid curves of the same colour in Figure~\ref{fig.12}.  This difference equals the fraction of MMPs that after achieving group status at some earlier epoch by virtue of accreting one or two luminous galaxies and just meeting the threshold criterion of three galaxies, lose one of them to a merger.  Group halos are most susceptible to such fluctuations in status while the number of hosted luminous galaxies is small.  The MMPs of the present-day high mass groups experience such fluctuations at relatively high ($z\approx 3-4$) redshifts.  At that time, the fraction of MMPs (of high mass groups) that qualify as groups is small, $\sim10\%$, but this fraction rises steeply and by $z=1$, nearly all the groups are well established and the fraction exceeds $90\%$.  In contrast, the fraction of  MMPs of the our low mass present-day groups that qualify as groups at $z=4$ is $0\%$, and even at redshifts as low as $z=0.5$, the fraction is only $\sim35\%$.  Most of the present-day low mass systems qualified (or re-qualified) as bona fide groups at the present as a result of late-time ($z<0.5$) mergers.  These mergers not only inject additional luminous galaxies into the MMPs but also contribute a significant fraction of MMPs final total masses, both biasing the groups' formation epochs towards the present \citep{Z06} and also providing a direct physical connection between $Z_{\rm group}$ and $Z_{\rm 0.5halo}$, which in turn accounts for a tighter relationship (\cf~Figure~\ref{fig.10}) between the two in the case of low mass groups.}


\section{METAL ENRICHMENT OF THE INTRAGROUP MEDIUM}
\label{Sec:5}

 \begin{figure*}
\centering
 \vspace{-4pt}
  \includegraphics[width=1.0\textwidth]{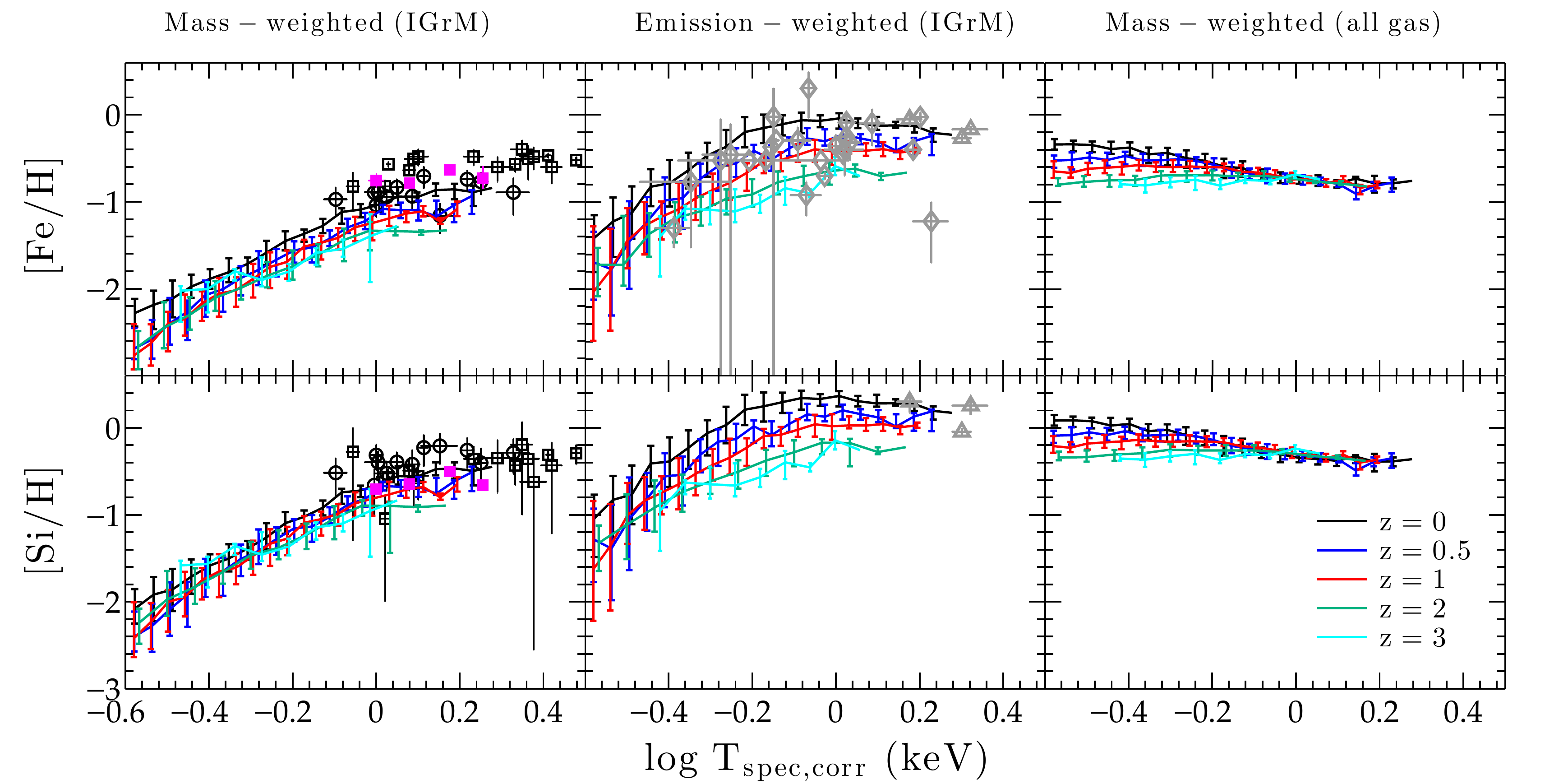}
\caption{Global iron (top row) and silicon (bottom row) abundances within $R_{500}$ of the group centers.   The left column shows the mass-weighted abundances in the IGrM;  the middle column shows the X-ray emission-weighted abundances in the IGrM; and the right column shows the global mass-weighted abundances of all the gas, including the cold gas within individual group galaxies.  The coloured lines and the corresponding error bars show the median values  and the 1-$\sigma$ dispersion for group populations in the simulation volume at $z=0$ (black),  $z = 0.5$ (blue), $z = 1$ (red), $z = 2$ (green) and  $z=3$ (cyan).   The open black circles, the open black squares, and the filled magenta squares in the left column show measurements from  \citet{RP09}, \citet{FU98} and \citet{S14}, respectively.   The grey diamonds and triangles are results from \citet{HP00} and \citet{TP03}, respectively.
}
\label{fig.13}
\end{figure*}

In addition to altering the thermal and baryonic properties of galaxy groups, large-scale galactic outflows also transport metals from  the galaxies to the intergalactic space.   Such outflows are key to explaining the widespread enrichment of the intergalactic medium (IGM) as early at $z\sim 5$ \citep{OP06,O09} and the observed mass-metallicity relation in galaxies, both today and at higher redshifts \citep{FD08,RD11,HI13,SD14}.    In this section,  we focus specifically on the hot, diffuse, X-ray emitting, intragroup medium.  The observed iron and silicon abundances ranging from $\sim 0.1$ to $\sim 0.7$ solar offers clear evidence that a significant fraction of the metals produced in galaxies escapes from these systems.   We determine the level of iron, oxygen and silicon enrichment in the IGrM that can be attained via our momentum-driven outflows model and assess how these compare with the latest observations.   We also show how the abundances and abundance ratios evolve with time.   And, we discuss how our metal abundances would change if the global stellar mass in our simulated groups were to be reduced by a factor of $\sim 2$ to reconcile the model results with the observations (\cf~Figure \ref{fig.8}).   

 \begin{figure*}
\centering
 \vspace{-4pt}
  \includegraphics[width=1.0\textwidth]{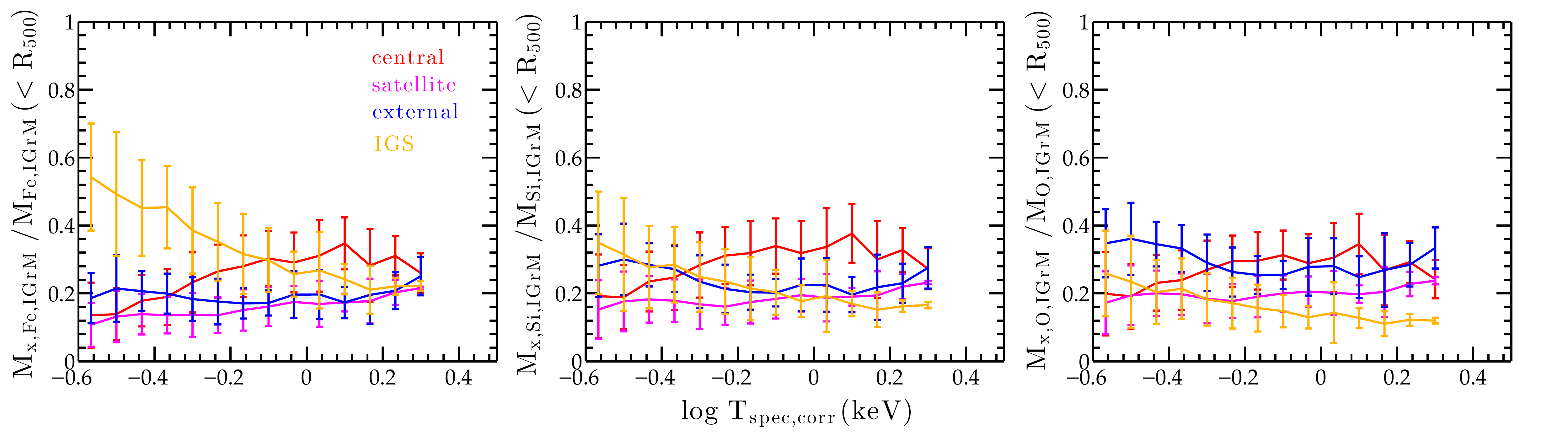}
\caption{
The fraction of IGrM iron (left panel), silicon (middle panel) and oxygen (right panel) mass within $R_{500}$ in $z = 0$ groups, characterized by their $T_{\rm spec,corr}$, contributed by the central galaxies (red curve), the group satellite galaxies (magenta curve), the non-group external galaxies  (blue curve), and the intragroup stars (orange curve)  over cosmic time.  See \S 5.2 for our schema for classifying galaxies as central, satellite or external.  The error bars depict 1-$\sigma$ error.
}
\label{fig.14}
\end{figure*}

\subsection{The metallicity of the IGrM}

In Figure~\ref{fig.13}, we plot the global mass-weighted (left column) and emission-weighted (middle column)  iron and silicon abundances in the IGrM within $R_{500}$ of the simulated groups  (top and bottom rows, respectively), as well as  the global mass-weighted {abundances} of all the gas, including the cold gas within individual group galaxies (right column),  as a function of core-corrected spectroscopic temperature.  The coloured lines show the abundances at different epochs over the redshift range $0 \leq z \leq 3$.   {The fact that} we have chosen to show both mass- and emission-weighted curves may seem a bit excessive.   For the kind of comparisons we wish to carry out, mass-weighted data is preferable.  However, the available group {observations} (for comparison) {are} limited because spatially resolved, X-ray spectroscopy of galaxy groups, which is a prerequisite for mass-weighted abundance measurements, involves long observations and challenging analyses.   On the other hand, there has been a steady reporting of  group abundance measurements in the literature over the years but most of this data is  emission-weighted.   Given such circumstances, we have opted to leverage both types of measurements:
In the first column on the left, the black open squares show the core-corrected data from \citet{FU98}\footnote{As discussed in Appendix A3 of \citet{NAG05}, the core-corrected abundance measurements of each group in \citet{FU98} provide a reliable estimate of the global mass-weighted abundance of the groups under consideration.}, the open black circles show the results from \citet{RP07}, and the magenta filled squares  show the latest {\it Suzaku} results from \citet{S14}.   We focus on measurements from ``warm'' groups, \ie~groups with $T_{\rm spec,corr}\,\simgreat\, 0.8$ keV or $M_{500}\,\simgreat\, 1.7\times 10^{13}\;\rmn{M_{\odot}}$, because this data range has been studied by several independent groups and collectively, the measurements are more likely to be representative.  In the middle column, we show as grey diamonds the data from \citet{HP00} while the grey triangles show data from \citet{TP03}.  To facilitate comparison, all metal abundances, whether theoretical or observational, are normalized to the solar ``photosphere abundances" level from \citet{AG89} (see Section~\ref{Sec:2.3}).

Turning first to the mass-weighted IGrM  abundance in warm $z=0$ simulated groups, we find that $\left[{\rm Fe/H}\right]$ rises gently from $-1.2$ to  $-0.9$ with temperature and then flattens, while $\left[{\rm Si/H}\right]$ rises from $-0.8$ to $-0.45$ and then flattens.   The model results are in very good agreement with the observations although there is a hint that the observed iron abundance measurements might be  higher by $0.1-0.2$ dex.  This level of mismatch, if real (note that the latest {\it Suzaku} results are lower than the earlier {\it XMM-Newton} or {\it Chandra} results and hence, much more compatible with the model results), is not unexpected given the factor $\sim 2$ uncertainties in the adopted nucleosynthesis yields and supernovae rates.  The emission-weighted {observations and the simulation} results are also in agreement; however, this is not surprising given the large scatter in the observational measurements.   Interestingly, the emission-weighted silicon and iron IGrM abundances of  warm $z=0$ groups overestimate the ``true'', \ie~mass-weighted, abundances by 0.6-0.7 dex.  This is a consequence of emission-weighted results being biased towards the brighter (in X-ray) central cores of the groups, which   \textendash{} if recent observations are a fair guide  \textendash{} are expected to be more metal-rich.   We will be examining the metallicity profiles, and other related distributions, of our simulated groups in a follow-up paper.

In the cooler ($T_{\rm spec,corr}\simless\, 0.8$ keV) simulated groups, the iron and silicon abundances drop with decreasing IGrM temperature.  This trend is the consequence of the metal-rich diffuse gas in these systems dropping out of the IGrM more efficiently because, as we have mentioned previously, the gas sits closer to the broad peak in the cooling curve.   The total iron and silicon abundances of all the gas, including the cold gas inside group galaxies, within $R_{500}$ are, however,  broadly similar across both warm and cool groups (see the right column of Figure~\ref{fig.13}),  and at $z=0$, perhaps even {shows evidence of} a slight rise towards the coolest groups.  This latter trend is not surprising given that the $z=0$ stellar fraction is the largest in the coolest groups.  It is possible that the inclusion of AGN heating as well as turbulent diffusion, which is not included in our present simulation but is expected to play a role in transporting the chemical elements from regions of high metallicity within the IGrM to regions of low metallicity, may moderate the decline in the abundances towards lower temperatures.

The coloured lines in Figure~\ref{fig.13} show how the abundances grow with time.   On the whole, the iron abundance within $R_{500}$ increases by a factor of $\sim 2.5-3$ from $z=2$ to $z=0$, and the silicon abundance increases by a factor of $\sim 2$.   Both show a similar growth pattern, growing gently between $0.5 < z < 2$ and then somewhat more rapidly between $0 < z < 0.5$, with the iron abundance growing a bit faster than silicon.  This late growth is {fuelled} by the release of metals locked up in  AGB stars (iron and silicon) as well as the injection of iron by delayed Type Ia SNe.   We will return to this issue when we consider the evolution of the abundance ratios in Section~\ref{Sec:5.3}.

 \subsection{Sources of the IGrM Metals}

Having compared the metallicity of the IGrM in present-day simulated galaxy groups with available observations, we now examine where the metals in the IGrM originated.   We focus on the IGrM within $R_{500}$ and label the potential sites of metal production  based on their {\it status at the time of enrichment} as follows:
\begin{description}[labelindent=0.3cm,leftmargin=1.7cm, style=sameline]
\item[Central:]{The central galaxy of the present-day group or the central galaxy of the group's MMP at an earlier epoch.}
\item [Satellite:]{A non-central galaxy that is contained within the $z=0$ group halo or within the group's MMP. } 
\item [External:]{\ A galaxy that is neither a central nor a satellite at the time of enrichment.}
\item [IGS:]{Direct enrichment of the IGrM by intragroup stars, \ie~stars in the present-day group or any of its progenitor subhalos that are not bound to any of the skid-identified  galaxies}
\end{description}
The left, middle and right panels of  Figure~\ref{fig.14} show the contribution of each of these to the total iron, silicon and oxygen mass, respectively, in the IGrM within $R_{500}$ of the present-day simulated groups.   The figure shows the results for the full sample of simulated groups: warm and cool.   In the following, we will only discuss the    
warm  (\ie~$T_{\rm spec,corr}\,\simgreat\, 0.8$ keV) simulated groups whose metallicity we are able to compare directly with observations.

As illustrated in Figure~\ref{fig.14}, {the central galaxy is an important source of all three metal species in the warm groups.}  This component contributes $\sim 30\%$ of the iron mass, $\sim 32\%$ of the silicon mass, and $\sim 30\%$ of the oxygen mass.   In the case of oxygen, the external galaxies contribute about the same fraction, followed by the satellite galaxies, which contribute about $20\%$.   The IGS contribution is approximately $12\%$ and the balance ($\sim 8\%$) comes from `unresolved' galaxies
({\it i.e.} SKID-identified galaxies with {a} total mass in cold gas and stars $< 2.92\times 10^9\;\rmn{M_{\odot}}$; \cf~\S 2.2).    In the case of silicon, the satellites, the externals, and the IGS all contribute about the same amount ($\sim 20\%$) and again, the `unresolved' galaxies contribute $< 10\%$.   Since silicon and oxygen are both $\alpha$-elements, it may seem surprising that the relative contributions of the four categories to IGrM mass of these two elements are not identical.   However, as we elaborate in Section~\ref{Sec:5.3}, while the two are produced via the same mechanisms, they are processed differently by AGB stars. {For} iron, the second most important source in the IGrM, after the central galaxy, is the IGS component.   This component contributes the same amount of iron ($30\%$) as the central galaxy in groups with $T_{\rm spec,corr}\approx 0.8$ keV but fraction drops with increasing IGrM temperature to $20\%$ in groups with $T_{\rm spec,corr}\approx 2$ keV.   The satellites and the externals both contribute the same amount:  $16-20\%$, with the unresolved systems making up the rest ($<10\%$).  One important take-away is that the central and the satellite galaxies within the group halo or its MMP (\ie~{\it in-situ} wind enrichment) produce nearly half of the total mass of all three metal species in the IGrM and are more important than the low-mass galaxies responsible for the early enrichment of the intergalactic medium.

Knowing where the metals are produced allows us to ascertain the extent to which the estimates of [Fe/H] and [Si/H] that we {compare with} the observations in Figure~\ref{fig.13} are affected by the overproduction of stars in the group galaxies.   To do so, we have determined how much of the iron and silicon mass in the warm groups originates in galaxies whose stellar mass exceeds  $10^{11}\;\rmn{M_{\odot}}$  at the time of ejection.    In the case of both iron and silicon, approximately $30\%$ of their IGrM mass is ejected from `super-sized' galaxies.  If, as an exercise in post-processing, we assume that some mechanism ({\it e.g.}~AGN heating) quenches star formation in massive galaxies, limiting their stellar mass to a maximum of $10^{11}\;\rmn{M_{\odot}}$, then the metals that were {ejected} of the galaxies after they evolved into `super-sized' galaxies would either have not been made or would have remained locked up in the galaxies.   In this case, {the} [Fe/H] and [Si/H] of warm groups in Figure~\ref{fig.13} would decrease by $0.15$, or less than space between consecutive tick marks in the plot.  {The} [Si/H] would lie right on top of the {\it Suzaku} observations while the [Fe/H] would drop below the {\it Suzaku} results, but still be consistent with the observations given the factor $\sim 2$ uncertainties in the nucleosynthesis yields and supernovae rates.   The main point of this exercise is to demonstrate that the metal abundance results shown in  Figure~\ref{fig.13}, and by extension the abundance {ratios} that we will discuss next, are relatively insensitive to any suppression of star formation in the massive galaxies invoked to bring the overall stellar mass fraction in the groups in alignment with the observations (\cf~Figure \ref{fig.8}).

\begin{figure}
 \vspace{-4pt}
  \includegraphics[width=0.48\textwidth]{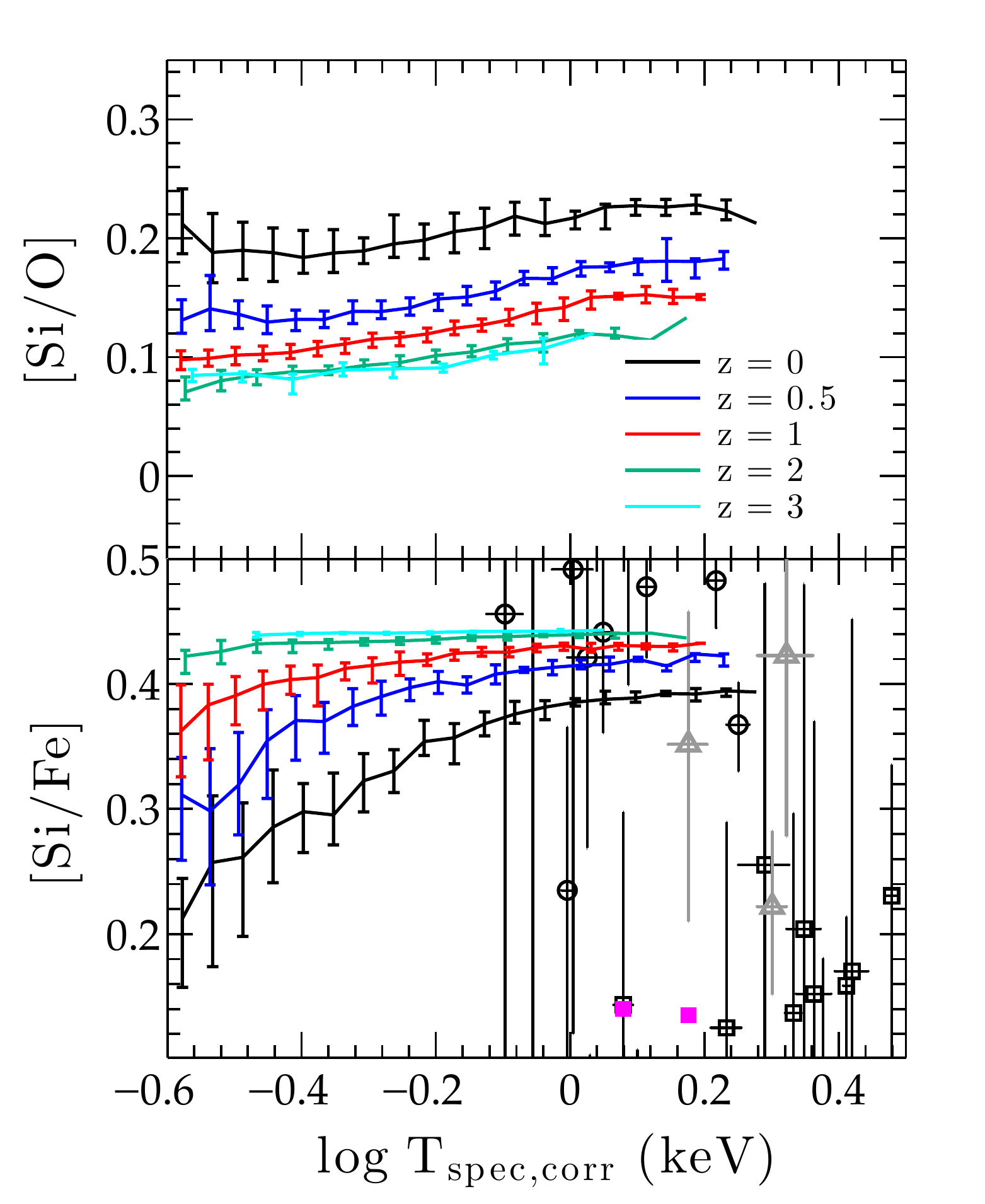}
\caption{Global silicon-to-oxygen  (top panel) and silicon-to-iron (bottom panel) abundance ratio within $R_{500}$.   The symbols show data from   \citet{RP09} (open black circles),  \citet{FU98} (open black squares), \citet{S14} (filled magenta squares), and \citet{TP03} (grey triangles).   The coloured lines and the corresponding error bars show the median values  and the 1-$\sigma$ dispersion for group populations in the simulation volume at $z=0$ (black),  $z = 0.5$ (blue), $z = 1$ (red), $z = 2$ (green) and  $z=3$ (cyan).  We point out that this $y$-axis scale is {\it not} the same as in  Figure~\ref{fig.13}.   We have deliberately zoomed in to highlight the differences between the curves.
}
\label{fig.15}
\end{figure}

\subsection{Abundance Ratios in the IGrM}
\label{Sec:5.3}

In Figure \ref{fig.15}, we examine the IGrM silicon-to-oxygen (top panel) and silicon-to-iron (bottom panel) abundance ratios within $R_{500}$ for the simulated groups at different epochs.   Silicon and oxygen are both $\alpha$-elements and are produced by core-collapse SNe.   As a result, in the absence of any other process, the silicon-to-oxygen abundance ratio would be expected to remain constant over time.   The ratio in our simulation, however, increases with time.   This is {a} consequence of silicon and oxygen being processed differently by AGB stars \citep{OD08}:   When AGB stars form, the silicon and oxygen present in the ISM is locked up in these stars.    Over their lifetime, these stars burn some of the oxygen while the silicon remains unaffected.  Consequently, when the AGB stars release their metals back into the ISM,  the amount of silicon is nearly the same as that locked up in the first place but the amount of oxygen returned is {reduced}.   The evolution in [Si/O] in Figure \ref{fig.15} {results from} this differential evolution.   As for the gentle rise in [Si/O] with increasing group temperature, this is a consequence of the four categories in  Figure~\ref{fig.14} not contributing identically to the silicon and oxygen mass.  We emphasize, however, that this increase with group temperature amounts to maximum change of $\Delta$[Si/O]$\approx 0.05$, which is insignificant.  For all {intents and purposes}, [Si/O] is independent of group temperature or mass.

In contrast, $z=0$ [Si/Fe] curve (lower panel) not only increases by $\Delta$[Si/Fe]$\approx 0.2$ in going from the coolest to the warmest groups in our simulation sample, this
change evolves from $\Delta$[Si/Fe]$\approx 0$ at $z=3$ to its present-day value while the value of [Si/Fe] in the warmest group drops slightly.   To zeroth order, the redshift evolution is due to delayed Type Ia SNe spewing new forged and winds from AGB stars spewing previously locked-up iron mass to their local environment.   Generally, this environment is the ISM within the group galaxies.   However, transporting this `late' iron out of the galaxies and into the IGrM is not straightforward.   As we have noted previously, metal-rich winds from the central galaxies in cool groups tend to behave more like galactic fountains and therefore, very little of the `late' iron production gets into the IGrM.   The central galaxies in the more massive groups are able to drive the iron into the IGrM but they are not outrightly dominant sources because they are running out of cold gas and hence, not forming stars as vigorously.  Moreover, because of the size of the galaxies,  the mass loading factor of the wind that is ejected is only a fraction of the star formation rate (\cf~Section~\ref{Sec:2.1}) and even in this case, the ejection of the `late' iron is inefficient.   However, winds are not the only way to enrich the IGrM.   Direct enrichment by the IGS component is another.  And as illustrated in Figure~\ref{fig.14}, the latter is the dominant source of iron mass in the cool groups  and also the reason why [Si/Fe] in cool groups evolves much more rapidly than in warm groups.

\begin{figure}%
         \centering
\captionsetup[subfigure]{labelformat=empty}
         \subfloat[][]{%
           \label{fig.15-a}%
\includegraphics[width=0.482\textwidth]{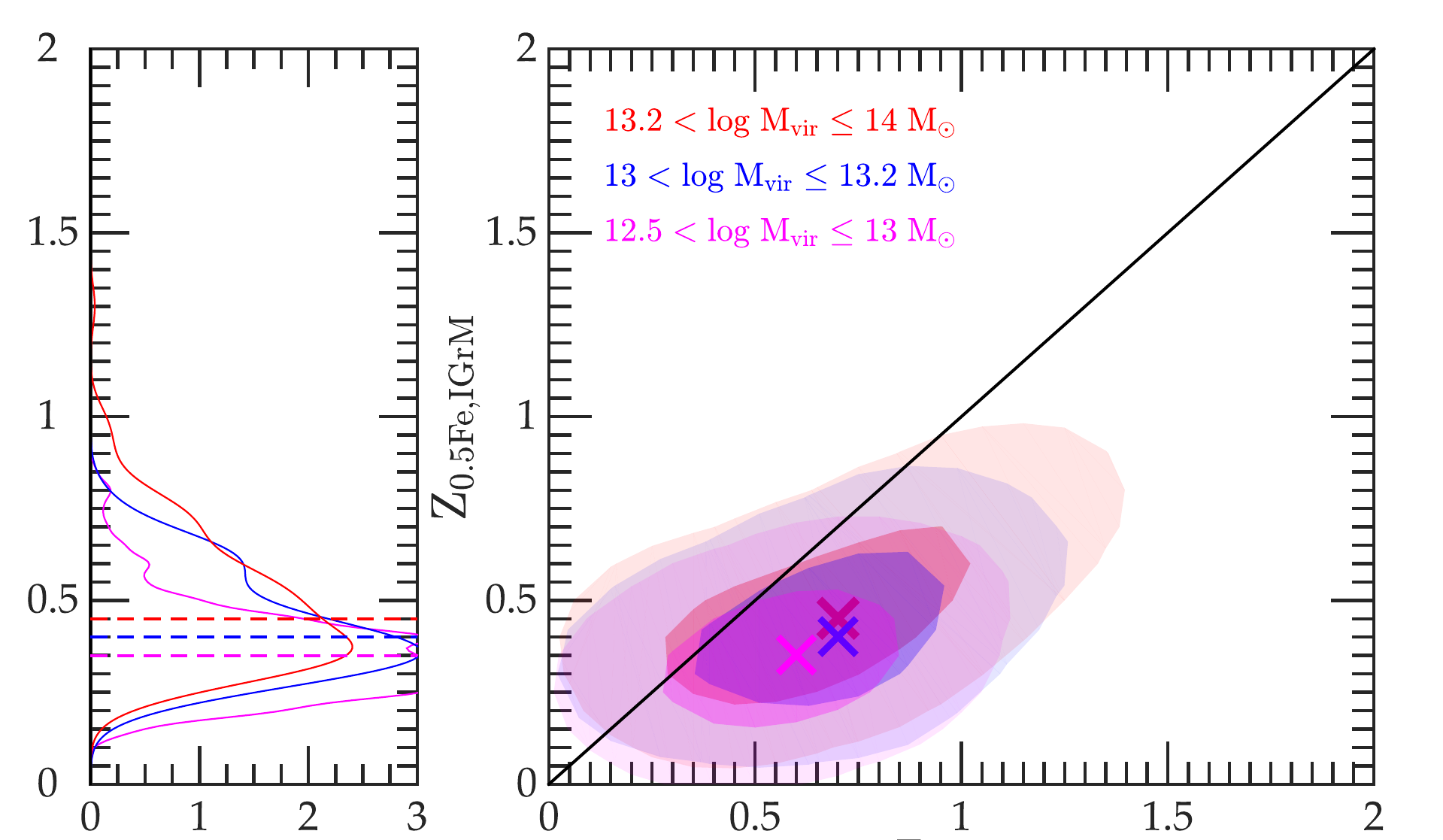}}\\
\vspace{-25pt}%
\captionsetup[subfigure]{labelformat=empty}
         \subfloat[][]{%
           \label{fig.15-c}%
\includegraphics[width=0.48\textwidth]{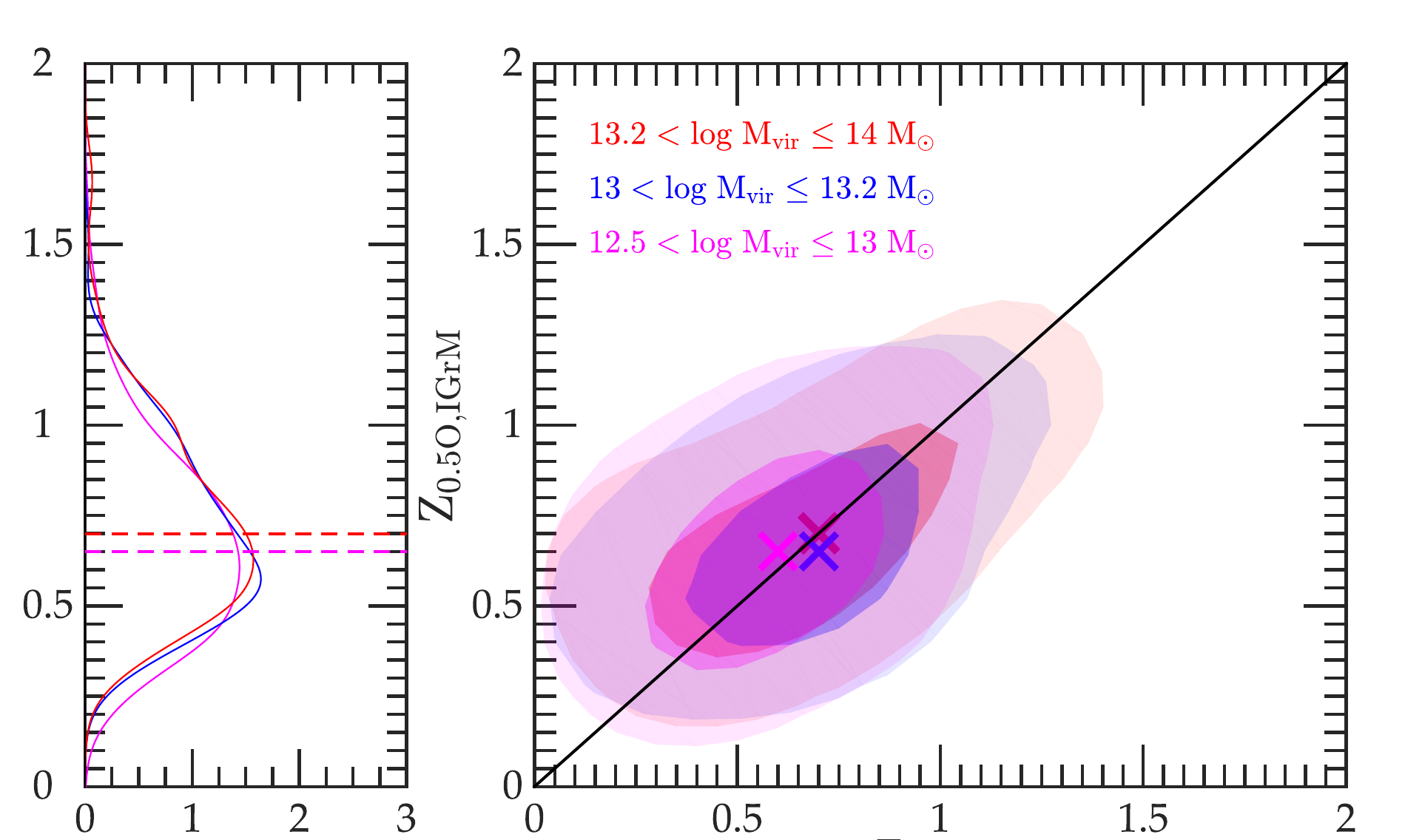}}\\
\vspace{-24pt}%
\captionsetup[subfigure]{labelformat=empty}
         \subfloat[][]{%
           \label{fig.15-e}%
\includegraphics[width=0.48\textwidth]{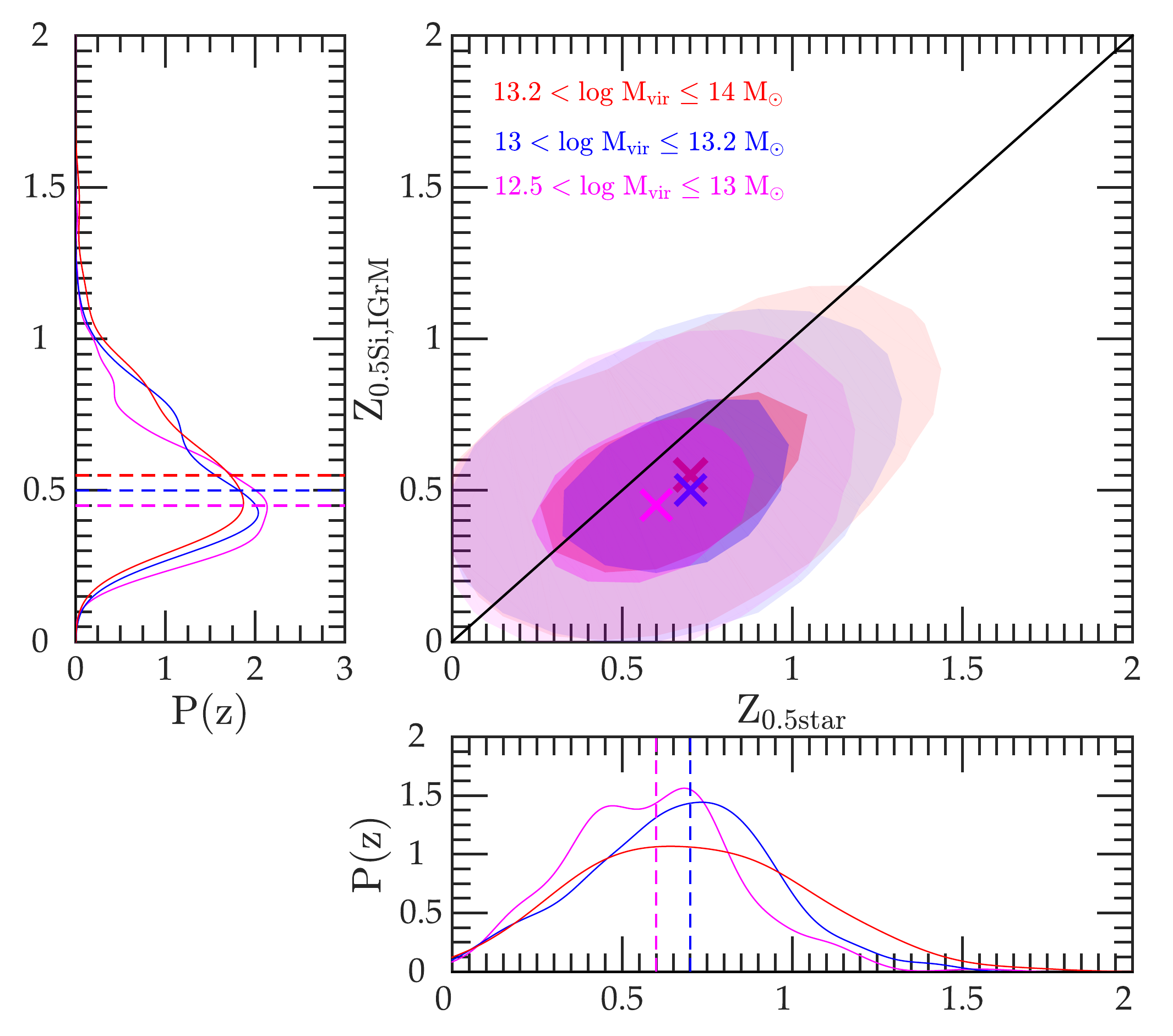}}%
\vspace{-19pt}%
         \caption{The joint distribution of $Z_{0.5\;  {\rm XX,IGrM}}$, the redshift by which half of the metals of species XX=$\{$Fe, O, Si$\}$ in a present-day group's IGrM  has been {\it forged} by the stars/supernova,  versus $Z_{0.5\;  {\rm star}}$, the distribution of redshifts by which half of the present-day group's stellar mass has been assembled in its MMP.    The contour plots show the 2D distribution of the redshifts for the low, intermediate and high mass groups \textendash{} \ie~$12.5 < {\rm log} \; M_{\rm vir} \leq 13.0 \;\rmn{M_{\odot}}$ (magenta), $13.0 < {\rm log} \;M_{\rm vir} \leq 13.2 \;\rmn{M_{\odot}}$ (blue), and $13.2 < {\rm log} \;M_{\rm vir} \leq14.0 \;\rmn{M_{\odot}}$ (red)  \textendash{} separately.   The inner and the outer contours of the shaded regions of each colour correspond to 1-$\sigma$ and 2-$\sigma$, while the $\times$ marks the median for the galaxies in each mass bin.   The panels to the left and below the contour plots show the normalized marginalized distributions of  $Z_{0.5\;  {\rm XX,IGrM}}$ (left), and  $Z_{0.5\;  {\rm star}}$ (below).    The different colour curves show the results for the low, intermediate and high mass groups, and the dashed lines indicate the median.
 }%
         \label{fig.16}%
      \end{figure}

\subsection{The Characteristic Timescales for Metal Enrichment of the IGrM}

We conclude our investigation of the metal enrichment of the IGrM by examining the redshifts by which half of the iron, silicon and oxygen mass in the $z=0$ IGrM within $R_{200}$ is produced by the stars, regardless of whether the metals are initially deposited in the ISM or introduced directly into the IGrM.   We refer to these characteristic redshifts as $Z_{0.5\;  {\rm Fe,IGrM}}$, $Z_{0.5\;  {\rm O,IGrM}}$ and $Z_{0.5\;  {\rm Si,IGrM}}$.   We show their distributions in Figure \ref{fig.16}, where we compare them to the redshifts by which half of the group's $z=0$ stellar mass has been assembled in its MMP ($Z_{0.5\;  {\rm star}}$).    There are several other characteristic redshifts, such as the halo formation redshift ($Z_{0.5\;  {\rm halo}}$) or the redshift at which the hot ($T>5\times 10^5\;$K) gas mass in the MMP is half of the group's final IGrM mass ($Z_{0.5\;  {\rm IGrM}}$),  {that we have compared $Z_{0.5\;  {\rm XX,IGrM}}$ (where XX=$\{$Fe, O, Si$\}$) against; however, we do not show these because they} do not offer any additional insights and Figure \ref{fig.10} offers a straightforward map between $Z_{0.5\;  {\rm star}}$ and the other potential redshifts of interest.

Examining the timescales in detail, the {most} significant feature {is} that in all but the most recently formed groups, the characteristic  `metal production' redshifts are lower than $Z_{0.5\;  {\rm halo}}$ or $Z_{0.5\;  {\rm IGrM}}$.   In other words, typically more than half of the iron, silicon and oxygen in the $z=0$ IGrM was forged after half of the groups' IGrM was already in place within the nascent groups.

The relationship between $Z_{0.5\;  {\rm XX,IGrM}}$, where XX=$\{$Fe, O, Si$\}$ and  $Z_{0.5\;  {\rm star}}$ is more nuanced.   In the case of iron (\cf~the top panel of Figure \ref{fig.16}), the major axis of the contours has a shallower slope relative to the one-to-one line.  Since $Z_{0.5\;  {\rm star}}$ is a measure of when stars first appear inside groups, the orientation of the contours indicates that the iron in the $z=0$ IGrM is typically made after half of the groups' final stellar mass is in place within the MMP.   This is not surprising.   As we have discussed previously, AGB stars and delayed Type Ia SNe play a key role in the build-up of iron in the IGrM and there is a lag between the formation of a stellar population and the start of enrichment by Type Ia SNe and AGB stars associated with that population.    However, if this was all, we would have expected the width of the contours at fixed $Z_{0.5\;  {\rm star}}$ to be fairly narrow.    Instead, as suggested by the left panel of Figure \ref{fig.14}, about 20\% of the $z=0$ IGrM iron  is forged by the stars/supernovae before the source galaxies become incorporated into groups.   In this case,  $Z_{0.5\;  {\rm star}}$ registers the stars in these external galaxies only after they fall in; \ie~if they fall in before $z=0$, while the metal production is registered whenever it happens.

Turning to the second panel of Figure \ref{fig.16}, we see that there {is} more of a one-to-one relationship between the timescale for oxygen production and $Z_{0.5\;  {\rm star}}$.  This alignment is the result of two effects acting in concert:  (i) Between 60-70\% of the oxygen in the $z=0$ IGrM is forged by stars that are already in the groups (\cf~the right panel of Figure \ref{fig.14}), and (ii) oxygen production by core-collapse Type II SNe and transport into the IGrM (by winds) are both tied directly to star formation, with no significant built-in lag between these events.  

Turning to the third panel of Figure \ref{fig.16}, we find that {the} orientation of the contours in the  $Z_{0.5\;  {\rm Si,IGrM}} - Z_{0.5\;  {\rm star}}$ plot is midway between that {of iron and oxygen}.    As discussed in the previous subsection, although silicon and oxygen are both produced in an identical manner, there is one significant difference.   AGB stars capture and retain a non-negligible amount of silicon present in the ISM at the time their progenitor stars form, and return it to their surroundings only after a lag time.  In effect, this means that silicon enrichment of the IGrM can proceed via two channels: the galactic wind and, as is the case with iron, via direct enrichment by the AGB stars in the IGS.   This late-time injection of silicon differentiates the evolution of  silicon from that of oxygen and shifts the distribution of $Z_{0.5\;  {\rm Si,IGrM}}$ slightly towards lower redshifts with respect to the distribution of $Z_{0.5\;  {\rm O,IGrM}}$.

\section{SUMMARY AND CONCLUSIONS}
\label{Sec:6}

There is a growing consensus that models of galaxy evolution \textendash{} or for that matter, models describing the formation and evolution of galaxy groups and clusters \textendash{} that do not allow for large-scale galactic outflows will fail to match the global evolutionary properties of galaxies.    Observations of both local as well as high-redshift galaxies indicate that not only are large-scale galactic outflows ubiquitous, but that they  have a profound impact on the conditions inside the galaxies as well as conditions in the wider environment around  galaxies.    For example, galactic outflows are thought to play a central role in establishing the observed mass-metallicity relation in galaxies, in promoting widespread enrichment of the IGM as far back as $z\sim 5$, and in accounting for the abundances and abundance ratios of $\alpha$ and iron-group elements observed in the hot diffuse X-ray emitting gas in galaxy groups and clusters.  Recent observations as well as theoretical models indicate that  AGNs and stars/SNe are both capable of driving powerful outflows and while definitive observational evidence outrightly favouring one over the other remains elusive, we argue that only stellar processes are capable of driving large-scale galactic outflows that can transport a significant fraction of the metal-enriched ISM out of the galaxies and into their halos and beyond.

In this paper, we use cosmological SPH simulations to document the impact of a well-studied galaxy formation model that incorporates stellar-powered, {galaxy-scale winds with momentum-driven scalings} \citep[see][and references therein]{SD14} on the global properties of galaxy groups over the redshift range $0 \leq z \leq 3$.   We look at  some of the commonly constructed X-ray scaling relations, the evolution of the hot gas, {the} stellar and the total baryon fractions, as well as the growth of {the} iron, silicon and oxygen abundances within the intragroup medium.  We examine the characteristic timescales for the emergence and the enrichment of this IGrM.   Since the present model does not include AGN feedback, we also take this opportunity to lay bare both the successes and the failings of stellar-powered winds so that we can identify precisely when, where and in what form AGN feedback {is} required, and we are using the resulting insights to guide the development of our own model of AGN accretion and feedback.  In this respect, the present study establishes a detailed baseline model of the galactic outflows as a prelude to a similar study including AGN feedback, although we expect that many of the conclusions will be robust to the inclusion of AGN feedback that quenches massive galaxies.

Our main findings are as follows:
\vspace*{-3.5mm}
\begin{description}[labelindent=\parindent,leftmargin=0.0cm, style=sameline]

\item[{\rm (i)}] {{The distribution of the groups' formation epochs can be reasonably approximated by a Gaussian with a median of $z\sim 1$ and $\sigma = 0.4$.  Moreover, the epoch when half of a present-day group's X-ray emitting,  intragroup medium ({\it i.e.}~the diffuse halo gas with $T > 5\times 10^5$ K) is in place is tightly correlated with the group's formation epoch.   Examining the emergence of the latter in more detail, we find that in halos with gravitational potential wells of a given depth, 
{the median IGrM mass fraction increases with time prior to $z\approx 1$ as the halos recapture the gas that was expelled out of the galaxy-scale halos at earlier epochs, (the ratio of baryons-to-dark matter of the infalling material is larger than the universal value)  and most of these baryons are shock-heated to roughly the virial temperature of the groups upon accretion.   After $z\approx 1$,  the IGrM fractions within $R_{200}$ cease to increase with time because the IGrM is sufficiently extended that the newly accreted (and thermalized) baryons primarily remain at  $r>R_{200}$ and become part of this extended halo gas distribution.  Apart from this trend with time, the IGrM mass fractions within $R_{200}$ also increase with halo mass at all epochs.  This increase is the result of the larger mass systems having deeper potential wells and higher virial temperatures.  Consequently, more of the diffuse gas is shock-heated to constitute the IGrM and the deeper potentials confine this gas more effectively. }}}

\item[{\rm (ii)}]{{Our stellar-powered,  momentum-driven wind model yields  X-ray scaling relations that are in excellent agreement with observed scaling relations ({\it e.g.}~X-ray luminosity-temperature, mass-temperature, entropy-temperature, etc.)  over much of the regime associated with galaxy groups despite the fact that the model does not include AGN feedback.  These scaling relations evolve self-similarly from $z=1$ to the present, as does $M_{\rm IGrM,200} / M_{200}$ versus $M_{200}$.  The hot, diffuse, X-ray emitting, intragroup gas is not subject to catastrophic cooling.  Typically only a percent or less of $z=0$ IGrM mass is lost via cooling over a Hubble time.  We do, however, see a tendency for the simulated galaxy groups to be slightly more X-ray luminous and/or  have slightly cooler X-ray spectroscopic temperature  than the observed groups on mass scales $M_{500} E(z) \simgreat 10^{14}\; \rmn{M_{\odot}}$.    At face value, these results collectively suggest that AGN feedback is not necessary to understand the properties of the hot diffuse gas in the simulated groups until the halos approach cluster-scale.}}

\item[{\rm (iii)}]{{Our simulation also successfully reproduces both the observed, spatially resolved, mass-weighted as well as the observed, unresolved, emission-weighted IGrM silicon and iron abundances within $R_{500}$.   This agreement also includes the observed trend of [Fe/H] and [Si/H] increasing with temperature until $\sim 1$ keV and then flattening.   Probing the origin of the metals in the IGrM in more details, we find that nearly $50\%$ of the IGrM silicon, oxygen and iron mass in our simulated groups are produced in the central and the satellite galaxies of a present-day group or its MMP, and infused into the IGrM via galactic outflows, while between $\sim 12\%$ (oxygen) to $\sim 30\%$ (iron) is transferred to the IGrM via direct enrichment by the IGS.}}

\item[{\rm (iv)}]{{Turning our attention to the group galaxies, we find that the stellar-powered, momentum-driven wind model results in a present-day stellar mass function for group galaxies that is in excellent agreement with the observations for $M_{*}<10^{11}\;\rmn{M_{\odot}}$; however, we also find galaxies \textendash{} typically, one per group and invariably, the group central galaxy \textendash{} that {have} much larger stellar {masses} than any observed galaxy.  These are the galaxies that are responsible for the elevated stellar and total baryonic mass fractions in our simulated groups.   Artificially reducing the stellar mass in only these `large' galaxies by a factor of $\sim 3$ reconciles the group stellar mass fractions with the observations across the entire mass range  $12.5 \leq \log(M_{*}) \leq 14.0$.}}

\item[{\rm (v)}]{{The excess stellar mass in these `large' group galaxies is due to galaxies no longer being able to moderate their star formation rates by depleting their cold gas mass via expulsion once they grow larger than $M_{*}\approx 4\times 10^{10}\;\rmn{M_{\odot}}$.   The deepening gravitational potential of these galaxies and a higher cross-section for hydrodynamic interactions between the galactic outflows and the shock-heated halo gas component, once the latter starts to form, confine the wind material within the circumgalactic region. Being metal-rich, most of this wind material cools down and falls back into the galaxy in a manner more akin to galactic fountains rather than outflows {\citep[\cf~][]{OD10}}.  The  high cold gas mass in our group central galaxies is largely a consequence of this, and the overproduction of stars is a byproduct.   The breakdown of the stellar-powered winds model in our giant group galaxies generally occurs {\it before} the galaxies are incorporated into bonafide groups and is the earliest indication that another feedback mechanism, like AGN feedback, is needed.}}

\item[{\rm (vi)}]{{We assert that in ÔlargeÕ galaxies, at least, AGN feedback cannot simply act to heat the halo gas just enough to offset the radiative cooling losses. This Òmaintenance-modeÓ or Òhot halo quenchingÓ feedback may represent a reasonable description of how AGN feedback operates in galaxy clusters and may even result in `large' galaxies with realistic stellar properties \citep[\cf~][and references therein]{GaD12,GD15}, but because this type of feedback is, in effect, only shifting baryons in our simulated groups from the galaxies to the IGrM component,  the total baryon fraction of the groups will not change.  Our simulated groups, however, already have too high a baryon fraction (\cf~ Figure~\ref{fig.8}).   At the same time, the IGrM mass fractions will become more discrepant and  a denser IGrM means that the simulated groups of a given mass will be  more X-ray luminous than their observed counterparts.  To ensure that both the stellar {masses of} the `large' galaxies and {the hot gas properties of} the groups agree with observations, AGN feedback (or for that matter, any new feedback mechanism or combination of mechanisms) must step in when stellar feedback starts to fail and continue to drive outflows beyond the galactic halos and perhaps even, beyond the low mass group halos.}}

\item[{\rm (vii)}]{{Finally, we emphasize that we do not expect the inclusion of AGN feedback, and the expected reduction in the stellar mass within the groups, to alter the agreement between our simulation results and the observed IGrM metal abundances (and abundance ratios) in galaxy groups.   Even after discounting the metals produced by 2/3 of the stars in the `over-sized' galaxies, the absolute abundances in the simulation are still consistent with the observations, especially when one accounts for the factor $\sim$2 uncertainty in the adopted nucleosynthesis yields and supernova rates.}}

\end{description}

\section*{Acknowledgments}
This research was enabled in part by support provided by WestGrid (www.westgrid.ca) and Compute Canada Calcul Canada (www.computecanada.ca), and by the National Science Foundation under Grant No. 1066293 to the Aspen Center for Physics.   AB, FD and LL acknowledge support from NSERC (Canada) through the Discovery Grant program.  RD acknowledges support from the South African Research Chairs Initiative, and from NASA grant NNX12AH86G.  NSK and MAF acknowledge support from NATA ATP grant NNX10AJ95G and NSF grant AST-1009652, respectively, to the University of  Massachusetts. {TQ was partially supported by NSF award AST-1311956.}  We would like to thank Shuiyao Huang and Jared Gabor for helpful discussions and  RD, AB and FD would like to acknowledge the simulating collaborative atmosphere of the Aspen Center for Physics, where this work was completed. {Finally, we would like to thank the anonymous referee for valuable comments that improved the clarity of the paper, as well as M.L. Balogh, J.M. Gabor and A.H. Gonzalez for providing us with access to detailed observational and simulation results in advance of publication. }

\bibliography{metal2}{}
\bibliographystyle{mnras}

\bsp
\label{lastpage}
\end{document}